\documentclass[showpacs,nofootinbib]{revtex4}
\usepackage{graphicx}
\usepackage{epsfig}

\begin{document}

\title{Cosmological constraints on non-standard inflationary quantum collapse models}

\author{Susana J. Landau}
\affiliation{Instituto de F{\'\i}sica de Buenos Aires, CONICET-UBA, Ciudad Universitaria - Pab. 1, 1428 Buenos Aires,
  Argentina} 
\author{Claudia G. Sc\'occola}
\affiliation{Instituto de Astrof{\'\i}sica de Canarias (IAC), C/V{\'\i}a L\'actea, s/n, E-38200, La Laguna, Tenerife, Spain. Dpto. Astrof{\'\i}sica, Universidad de La Laguna (ULL), E-38206 La Laguna, Tenerife, Spain}
\author{Daniel Sudarsky}
\affiliation{Instituto de Ciencias Nucleares, Universidad Nacional Aut\'onoma de M\'{e}xico, A. Postal 70-543, M\'{e}xico D.F. 04510, M\'exico}


\begin{abstract}
 We briefly review an important shortcoming --unearthed in previous
 works-- of the standard version of the inflationary model for the
 emergence of the seeds of cosmic structure.  We consider here some
 consequences emerging from a proposal inspired on ideas of Penrose
 and Di\'osi \cite{Penrose,Penrose01,Diosi87,Diosi89} about a
 quantum-gravity induced reduction of the wave function, which has
 been put forward to address the shortcomings, arguing that its effect
 on the inflaton field is what can lead to the emergence of the seeds
 of cosmic structure \cite{Perez2006}.

The proposal leads to a deviation of the primordial spectrum from the
scale-invariant Harrison-Zel'dovich one, and consequently, to a
different CMB power spectrum. We perform statistical analyses to test
two quantum collapse schemes with recent data from the CMB, including
the 7-yr release of WMAP and the matter power spectrum measured using
LRGs by the Sloan Digital Sky Survey. Results from the statistical
analyses indicate that several collapse models are compatible with CMB and LRG
data, and establish constraints on the free parameters of the models. The data put no restriction on the timescale for the
collapse of the scalar field modes.

\end{abstract}

\maketitle


\section{Introduction}

The great advances made in physical cosmology over the past few years
open a window for the consideration of issues long dismissed as
philosophical speculations. The agreement between theory and
observations of the spectrum of the cosmic microwave background (CMB)
anisotropies has strengthened the theoretical status of inflationary
scenarios among cosmologists. The inflationary paradigm is said to
account for the origin of all cosmic structure: the fluctuations
around homogeneity and isotropy that have a fundamentally quantum
nature are, according to these ideas, the seeds of galaxies, stars,
life and humans.  Hence, all structures in our universe emerge from a
featureless stage described by a background Friedmann-Robertson-Walker
(FRW) cosmology with a nearly exponential expansion driven by the
potential of a single scalar field\footnote{In the simplest models of
  inflation: $\phi$, the inflaton.}, and from its quantum fluctuations
characterized by a simple vacuum state.  In particular, the quantum
fluctuations transmute into the classical statistical fluctuations
that imprint their signature in the CMB photons, and eventually grow
into the structures we find in our universe today.

However, when this picture is considered thoroughly, an unavoidable
issue arises. According to these ideas, a completely homogeneous and
isotropic stage, somehow evolves, after some time, into an
inhomogeneous and anisotropic situation. Obviously, this is not simply
the result of quantum unitary evolution, since, in this case, the
dynamics does not break the initial symmetries of the system.  As
discussed in Ref.~\cite{Shortcomings}, and despite multiple claims to
the contrary, there is no satisfactory solution to this problem within
the standard physical paradigms.  While much of the focus of the
research in inflationary cosmology has been directed towards
elucidation of the details of the inflationary model, very little
attention has been given by the community to an issue of fundamental
principles such as the aforementioned one.

One proposal to handle this shortcoming has been developed in
Refs. \cite{Perez2006,Sudarsky06b,Sudarsky07}. That approach
attempts to deal with the problem by introducing a new ingredient into
the inflationary account of the origin of the seeds of cosmic
structure: \emph{the self induced collapse hypothesis}, i.e. a scheme
in which an internally induced collapse of the wave function of the
inflaton field\footnote{As shown in \cite{MukCollapso}, one can
  implement the collapse hypothesis at the level of the Mukhanov
  Sasaki variable.} is the mechanism by which inhomogeneities and
anisotropies arise at each particular length scale. That work was
inspired in early ideas by R. Penrose and L.  Di\'osi
\cite{Penrose,Penrose01,Diosi87,Diosi89} which regarded the collapse
of the wave function as an actual physical process (instead of just an
artifact of the description of Physics) and it is assumed to be caused
by quantum aspects of gravitation. We will not recapitulate the
motivations and discussion of the original proposal but instead refer
the reader to the above mentioned papers.

The way we treat this process is by assuming that at a certain stage
in cosmic evolution there is an induced jump in a state describing a
particular mode of the quantum field, in a manner that is similar to
the quantum mechanical reduction of the wave function associated with
a measurement, but with the difference that in our scheme no external
measuring device or observer is called upon as triggering such
collapse (as there is nothing in the situation we are considering that
could be called upon to play such a role).  The issue that then arises
concerns the characteristics of the state into which such jump occurs.
In particular, what determines the expectation values of the field and momentum
conjugate variables for the after-collapse state.  Previous works
by people in our group have extensively discussed both the conceptual
and formal aspects of that problem, and the present manuscript will
not dwell further into those aspects, except for a very short review.
   
In Ref.~\cite{Perez2006} two schemes were considered; one in which,
after the collapse, both expectation values are randomly distributed
within their respective ranges of uncertainties in the precollapsed
state, and another one in which it is only the conjugate momentum that
changes its expectation value from zero to a value in its
corresponding range, as a result of the collapse.  We will discuss the
motivations and detailed characterizations of the two processes in the
following sections.  As reported in \cite{Perez2006}, the different
collapse schemes give rise to different characteristic departures from
the conventional Harrison-Zel'dovich (HZ) flat primordial spectrum.
   
These aspects have been preliminarily analyzed in
Ref.~\cite{Unanue2008} using a simple approach which ignores the
late-time physics effects and simply considers what can be a
reasonable expectation for the allowed deviation from the conventional
flat spectrum.  The main objective of this article is to consider in
detail the shape of the primordial spectrum emerging from such a
scheme, and in particular to explore with precision the deviations
that should be expected once one takes into account the physics
associated with plasma dynamics, which is responsible for the
generation of the acoustic peaks, and other modifications to the
spectrum associated with well established physics. All of these must
be taken into account so that one can compare directly the predictions
corresponding to a particular collapse model to recent data from the
CMB fluctuation spectrum and the matter power spectrum from recent
galaxy surveys, and thus be able to put bounds on the model's
parameters.

The paper is organized as follows. In Sect. \ref{models}, we
describe the theoretical framework in which to study the models to be
tested.  Section \ref{Description_schemes} gives some specifications
about the proposed collapse schemes together with a brief physical
motivation for them, while Sect. \ref{Comparing_data} is dedicated
to general predictions of these models, and a broad comparison to
observational data. We define a fiducial model which is determined by
the best fit values obtained by the WMAP collaboration with the same
data used in this paper and without assuming any collapse of the wave
function. It should be noted that the fiducial model has a value of the
spectral scalar index $n_s$ different than 1, while calculations we have carried out here of
the primordial fluctuation spectrum for models including collapse are
restricted to the case $n_s=1$.  In Sect. \ref{resultados}, we use
data from WMAP 7-year release and the power spectrum of the Sloan
Digital Sky Survery DR7 LRG, to put bounds on the free parameters of
the collapse models, allowing also other cosmological parameters to
vary. In Section \ref{discusion}, we discuss our results and show our
conclusions.

\section{The Formalism/ Theoretical Model}
\label{models}

The starting point is the same as for the standard approaches, and in
particular we will focus on one of the simplest inflationary models
corresponding to a single scalar field, minimally coupled to gravity,
with an appropriate potential. The action for the theory is
\begin{equation}
  \label{eq:action}
  S[\phi, g_{ab}] = \int d^4x\sqrt{-g}\,\Big(
  \frac{1}{16\pi G} R[g_{ab}]-
  \frac{1}{2}\nabla_a\phi\nabla_b\phi g^{ab} -V(\phi)\Big).
\end{equation}
 As it is customary in such studies, we separate the fields into a
 ``background'' part, taken to be homogeneous and isotropic, FRW
 universe driven by an equally homogeneous and isotropic configuration
 of the inflaton field, and the perturbations (or
 ``fluctuations'')\footnote{In fact, the proposal has been developed
   to fit within a semiclassical treatment where gravitation is
   treated at the classical level but the scalar fields are completely
   described using quantum field theory on curved space-time in a
   self-consistent approach adapted to incorporate the collapse
   hypothesis. For a detailed description of this approach see
   \cite{Alberto}. }. In this way, the metric and the scalar field are
 written as: $g = g_0 + \delta g$ and $\phi = \phi_0 +
 \delta\phi$. This perturbative treatment requires, as usual, to deal
 with the gauge freedom, which we do by fixing the gauge (conformal
 Newton gauge). In the present work, we will ignore the vector and
 tensor parts of the metric perturbations. The space-time metric is
 described as: $\label{eq:conformal-newton} ds^2 = a(\eta)^2\left[
   -(1+2\phi)d\eta^2 + (1-2\Psi) \delta_{ij}\, dx^i dx^j \right], $
 where $\Psi$ and $\phi$ are the  \emph{Newtonian potentials}. In this case, as  it is  well-known \cite{Perez2006} the  field equations  imply $\phi = \Psi$ . The scale factor $a$ is normalized so that today $ a=1$.

In order to further specify the setting of the problem, let us recall
the standard inflationary version of cosmology.  The universe is
thought to go through the following epochs: 1) A quantum gravity
regime which leads to 2) a preinflationary classical space-time which
in turns enters 3) an inflationary regime with $a= a_{BI}$ (BI stands
for ``beginning of inflation'') and that ends when $ a=a_{EI}$ (EI
stands for ``end of inflation''), passing through a rapid 4) reheating
period, taken in our approximation to occur instantaneously. The next
epoch is 5) a radiation-dominated era starting at $ a=a_{EI}$, with
the radiation characterized by a temperature $ T=T_{GUT}$, and ending
at the transition $a_{eq}$ where 6)  the matter (and later also
  the dark energy) era begins.
 Particularly interesting for us will be the recombination time
 occurring during the radiation-dominated regime, with $ a=a_D$ and
 whose intersection with our past light-cone defines the surface of
 last scattering that is observed today in the CMB.

We will use a different conformal time coordinate for each of the
eras.  Thus, we will use $\eta$ as the conformal time during
inflation, $\tilde \eta$ the conformal time during radiation epoch and
$\bar \eta$ the conformal time during the matter and dark energy
epochs, and therefore the values of these coordinates at the
transition points will generally differ.

We choose the origin of $\eta$ so that during inflation we have
$a=\frac{-1}{H_I \eta}$, thus inflation will end at $ \eta=
\eta_{EI}=\frac{-1}{H_I a_{EI} } <0$.  On the other hand, we choose
the radiation-dominated era to start at $\tilde \eta =0$ so that
during the radiation epoch $ a= C \tilde \eta +B$, with $C=
[\frac{8\pi G}{3} ( \rho_{Rad} a^4)]^{1/2}$ and $B = a_{EI} $, where
$\rho_{Rad}$ is the radiation energy density so that $\rho_{Rad} a^4$
is constant during the era of radiation domination. Note that we are
considering the reheating to occur instantaneously and hence the end
of inflation (at $\eta= \eta_{EI}$) and the start of radiation
dominated era (at $\tilde \eta =0$) coincide, so that $ B= a_{EI}$.
  
 The study will rely on Einstein Field Equations (EFE) to zeroth and
 first order. The zeroth order gives rise to the standard solutions in
 the inflationary stage, where $a(\eta) = -\frac{1}{H_I\eta}$, with
 $H_I^2 \simeq (8\pi/3)GV$
with the scalar field $\phi_0$ in slow-roll regime,
so that $\phi_0' \simeq -\frac{1}{3H_I}\frac{dV}{d\phi}$;
and the first order EFE leads to an equation relating the
gravitational perturbation and the perturbation of the
field:
\begin{equation}
  \label{eq:classical_fundamental} \nabla^2\Psi = 4\pi
  G\phi_0'\delta\phi' \equiv s\delta\phi', 
\end{equation}
where $s \equiv 4\pi G\phi_0' $.  Next, we must consider the
quantization of the inflaton perturbation field.  It is convenient to
work with the rescaled field $y = a\delta\phi$. In order to avoid
infrared problems, we consider the system restricted to a box of side
$L$, where we impose, as usual, periodic boundary conditions. We thus
write the fields as
\begin{equation}
  \hat{y}(\eta,
  \vec{x}) = \frac{1}{L^3}\sum_k e^{i\vec k \cdot\vec x} \hat
  y_k(\eta), \quad \hat{\pi}(\eta, \vec{x}) = \frac{1}{L^3}
  \sum_k e^{i\vec k \cdot\vec x}\hat \pi_k(\eta),
\end{equation}
where $\hat\pi_k$ is the canonical momentum of the scaled field. The
wave vectors satisfy $k_iL = 2\pi n_i$, with $i = 1,2,3$. Also, as
usual, we write the field operators in terms of the time-independent
creation and annihilation operators, $\hat y_k(\eta) \equiv
y_k(\eta)\hat a_k + \bar y_k(\eta) \hat a_k^\dag$, and $\hat
\pi_k(\eta) \equiv g_k(\eta) \hat a_k + \bar g_k(\eta)\hat a_k^\dag$.
The functions $y_k(\eta), g_k(\eta)$ reflect the selection of the
vacuum state, and here we again proceed as in the standard approaches
and choose the so called Bunch-Davies vacuum:
\begin{equation}
  \label{eq:Bunch-Davies} y_k(\eta) = \frac{1}{\sqrt{2k}}
  \left( 1 - \frac{i}{\eta k }\right) e^{ -i k \eta }, \quad g_k(\eta)
  = -i \sqrt {\frac{k}{2}} e^{- i k \eta }.
\end{equation}

The vacuum state, defined by $\hat a_k |0\rangle =0$ for all $k$, is
exactly homogeneous and isotropic.  We assume that
at a certain time $\eta^c_k$ the part of the state characterizing the
mode $\vec k$ (we must of course be aware that the state of the field
is a collective state of all modes but taking here in this loose sense
will do no harm) jumps to a new state, which is no longer homogeneous
and isotropic.  The detailed description of this process is as
follows.

We decompose the fields into their hermitian part as: $\hat y_k = \hat
y_k^R(\eta) + i \hat y_k^I(\eta)$, and $\hat \pi_k = \hat
\pi_k^R(\eta) + i \hat \pi_k^I(\eta)$. We note that the vacuum state
$|0\rangle$ is characterized in part by the following: its expectation
values are $\langle \hat y_k^{R,I}(\eta)\rangle = \langle \hat
\pi_k^{R,I}(\eta) \rangle=0$, and its uncertainties are $\Delta\hat
y_k^ {R,I} = 1/2|y_k|^2(\hbar L^3)$ and $\Delta\hat \pi_k^{R,I} =
1/2|g_k|^2(\hbar L^3)$.

According to the collapse hypothesis, at some appropriate time
$\eta^c_{\vec k}$, the state undergoes an instantaneous jump to a
different state so that the mode $\vec k$ is not longer in its vacuum
state.

For any state of the field $|{\Omega}\rangle $,
we introduce the quantity $d_k \equiv \langle \Omega | \hat
a_k^{R,I}|\Omega \rangle \equiv |d_k^{R,I}| e^{i \alpha_k}$
so that, for that state,  we  have
\begin{equation}
  \label{eq:expectation_values} \langle \hat
  y_k^{R,I}\rangle = \sqrt 2 \, \Re \, (y_k d_k^{R,I}), \quad
  \langle \hat \pi_k^{R,I} \rangle = \sqrt 2 \, \Re \, (g_k
  d_k^{R,I} ),
\end{equation}
indicating that it specifies the main quantity of interest
in characterizing the state of the field.

The analysis proceeds with the specification of the scheme of collapse
determining the state of the field after the collapse\footnote{ At
  this point, in fact, all we require is the specification of the
  expectation values of certain operators in this new quantum state.}.
The detailed characterization of the schemes under consideration is
the main purpose of the next section.  With such a collapse scheme at
hand, one then proceeds to evaluate the perturbed metric using a
semiclassical description of gravitation in interaction with quantum
fields as reflected in the semiclassical EFE's: $G_{ab} = 8\pi G
\langle T_{ab} \rangle$ (see Eq. (\ref{eq:classical_fundamental}) for the classical first order version).  To lowest order, and for the quantity of
interest, this set of equations reduces to
\begin{equation}
  \label{eq:semiclassical-fundamental} \nabla^2\Psi_k =
  s\langle \delta\hat\phi'_k \rangle_\Omega,
\end{equation}
where $\langle \delta\hat\phi'_k \rangle_\Omega $ is the
expectation value of the momentum field $\delta\hat\phi_k' =
\hat\pi_k/a(\eta)$ on the state $|{\Omega}\rangle$
characterizing the quantum part of the inflaton field.  It
is worthwhile emphasizing that \emph{before} the collapse
 has occurred  \emph{there are no} metric perturbations,
i.e. the r.h.s. of the last equation evaluated on the vacuum state is zero, so, it is only
\emph{after} the collapse that the gravitational
perturbations appear, i.e. the collapse of each mode
represents the onset of the inhomogeneity and anisotropy at
the scale represented by the mode.  Another point we must
stress is that, according to   our views, at all times,
the \emph{Universe would be defined by a single state
  and not by an ensemble of states. The  precise  state  at  
  any  time, could  be  written if we  knew  the 
 modes  that  have collapsed  up to that time,  
 and the post-collapse   states for  each of these modes.}  The
statistical aspects arise once we note that we do not
measure directly and separately each of the modes with
specific values of $\vec k$, but rather the aggregate
contribution of all such modes to the spherical harmonic
decomposition of the temperature fluctuations of the
celestial sphere (see below).

In order to be able to compare to observations we note that the quantity that
is experimentally measured (for instance by WMAP satellite) is the
anisotropy in the temperature, $\Delta T(\theta, \varphi)/ T $, which
is expressed in terms of its spherical harmonic decomposition
as $\sum_{lm} a_{lm}Y_{lm}(\theta,\varphi)$.  The link to theoretical
calculations is made through the theoretical estimation of the value
of the $a_{lm}$'s, which are expressed in terms of the Newtonian
potential on the 2-sphere corresponding to the intersection of our
past light cone with the last scattering surface (LSS): $\Psi(\eta_D,
\vec x_D)$, $a_{lm} = \int \Psi(\eta_D, \vec x_D) Y_{lm}^*
d^2\Omega$. We must then consider the expression for the Newtonian
potential (Eq.~(\ref{eq:semiclassical-fundamental})) at those points:
\begin{equation}
  \label{eq:fundamental}
  \Psi(\eta, \vec x) = \sum_k \frac{s \mathcal{T}(k)}{k^2L^3} \langle
  \delta\hat\phi'_k \rangle
  e^{i\vec k \cdot \vec x},
\end{equation}
where we have introduced the factor $\mathcal{T}(k)$ to
represent the physical effects of the period between
reheating and decoupling.

Writing the coordinates of the points of interest on the
surface of last scattering as $\vec x = R_D(\sin\theta
\sin\phi, \sin\theta \cos\phi, \cos\theta)$, where $R_D$ is
the comoving radius to that surface, and ($\theta$, $\phi$) are
the standard spherical coordinates on the sphere, and using
standard results connecting Fourier and spherical expansions,
we obtain
\begin{equation}
\label{alm}
  a_{lm} = \sum_k\frac{s \mathcal{T}(k)}{k^2L^3} \int
  \langle\delta\hat\phi_k'\rangle e^{i\vec k \cdot \vec x}
  Y_{lm}(\theta,\phi) d^2\Omega.
\end{equation}
As indicated above, statistical considerations arise when noting that
Eq.~(\ref{alm}) indicates that the quantity of interest is
in fact the result of a large number (actually infinite) of harmonic
oscillators, each one contributing with a complex number to the sum,
leading to what is in effect a two dimensional random walk whose total
displacement corresponds to the quantity of observational interest.
Note that this part of the analysis is substantially different from
the corresponding one in the standard approach.  In order to obtain a
prediction, we need to find the {\it magnitude} of such total
displacement, i.e. $|a_{lm}|^2$, and given that by assumption there are
multiple random processes involved, the best that can be obtained is
the ``most likely'' value for this quantity (just as in a random walk
one can not expect to predict the actual value but at best its most
likely value). We do this with the help of the {\it imaginary}
ensemble of universes\footnote{This is just a mathematical evaluation
  device and no assumption regarding the existence of such ensemble of
  universes is made or needed. These aspects of our discussion can be
  regarded as related to the so called cosmic variance problem.}  and
the identification of the most likely value with the ensemble mean
value.  We should emphasize however that if we knew the specific
values taken by the random numbers $x$ we will specify below, we would
be able to compute explicitly the value of each $a_{lm}$ including the
phase.

As we will see, the ensemble mean value of the product
$\langle\delta\hat\phi_k\rangle \langle\delta\hat\phi_{k'}\rangle^*$,
evaluated in the post-collapse states\footnote{ Note here again the
  difference with the standard treatment of this part of the
  calculation, which calls for the evaluation of the expectation value
  $\langle\delta\hat\phi_k\delta\hat\phi_{k'}\rangle^*$ on the vacuum
  state which as already emphasized is completely homogeneous and
  isotropic.}, results in a form $\kappa C(k) \delta_{\vec{k}\vec{k}'}
$, where $\kappa = \hbar L^3 k / (4 a^2)$ and $C(k)$ is an
adimensional function of $k$ which encodes the relevant information
of detailed aspects of the collapse scheme. Then, using the expression
for $|a_{lm}|^2$, writing the sum as an integral, and doing a change
of integration variable $x = k R_D$, we arrive to the following
expression for the most likely (ML) value of the quantity of interest:
\begin{equation}\label{eq:contacto_obs}
  |a_{lm}|^2_{ML}  = \frac{s^2\hbar}{2 \pi a^2}
  \int   \frac{C(x/R_D)}{x} \mathcal{T}(x/R_D)^2
  j_l^2(x)  dx.
\end{equation}
With this expression at hand, we can compare the expectations from
each of the schemes of collapse against the observations. We note, in
considering the last equation, that the standard form of the spectrum
corresponds to replacing the function $C$ by a constant. In fact if
one replaces $C$ by $1$ and, furthermore, one takes the function
$\mathcal{T}$ which encodes the late-time physics including the plasma
oscillations which are responsible for the famous acoustic peaks, and
substitutes it by a constant, one obtains the characteristic signature
of a scale invariant spectrum: $ |a_{lm}|^2_{ML} \propto
\frac{1}{l(l+1)}$. Maintaining the appropriate function $\mathcal{T}$
leads of course to the well-known spectral shape that fits the data
quite well once some basic cosmological parameters have been
appropriately adjusted.

In the remainder of the paper, we will focus on the effects that a
nontrivial form of the function $C$ has on the predicted form of the
observational spectrum, and on using data to constrain aspects of the
collapse models.

\section{Description of the proposed collapse schemes}

\label{Description_schemes}

A collapse scheme is a recipe to characterize and select the state
into which each of the modes $\vec k$ of the scalar field jumps at the
corresponding time of collapse $ \eta^c_{\vec k}$.

As we have clearly stated, we do not know exactly what kind of
physical mechanism would lie behind what, at the semiclassical level
we are working, looks like a spontaneous collapse of the wave
function. Thus, specifying the collapse scheme should be taken as
being at this point purely guesswork, which we hope to address with as
much physical intuition as possible.  The aspect that makes such
efforts worthwhile is the fact that, as we shall see, the various
specific collapse schemes lead generically to different generic
patterns of deviations of the form of the primordial spectrum from the
standard HZ scale-free shape.

The differences in the form of the primordial spectrum lead in turn to
characteristic deformations for the predictions of the observational
spectrum which is the result of late-time and well understood physical
effects on the primordial spectrum. Thus, observations can help us to
determine which one of the naively guessed collapse schemes is favored
by the data.

In this paper, we focus on two schemes which we characterize simply in
terms of the expectation values of the fundamental field operators
(field and momentum conjugated) for the various modes in the state
just after the corresponding collapse. It turns out that this limited
characterization of the state (together with the times of the
collapses) is all one needs to compute the shape of the expected
spectrum. The higher level characterization of the states, such as the
uncertainties and field-momentum correlations, turns out to be relevant
only when one considers the possibility of multiple collapses per mode
(see \cite{multiples}) something that lies outside the scope of the
present work.
\\
{ }

\noindent{\bf Scheme I:} It is the scheme where both the expectation
  value of the field and the expectation value of the conjugated
  momentum change as a result of the collapse in a random uncorrelated
  manner dictated by the uncertainties of the precollapse state, so
  that, immediately after the collapse time $\eta_k^c$, the
  expectation values are determined by
\begin{equation}
  \label{eq:collapse_scheme_1}
  \left\langle \hat
    y_k^{(R,I)}\left(\eta_k^c\right)\right\rangle_\Omega =
  x_1^{(R,I)}
  \sqrt{\left(\Delta y_k^{(R,I)}\right)_0^2}, \qquad   \left\langle
    \hat \pi_k^{(R,I)} \left(\eta_k^c\right)
  \right\rangle_\Omega =   x_2^{(R,I)}
  \sqrt{\left(\Delta \pi_k^{(R,I)}\right)_0^2} \, ,
\end{equation}
where $x_{1,2}^{(R,I)}$ are random variables, characterized by a
Gaussian distribution centered at zero with dispersion equal to 1, $\Delta y_k^{(R,I)}$ and $\Delta \pi_k^{(R,I)}$ are the uncertainties of $y_k^{(R,I)}$ and $\pi_k^{(R,I)}$ respectively in the precollapse state.
This is, in a sense, a very simple scheme and one could argue that it
represents a natural prescription that treats the field and conjugate
momenta on an equal footing.  However, a close examination
(specifically, a close look at Eq.~(\ref{eq:classical_fundamental}))
indicates that field and momentum conjugate play rather different
roles in determining the gravitational perturbation that results after
the collapse.  There we see that the momentum conjugate is the
quantity that determines the Newtonian potential.

This leads us to consider a collapse scheme in which it is only the
expectation value of the conjugate momentum the one that changes as a
result of the collapse (we would be taking here a view according to
which, heuristically speaking, the uncertainty of the source of the
metric perturbation, the Newtonian potential, is somehow connected to
the triggering of the collapse of the source).  See a more detailed
discussion of these ideas in \cite{Perez2006} and the motivating ideas
in \cite{Penrose,Penrose01,Diosi87,Diosi89}.  Thus we define: 
\\ { }

\noindent{\bf Scheme II:} It is the scheme where it is only the
expectation value of the conjugated momentum the one that changes as a
result of the collapse, and it does so in a random way, dictated by
the corresponding uncertainties of the precollapse state, so that,
immediately after the collapse time $\eta_k^c$, the expectation
values are determined by
\begin{equation}
  \label{eq:collapse_scheme_2}
  \left\langle \hat
    y_k^{(R,I)}\left(\eta_k^c\right)\right\rangle_\Omega =
 0,  \qquad  \qquad  \left\langle
    \hat \pi_k^{(R,I)} \left(\eta_k^c\right)
  \right\rangle_\Omega =   x^{(R,I)}
  \sqrt{\left(\Delta \pi_k^{(R,I)}\right)_0^2} \, ,
\end{equation}
where $x_{1,2}^{(R,I)}$ are random variables, characterized by a
Gaussian distribution centered at zero with 
dispersion of 1, and $\Delta \pi_k^{(R,I)}$ is the uncertainty of $\pi_k^{(R,I)}$ in the precollapse state.

We should note that an additional and crucial piece of information
would be required to determine the primordial spectra: the exact
values of the collapse times for each mode.  Again, having no knowledge
of the physics behind the collapse, all we can do is to assume that
whatever that is, it has no intrinsic preferential directionality and
thus that $\eta^c_{\vec k}$ is only a function of $k=||\vec k||$, but
otherwise take that function as unknown.  We will then
parametrize in simple ways our ignorance about this function and
attempt to extract constraints from the data.

There are of course many more possibilities of collapse schemes but we
 limit our consideration to only those two, because they seem simple
and quite natural.  Next, we investigate these two schemes in
detail at a quantitative level.

\section{Comparing with observational data}

\label{Comparing_data}

Given the schemes proposed, and after some lengthy algebraic
manipulations, one can determine (see ref. \cite{Unanue2008}) the functions $C(k)$'s and then use
them to compare the resulting predictions for the shape of the
spectrum to observations. It was shown in Refs. \cite{Perez2006,Unanue2008} that we can recover the power spectrum of initial fluctuations of the standard cosmological model if $C(k)=1$. Therefore, the power spectrum of the collapse models can be written as:

\begin{equation}
P(k) = A_s k C(k)
\end{equation}

where $A_s$ is the amplitude of the scalar fluctuations.

The function $C(k)$ resulting
from Eq.~(\ref{eq:collapse_scheme_1}) for scheme I has the following form:
\begin{equation}\label{C1_sudarsky}
 C_{\rm I}(k) = 1 + \frac{2}{z_k^2}\sin^2\Delta_k +
 \frac{1}{z_k}\sin(2\Delta_k),
\end{equation}
whereas the one corresponding to the scheme II is:
\begin{equation}\label{C2_sudarsky}
  C_{\rm II}(k) = 1 + \left(1 - \frac{1}{z_k^2}\right) \sin^2\Delta_k 
  - \frac{1}{z_k}\sin (2\Delta_k), 
\end{equation}
where $z_k \equiv k \eta_k^c$, $\Delta_k = k (\eta_{EI} -\eta_k^c)$ and
$\eta_{EI}$ refers to the conformal time at the end of the inflationary period \footnote{This expression is different from the one in previous papers \cite{Perez2006,Unanue2008}, because in those, the work relied in an aproximation in which  the effects of the plasma physics  relevant to the era between the end of inflation and decoupling are ignored. }. We refer to them
hereafter as Model I and Model II, respectively.

 In Ref.~\cite{Unanue2008} a preliminary study of these two schemes and one
 additional scheme was performed. That was a relatively simple
 analysis concentrating on the main features of the resulting
 spectrum, but ignoring the late-time physics corresponding to the
 effects of reheating and acoustic oscillations (represented by
 $\mathcal{T}(k)$).  Thus, the actual comparison with empirical data
 was not possible except for obtaining order-of-magnitude estimates.
 A detailed comparison with the very precise data available, requires
 a much  more complex analysis,  such as the one we will conduct in this work.

\vspace{0.3cm} We recall that the standard form of the predicted
spectrum is recovered if one replaces $C(k)$ by a constant.  We will explore the sensitivity for small
deviations of the ``$z_k$ independent of $k$ pattern'' by considering
a linear departure from the situation in which $z_k$ is independent of
$k$. That will be characterized by $z_k$ as $z_k = A + B k $.  This
will allow us to examine the robustness of the collapse scheme in as
far as predicting the standard spectrum. In such way, the times of
collapse for each mode $k$ can also be written in terms of $A$ and $B$
as follows: $\eta_k^c = A/k + B$.  We can use this formula to compute
the collapse time for the relevant modes we observe in the CMB, namely
those that cover the range of the multipoles of interest, $1 \le l \leq
2600$. We can made use of the approximate relation\footnote{The
  relation between the angular scale $\theta$ and the multipole $l$ is
  $\theta \sim \pi/l$. The comoving angular distance, $d_A$, from us
  to an object of physical linear size $L$, is $d_A =
  L/(a\theta)$. $L/a \sim 1/k$, $d_A = R_D$ if the object is in the
  LSS, and using the first expression in this footnote, we get $l \sim
  \pi k R_D $.} $\,l \sim \pi k R_D$, where $R_D$ is the \emph{comoving} radius
of the last scattering surface, in order to interpret heuristically the
result of the analysis and to set reasonable viability constraints on
the parameters of the model. In standard cosmology\footnote{ In the
  present work we will be ignoring the effects of the late time
  acceleration associated with the so called ``Dark Energy" as this
  complication is thought not to impact on the results in any
  substantial way.} it is given by $R_D = \frac{2}{H_0}\left( 1 -
\sqrt{a_D}\right)$, where we have normalized the scale factor to be
$a_0 = 1$ today, so $a_D \equiv a(\eta_D) \simeq 10^{-3}$, and $H_0$
is the Hubble constant today. Its numerical value is $R_D = 5816.31
\,\ h^{-1}$ Mpc. Thus, the relevant modes for the CMB are those in the
range $4 \times 10^{-5}$ Mpc$^{-1}$ $\leq k \leq 0.11$ Mpc$^{-1}$.
The collapse times for these modes can be regarded as the times in
which inhomogeneities and anisotropies first emerged at the
corresponding scales.

\subsection{Physical meaning of the obtained results}
\label{priors}

The analysis carried out in previous works on this approach
\cite{Perez2006, Unanue2008} does not take into account the evolution
of the perturbation beyond the end of inflation and thus their
results, even though they provide certain general qualitative and
quantitative information about the modifications the general approach
leads to, cannot be considered as leading to actual bounds on the
parameters extracted from data.  In this work we want to extract
actual bounds that could, in the future, be taken as clues on the
nature of the physics that might lie behind what we represent at the
phenomenological level by the collapse mechanism.  In particular, we
need to ensure that when considering the collapse as being described
within the inflationary regime, the times where the relevant collapses
occur do indeed fall within that regime.  Hence, we will impose { \it
  a priori} limits on the values of $A$ and $B$, so that the collapse
of the wave function can occur only at a time between the beginning
($t_{BI}$) and the end ($t_{EI}$) of the inflationary epoch. In terms
of conformal times:
\begin{equation}
\eta_{BI} < \eta < \eta_{EI}.
\end{equation}

The period of inflationary expansion prior to the radiation dominated
era, corresponds to negative conformal time. The initial singularity
(or more precisely, the quantum gravity regime) is pushed back into
large and negative values of the conformal time and can be pushed
arbitrary far depending on the duration of inflation \cite{Kinney09}.

 Now, we estimate the time scales for the beginning and the end of the
 inflationary epoch.  The value of the conformal time at the end of
 inflation, is determined from the "temperature" of the radiation era
 viewed as a function of the scale factor: $ a(t_0) T(t_0) = a(t_{EI})
 T(t_{EI}) $ where $t_0$ is the present time, $a(t_0)=1$, $t_{EI}$ is
 the time at the end of inflation, and $T_0=2.728 K$.  Taking the
 usual values for inflationary models, $T({t_{EI}}) \simeq 10^{15}
 {\rm GeV}$, and consequently $a(t_{EI})=2.35 \times 10^{-28}$. We
 then use the relation between conformal time and scale factor during
 the inflationary era, $a(\eta) = -\frac{1}{H_I \eta}$, where $H_I$ is
 the value of the Hubble constant during this epoch. In most
 inflationary models, $H_I$ can be estimated as: $ H_I^2 = \frac{8 \pi
   G}{3} V_I $ where $V_I$ is the inflationary potential. Let us
 assume that $V_I \simeq 10^{60}\, {\rm GeV^4}$, then we obtain $H_I =
 2.37\, \times 10^{11}\, {\rm GeV} $ and thus
\begin{equation}
\eta_{EI} = -1.8\, \times 10^{16}\, {\rm GeV^{-1}} = - 7.2  \times 10^{-22}\, {\rm Mpc}.
\end {equation}
We express the results in ${\rm Mpc}$ because these are the
common units  for time and length used in Boltzmann codes that solves the
recombination equations, such as CAMB \cite{LCL00}. In order to solve
the ``horizon'' and ``flatness'' problem, inflationary models require
$80$ e-folds of inflation:
\begin{equation}
\log{\left(\frac{a(t_{EI})}{a(t_{BI})}\right)}= H_I (t_{EI} - t_{BI}) = 80.
\end{equation}
Thus, we obtain: $a(t_{BI})= 4.24 \times 10^{-63}$, and using the
relation between the scale factor and conformal time during inflation,
we get:
\begin{equation}
\eta_{BI}= -9.9 \, \times 10^{50} {\rm GeV^{-1}}=  -1.6\, \times 10^{14} {\rm Mpc}.
\end{equation}

Let us recall now that in the models that are being studied, each mode  is  thought to
collapse at a different time  given  by 
\begin{equation}
\eta_k^c = \frac{A}{k} + B,
\label{tcolapse}
\end{equation}
where $A$ and $B$ are constants.  Therefore, for the collapse of the
wave function to occur during the inflationary period, the values of
$A$ and $B$ should satisfy the following relation for all values of
$k$:
\begin{equation}
\eta_{BI} - \frac{A}{k} < B < \eta_{EI} - \frac{A}{k}. 
\label{Bphys}
\end{equation}
 Demanding that these inequalities hold for all the relevant modes,
 leads to the desired {\it a priori} bounds on the model parameters.

\subsection{Characteristic  signatures of the models on the CMB fluctuation spectrum}
\label{effects}

 In order to analyze the effects on the CMB fluctuations power
 spectrum, let us first define the fiducial model, which will be taken
 just as a reference to discuss the results we obtain for the collapse
 models. The fiducial model is a $\Lambda$CDM model with the following
 cosmological parameters: baryon density in units of the critical
 density, $\Omega_B h^2=0.02247$; dark matter density in units of the
 critical density, $\Omega_{CDM} h^2=0.1161$; Hubble constant in units
 of ${\rm Mpc^{-1} km \ \ s^{-1}}$, $H_0=68.7$; reionization optical
 depth, $\tau=0.088$; and the scalar spectral index,
 $n_s=0.959$. These are the best-fit values presented by the WMAP
 collaboration using the final 7-year release data \cite{wmap7} and
 the power spectrum from Sloan Digital Sky Survery DR7 LRG
 \cite{Reid09}.

\begin{figure}[ht!]
\begin{center}
\includegraphics[scale=0.3,angle=-90]{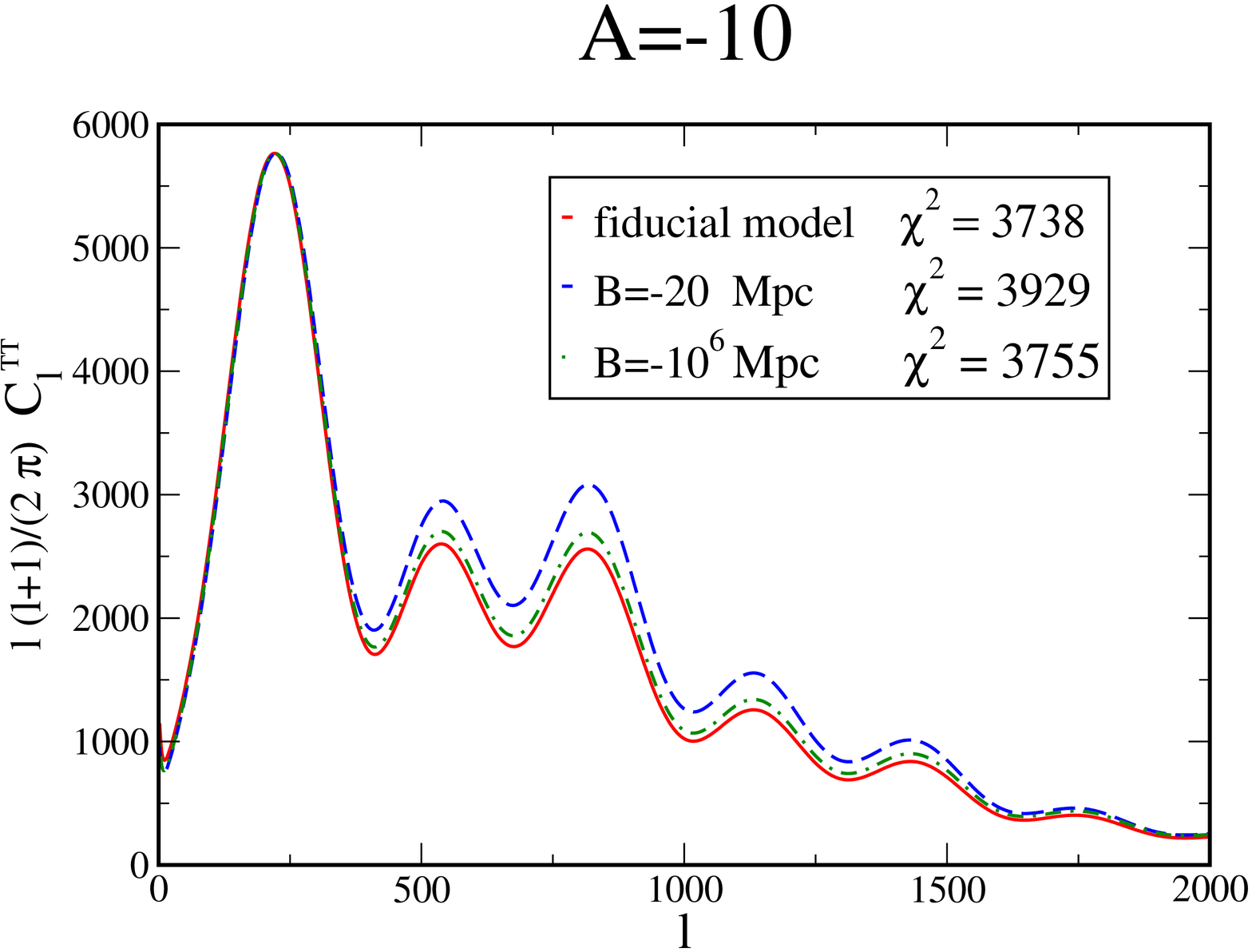}
\includegraphics[scale=0.3,angle=-90]{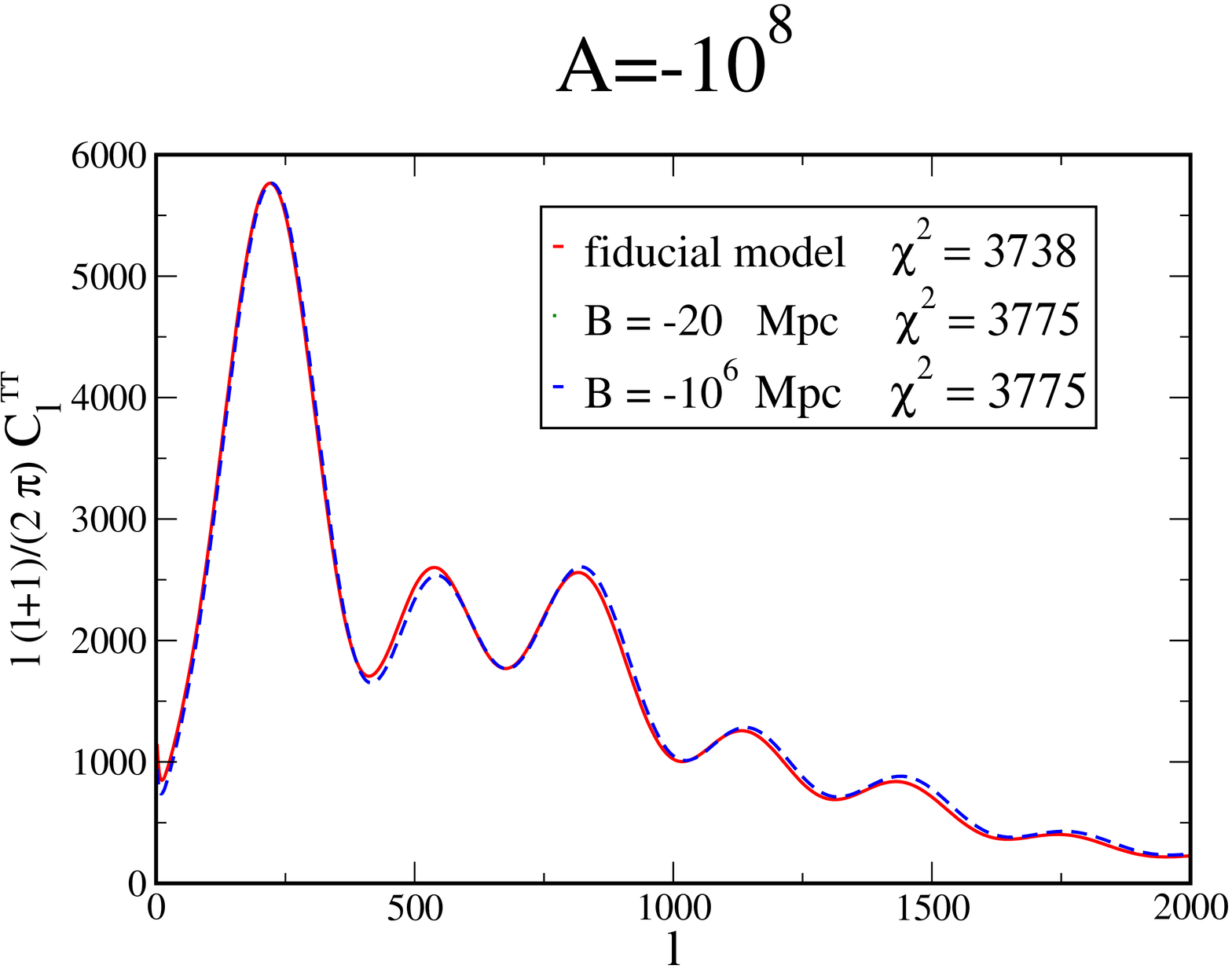}
\end{center}
\caption{The temperature autocorrelation (TT) power spectrum for
  Model I (Left: $A=-10$, Right: $A=-10^8$). All models are normalized
  to the maximum of the first peak of the fiducial model. The value of
  $\chi^2$ is calculated using only WMAP 7-year release data. The solid line corresponds to the fiducial model.}
\label{cttmodelo1}
\end{figure}

\begin{figure}[ht!]
\begin{center}
\includegraphics[scale=0.3,angle=-90]{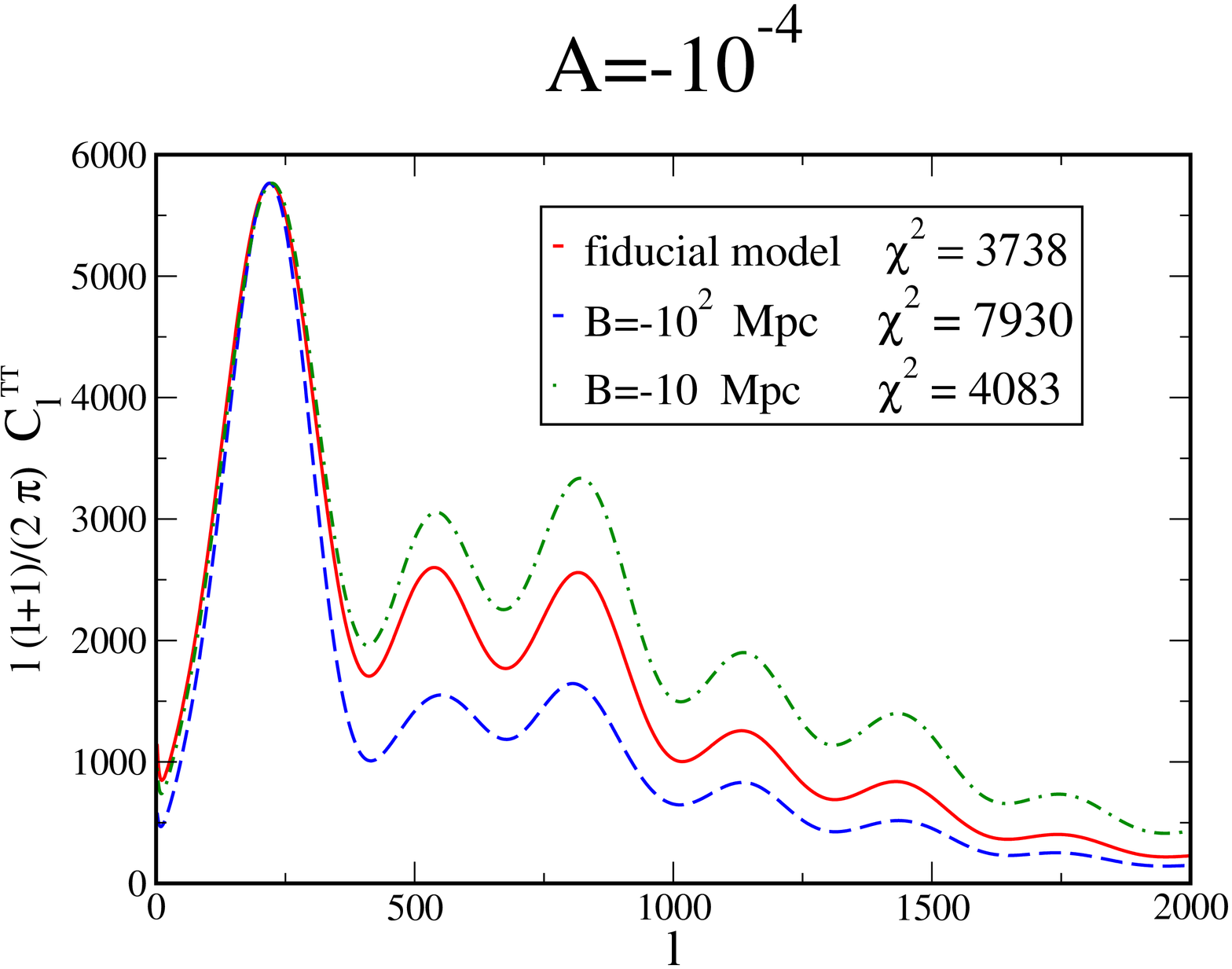}
\end{center}
\caption{The temperature autocorrelation (TT) power spectrum for
  Model I ($A=-10^{-4}$). All models are normalized to the maximum of
  the first peak of the fiducial model. The value of $\chi^2$ is
  calculated using only WMAP 7-year release data (both temperature and temperature-polarization power spectrum are included). The solid line corresponds to the fiducial model.}
\label{cttmodelo1v2}
\end{figure}

\begin{figure}[ht!]
\begin{center}
\includegraphics[scale=0.3,angle=-90]{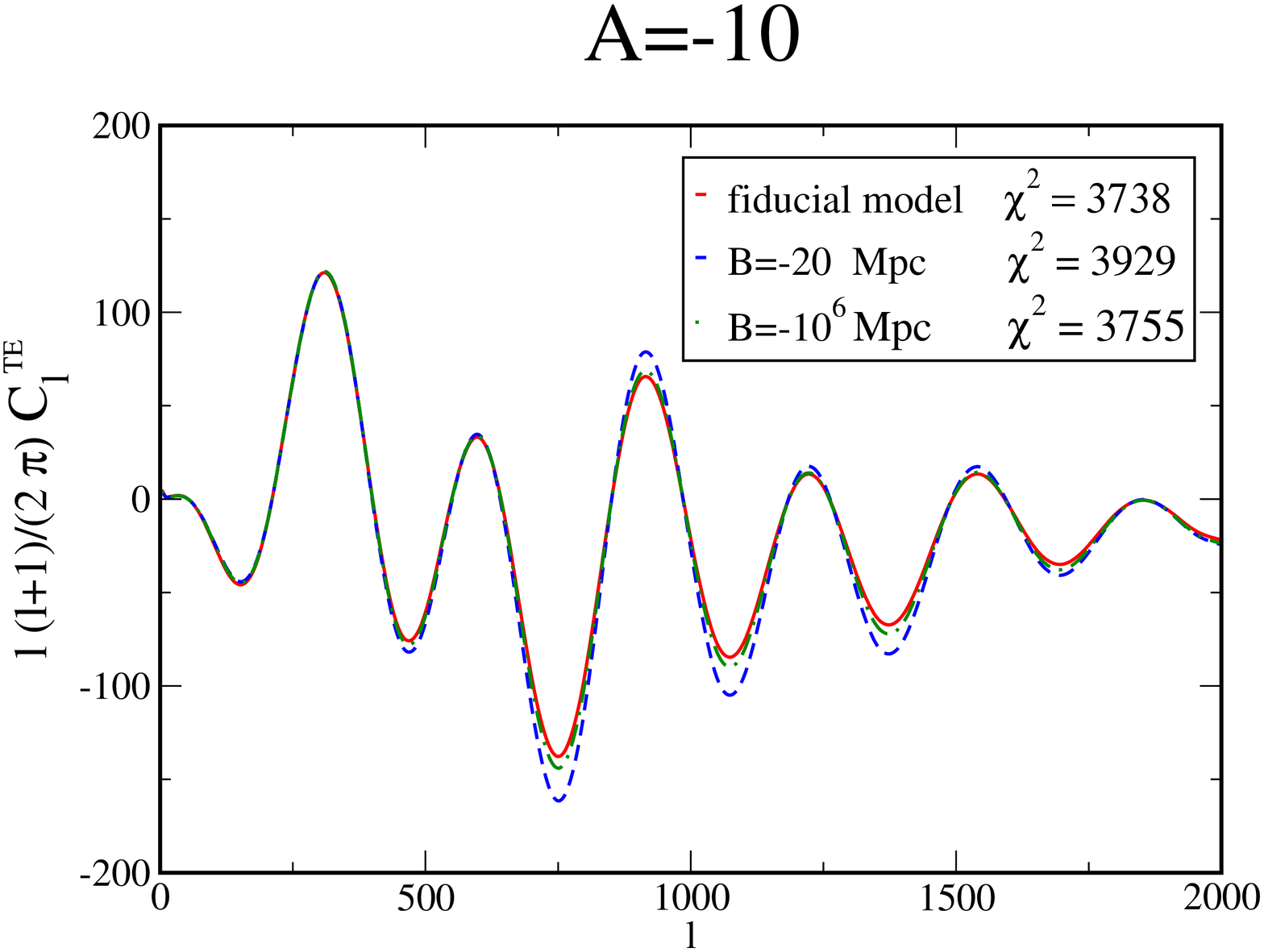}
\includegraphics[scale=0.3,angle=-90]{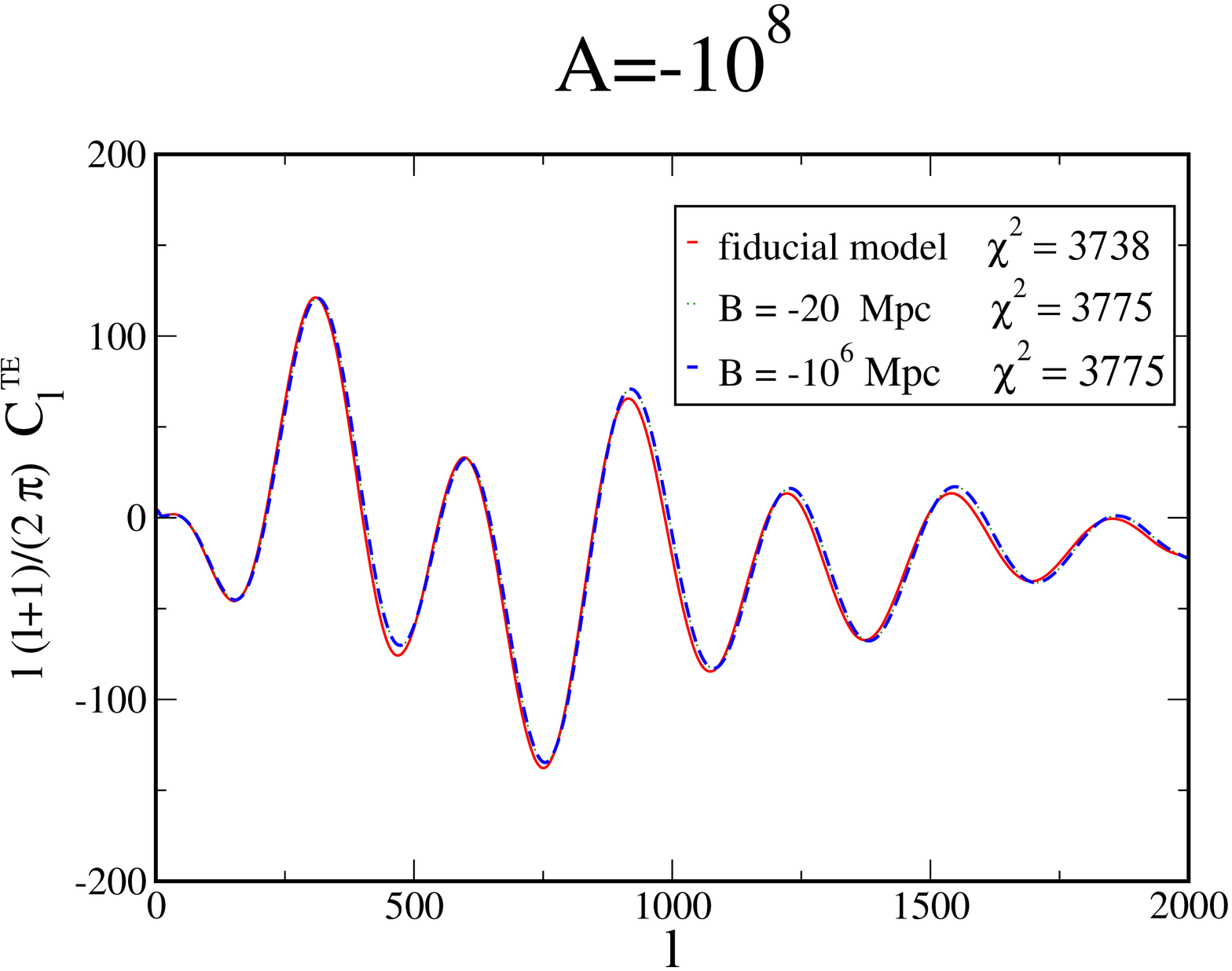}
\end{center}
\caption{The temperature-polarization (TE) cross-power spectrum for
  Model I (Left: $A=-10$, Right: $A=-10^8$). All models are normalized
  to the maximum of the first peak of the fiducial model. The value of
  $\chi^2$ is calculated using only WMAP 7-year release data (both temperature and temperature-polarization power spectrum are included).  The solid line corresponds to the fiducial model.}
\label{ctemodelo1}
\end{figure}

\begin{figure}[ht!]
\begin{center}
\includegraphics[scale=0.3,angle=-90]{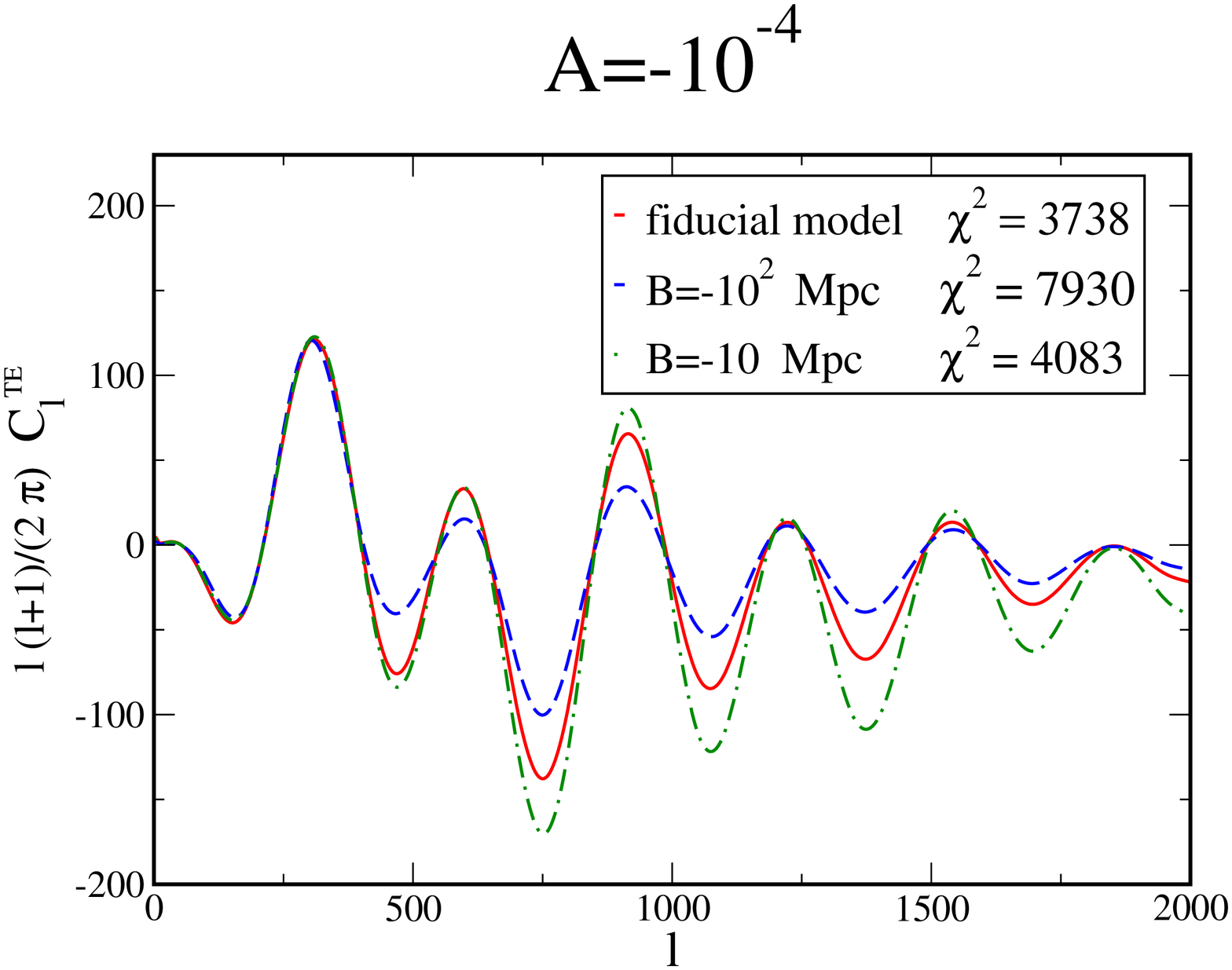}
\end{center}
\caption{The temperature-polarization (TE) cross-power spectrum for
  Model I ($A=-10^{-4}$). All models are normalized to the maximum of
  the first peak of the fiducial model. The value of $\chi^2$ is
  calculated using only WMAP 7-year release data (both temperature and temperature-polarization power spectrum are included). The solid line corresponds to the fiducial model.}
\label{ctemodelo1v2}
\end{figure}

\begin{figure}[ht!]
\begin{center}
\includegraphics[scale=0.3,angle=-90]{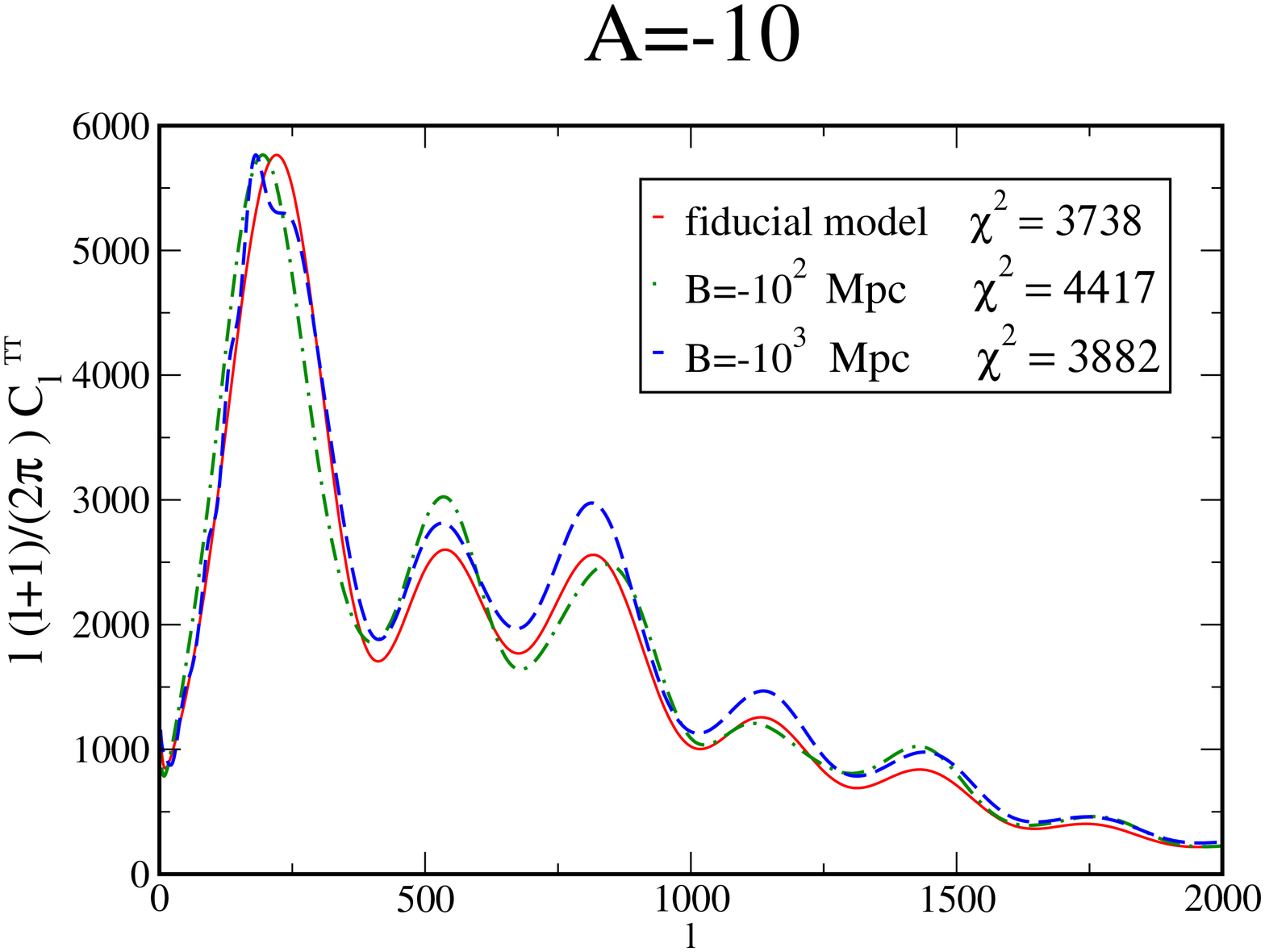}
\includegraphics[scale=0.3,angle=-90]{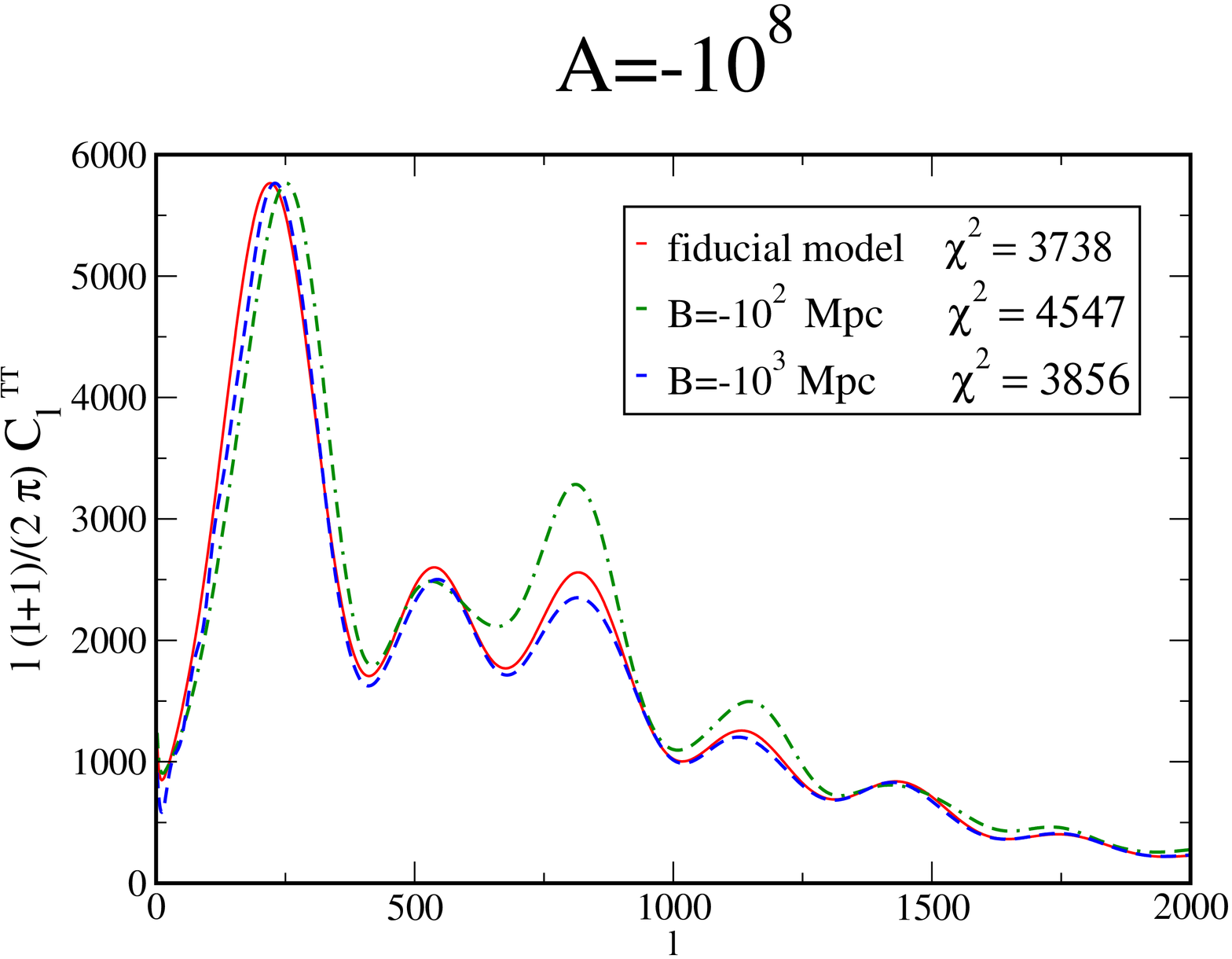}
\end{center}
\caption{The temperature autocorrelation (TT) power spectrum for
  Model II (Left: $A=-10$, Right: $A=-10^8$). All models are
  normalized to the maximum of the first peak of the fiducial
  model. The value of $\chi^2$ is calculated using only WMAP 7-year
  release data (both temperature and temperature-polarization power spectrum are included). The solid line corresponds to the fiducial model.}
\label{cttmodelo2}
\end{figure}

\begin{figure}[ht!]
\begin{center}
\includegraphics[scale=0.3,angle=-90]{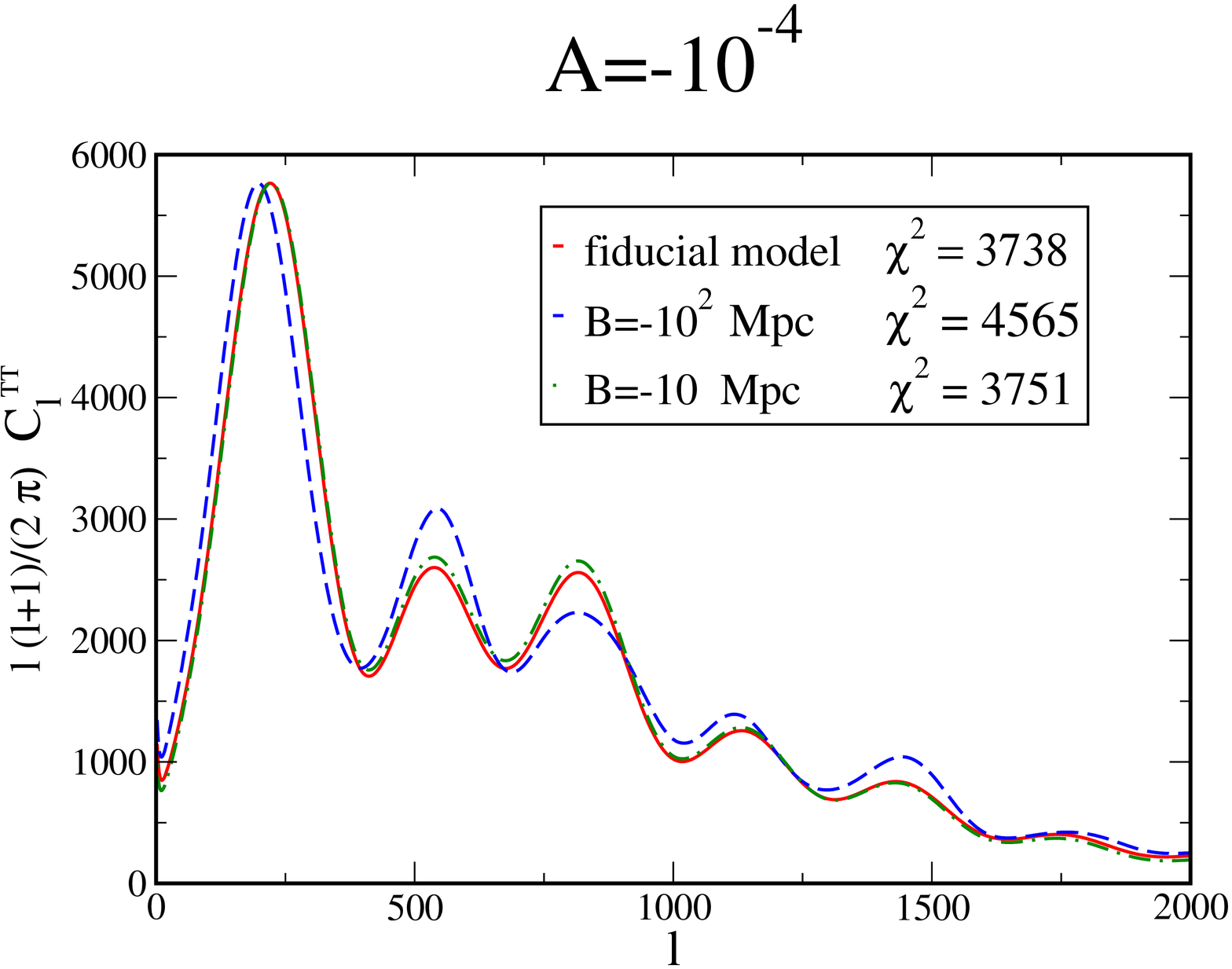}
\end{center}
\caption{The temperature autocorrelation (TT) power spectrum for
  Model II ($A=-10^{-4}$). All models are normalized to the maximum of
  the first peak of the fiducial model. The value of $\chi^2$ is
  calculated using only WMAP 7-year release data (both temperature and temperature-polarization power spectrum are included). The solid line corresponds to the fiducial model.}
\label{cttmodelo2v2}
\end{figure}

Figures \ref{cttmodelo1}, \ref{cttmodelo1v2}, \ref{cttmodelo2} and
\ref{cttmodelo2v2} show the prediction for the temperature fluctuation
spectrum for Models I and II for various values of the parameters $A$
and $B$. The value of $\chi^2$ in the figures is calculated using both
temperature and temperature-polarization fluctuation data from WMAP
7-year release.  We can see that the different collapse models have
distinct effects on the CMB temperature fluctuation spectrum, and
furthermore, that these effects depend rather strongly on the values
of the parameters $A$ and $B$ which determine the collapse times of
the various modes.

 Let us first analyze the case of Model I: When fixing $A=-10$ or
 $A=-10^{-4}$, we find that there is no change within the range of $B$ studied here (with respect to the
 fiducial model) in the position of the Doppler peaks or in the height
 of the first peak. On the other hand, we note an increase in the
 height of the secondary peaks, with the amplitude depending on the
 value of the free parameter $B$. Next, we set $A=-10^8$ (see
 Fig.~\ref{cttmodelo1}, right panel); in this case there is no change
 in the first peak (with respect to the fiducial model), while there
 is a extremely small increase in the height of the second peak and a
 a similar decrease in the height of the third peak.  In fact one does
 not observe any important change in the spectrum for different values
 of $B$ (other values of B in the explored range give the same
 spectrum). The small difference between the spectrum of the collapse
 models and the fiducial model is due to the fact that the former uses
 the canonical value for the spectral index ($ n_s=1$) while the
 latter is the best fit value to the data found by the WMAP
 collaboration assuming the standard cosmological model (which differs
 slightly from $ 1$). We have also studied the results for other
 various large negative values of $A$ (not shown here) and found that
 the behavior is similar to that of the case $A=-10^8$.  This is the
 result of a simple preliminary analysis; the exact value of $A$ at
 which the behavior of the CMB spectrum changes becomes important,
 phenomenologically speaking, will be determined from the statistical
 analysis (see Section \ref{resultados}). As the most accurate
 observational data corresponds to the first peak, it is expected that
 the collapse Model I with large values of $A$ will be
 phenomenologically indistinguishable from the fiducial model, as seen
 in Fig.~\ref{cttmodelo1} (right panel). Indeed, the difference in the
 value of the quality-of-fit estimator $\chi^2$, is found to be not
 significant within $1 \sigma$.

Figs.~\ref{ctemodelo1} and \ref{ctemodelo1v2} show the prediction for
the temperature-polarization cross-power spectrum for Model I. Here
again, the models with $A=-10^8$ differ very little from the fiducial
model and show no relevant dependence of the results on values of $B$
within the explored range.  Regarding the cases of $A=-10$ and
$A=-10^{-4}$, we find (in comparison with the fiducial model) an
increase in the values of the secondary peaks and a decrease in the
amplitude of the valleys, but the relative change is less relevant
than that in the temperature fluctuation power spectrum.

\begin{figure}[ht!]
\begin{center}
\includegraphics[scale=0.3,angle=-90]{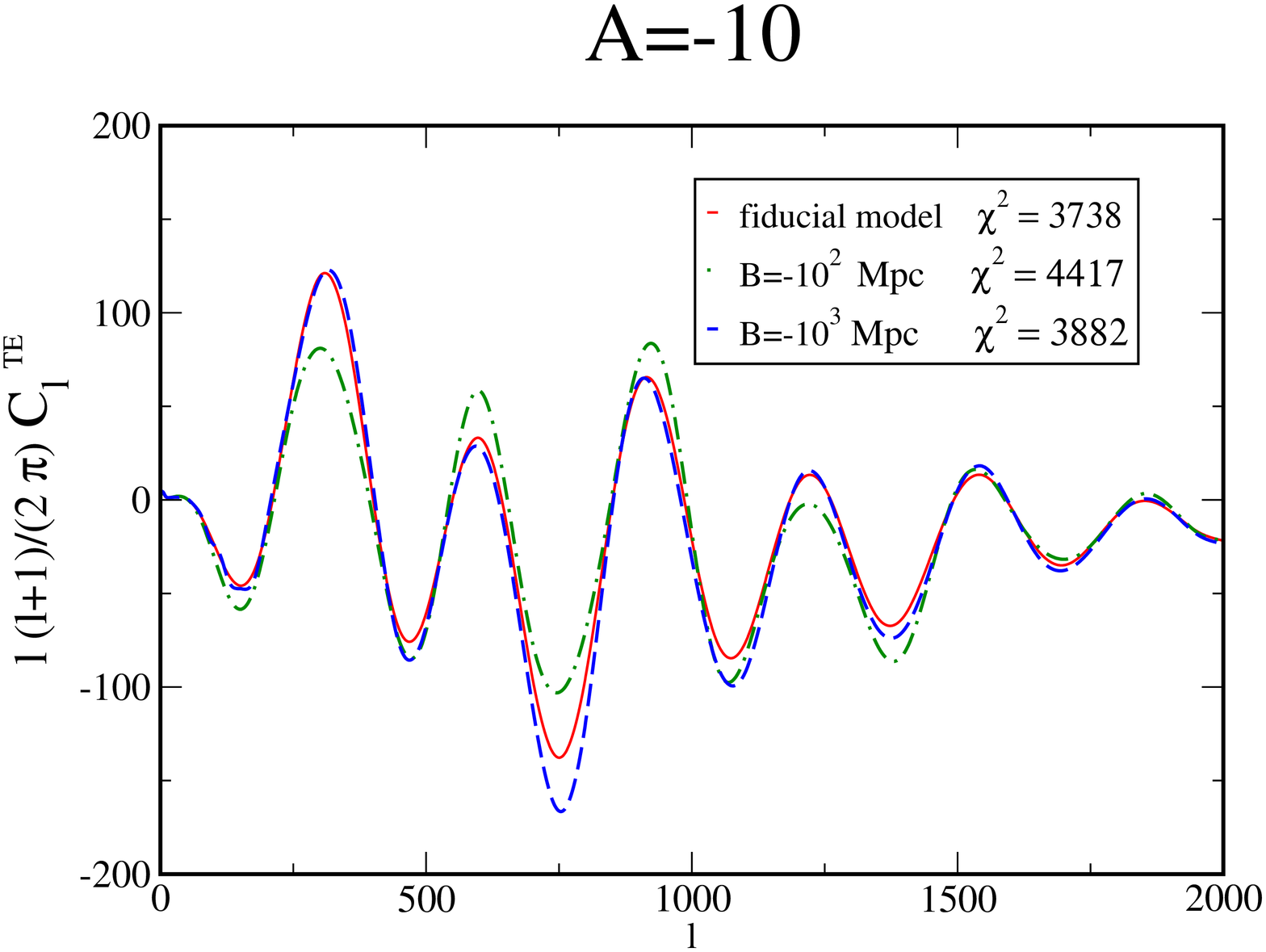}
\includegraphics[scale=0.3,angle=-90]{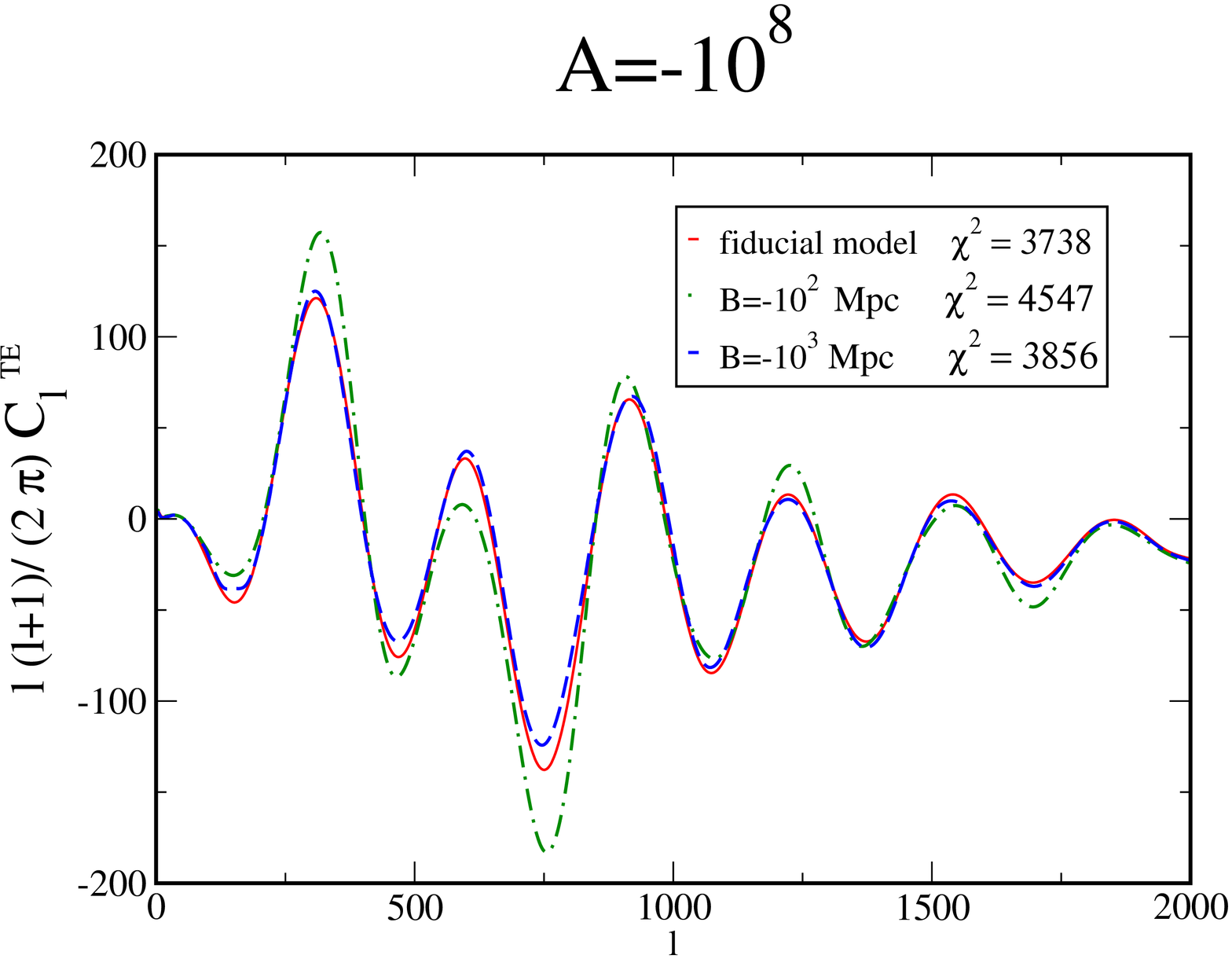}
\end{center}
\caption{The temperature-polarization (TE) cross-power spectrum for
  Model II (Left: $A=-10$, Right: $A=-10^8$). All models are
  normalized to the maximum of the first peak of the fiducial
  model. The value of $\chi^2$ is calculated using only WMAP 7-year
  release data (both temperature and temperature-polarization power spectrum are included). The solid line corresponds to the fiducial model.}
\label{ctemodelo2}
\end{figure}

\begin{figure}[ht!]
\begin{center}
\includegraphics[scale=0.3,angle=-90]{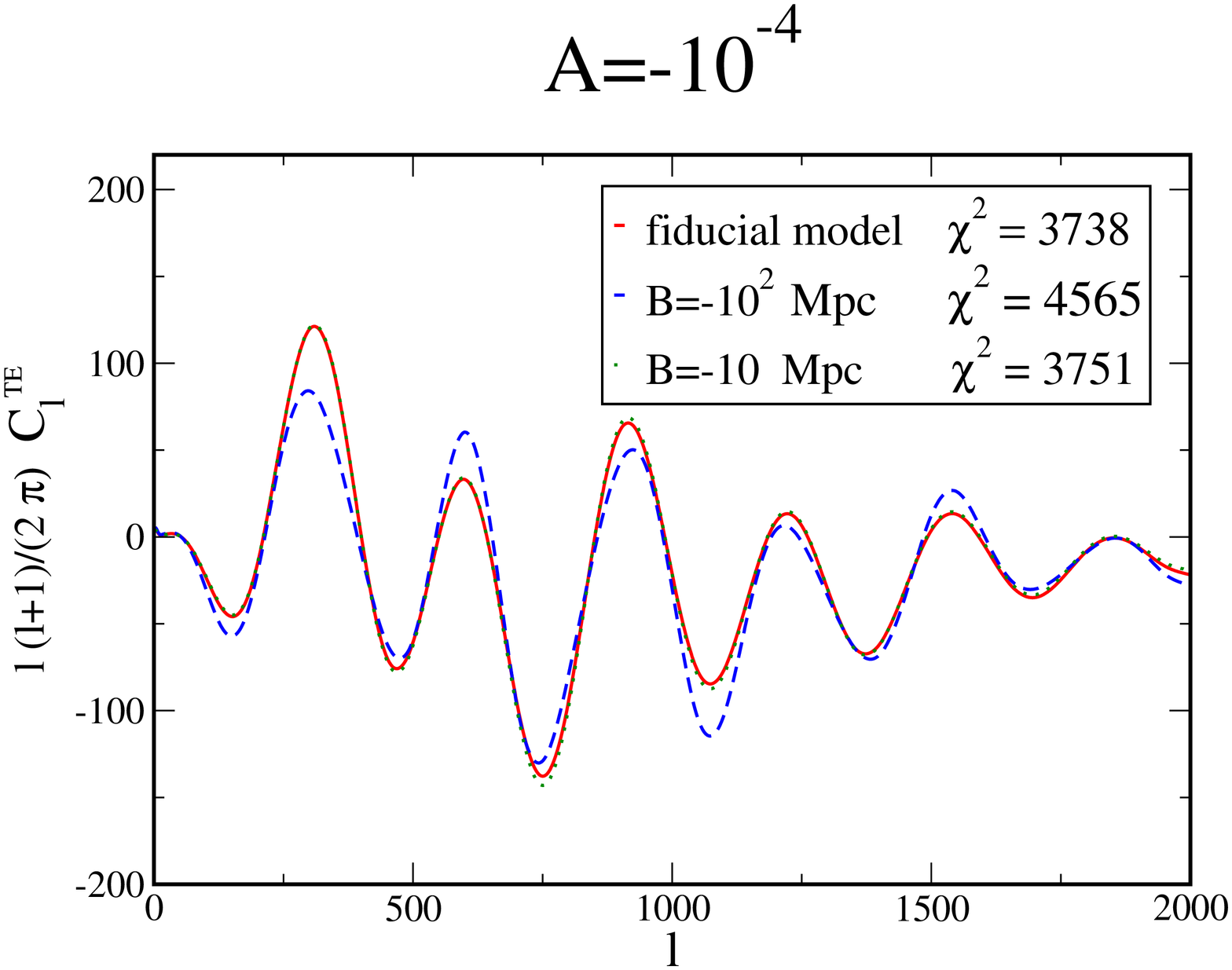}
\end{center}
\caption{The temperature-polarization (TE) cross-power spectrum for
  Model II ($A=-10^{-4}$). All models are normalized to the maximum of
  the first peak of the fiducial model. The value of $\chi^2$ is
  calculated using only WMAP 7-year release data (both temperature and temperature-polarization power spectrum are included).}
\label{ctemodelo2v2}
\end{figure}

Let us now focus on Model II (see Fig.~\ref{cttmodelo2}). Here again,
the effect is different for different values of $A$: Considering
firstly the case of $A=-10$, we find that there is a shift in the
position of the peaks, and in one case ($B=-10^3$), the first peak is
replaced by two peaks. On the other hand, when setting $A=-10^8$ or
$A=-10^{-4}$ , we find generically a shift in the position of the
first peak. In both cases, there is also a change in the height of the
secondary peaks, with the magnitude of the change depending on the
values of $A$ and $B$. Figs.~\ref{ctemodelo2} and \ref{ctemodelo2v2}
show the prediction for the temperature-polarization cross-power
spectrum for Model II. Here again, we observe (in comparison with the
fiducial model) an increase in the value of the peaks and a decrease
in the values at the valleys for all cases, with the magnitude of the
changes depending on the value of $B$. Therefore, in this case,
collapse models can be clearly distinguished from the fiducial
model. The difference in behavior of the two models can be understood
by looking at Eqs.~(\ref{C1_sudarsky}) and (\ref{C2_sudarsky}): for
sufficiently large values of $A$ and/or $B$, $C_1(k)$ becomes
independent from $k$, while the second term of $C_2(k)$ does not
vanish in the limit of large values of $k\eta^c_k$.  We would like to
emphasize the importance of this preliminary analysis on the behavior
of the collapse schemes, to determine the appropriate method for the
statistical analysis to explore the various relevant regions of
parameter space, as we described in detail in Sect.
\ref{resultados}.


\section{Results}
\label{resultados}

The observational data used for the analysis are the temperature and
temperature-polarization power spectra obtained from the final WMAP
7-year release \cite{wmap7}, and other CMB experiments such as CBI
\cite{CBI04}, ACBAR \cite{ACBAR02}, BOOMERANG
\cite{BOOM05_polar,BOOM05_temp}, BICEP \cite{BICEP} and QUAD
\cite{QUAD}, together with the matter power spectrum traced by LRGs as
measured in the Sloan Digital Sky Survey DR7 \citep{Reid09}. We
consider a spatially-flat cosmological model with adiabatic density
fluctuations, in which we add the effect of the collapse models in the
power spectrum of the initial fluctuations. The parameters allowed to
vary are:
\begin{equation}
P=\left(\Omega_B h^2, \Omega_{CDM} h^2, \Theta, \tau, A_s, A, B\right),
\end{equation}
where $\Theta$ is the ratio of the comoving sound horizon at
decoupling to the angular diameter distance to the surface of last
scattering, $\tau$ is the reionization optical depth, $A_s$ is the
amplitude of the primordial density fluctuations, and $A$ and $B$ are
the model parameters related to the conformal time of collapse of each
mode (see Eq.~(\ref{tcolapse})). Given that  the primordial power spectrum of density
fluctuations for the collapse models was computed only for the scale
invariant case \cite{Perez2006,Unanue2008}, the scalar spectral index
of density fluctuations ($n_s$) is fixed to $1$ in this
paper. However, we should keep in mind that the collapse models allow
different values for $n_s$ and the corresponding power spectrum will
be studied in future works.

 In order to place constraints on the parameters of the quantum
 collapse models, we modified the primordial power spectrum according
 to the schemes described in Section \ref{models}. We performed our
 statistical analysis by exploring the parameter space with Monte
 Carlo Markov chains generated with the publicly available CosmoMC
 code of Ref.~\cite{LB02} which uses the Boltzmann code CAMB
 \cite{LCL00} to compute the CMB power spectra.

 In a first trial, we performed the statistical analysis without
 imposing any prior on the baryon density  and found that  the confidence interval
 obtained for $\Omega_B h^2$ turned out to be  very large compared to the one
 obtained by the WMAP collaboration (i.e. without considering the
 collapse scheme). Therefore, we introduced a gaussian prior on the baryon
 density, using an independent data, namely the Big Bang
 Nucleosynthesis  (BBN) bound  \cite{MP09,Steigman10}.

Initially, we intended to perform the statistical analysis allowing
the two parameters of the model, $A$ and $B$, to vary
independently. However, after the first runs we realized that for a
very large range of values of $A$, there was always a range of values
of $B$ providing a good fit to the data.  Furthermore, in order to
allow the Markov chains to explore various orders of magnitude in
those parameters, we used the following reparametrization: $a = \sinh
(A)$ and $b = \sinh (B)$ (another possible reparametrization is $a=
\log(A)$ and $b=\log (B)$, but we are interested in exploring both
positive and negative values of $A$ and $B$).  We found that the
Markov chains did not converge when both $a$ and $b$ were allowed to
vary independently.  The reason for this is simply the fact that there
are several maxima of the probability function in the parameter space,
but the method of Markov chains does not allow to explore all maxima
at the same time. In contrast, we found that when we fixed values of
$A$ within a rather large range of values, convergence of the Markov
chains is found.
Moreover, let us recall that the primordial power spectrum in collapse
models is a function of $z_k = A + B k$; by taking $B=0$, we recover
the spectrum corresponding to the standard $\Lambda$CDM model with
spectral index $n_s=1$, up to an overall normalization factor. Thus,
the parameter $B$ gives a measure of the departure of the primordial
power spectrum with respect to that of the standard model, while
phenomenologically $A$ behaves as a normalization factor that is
degenerated with $A_s$.  We should note, however, that although the
effect on the spectrum of $A$ and $A_s$ is very closely connected,
their physical interpretation is quite different.  The results of the
statistical analysis, for fixed values of $A$ are shown in
Tables~\ref{tablam1}, \ref{tablam2neg} and \ref{tablam2pos} for
collapse models I and II.
\begin{table}[ht!]
\begin{center}
\renewcommand{\arraystretch}{1.3}
\caption{Model I. Mean values and $1 \sigma$ error for $b = \sinh B$
  considering different fixed values of parameter $A$. The $\chi^2$
  estimator is computed for the whole data set. For the fiducial
  model, $\chi^2_{\rm min}=3909$. Column 4 refers to the viability of
  the model (collapse times happen during inflation), and column 5
  shows whether there is any restriction to the viability in the range
  shown by column 2.}
\label{tablam1}
\begin{tabular}{|c|c|c|c|c|}
\hline 
  $A$ & Mean value and 1 $\sigma$ error & $\chi^2_{\rm min}$ & Viability & Restricted \\ \hline
 $-17.5$ & $2.74^{+0.28}_{-0.29} $& $3910$ & Y & N \\ \hline
$-15$ & $-3.11 \pm 0.19 $& $3909$ & Y & N\\ \hline
$-12.5$ & $-2.40 \pm 0.36 $& $3910$ & Y & N\\ \hline
$-10$ & $-0.87^{+0.79}_{-0.74} $& $3915$ & Y & N\\ \hline
$-7.5$ & $1.83^{+0.46}_{-0.38} $& $3911$ & Y & N\\ \hline
$-5$ & $1.95^{+0.24}_{-0.22} $& $3909$ & Y & N\\ \hline
$-2.5$ & $-1.56^{+0.23}_{-0.14} $& $3907$ & Y & N\\ \hline
$-1$ & $1.65 \pm 0.24 $& $3908$ & Y & N\\ \hline
$-10^{-1}$ & $0.48^{+0.50}_{-0.51} $& $3915$ & Y & N\\ \hline
$-10^{-2}$ & $0.03 \pm 0.55 $& $3916$ & Y & Y\\ \hline
$-10^{-3}$ & $0.005 \pm 0.54 $& $3916$  & Y & Y\\ \hline
$-10^{-4}$ & $0.02^{+0.54}_{-0.55} $& $3916$  & Y & Y\\ \hline
$10^{-8}$ & $-0.006^{+0.54}_{-0.55} $& $3916$ & Y & Y \\ \hline
$10^{-7}$ & $0.0005^{+0.56}_{-0.57} $& $3916$  & Y & Y\\ \hline
$10^{-6}$ & $-0.016\pm 0.56 $& $3916$  & Y & Y\\ \hline
$10^{-5}$ & $-0.006^{+0.54}_{-0.55} $& $3916$ & Y & Y \\ \hline
$10^{-4}$ & $0.72 \times 10^{-4} \pm 0.55 $& $3916$& N & - \\ \hline
$10^{-3}$ & $-0.01^{+0.51}_{-0.56}  $& $3916$& N & -  \\ \hline
$10^{-2}$ & $-0.05^{+0.54}_{-0.65}  $& $3916$& N & -  \\ \hline
$10^{-1}$ & $0.48^{+0.50}_{-0.51}  $& $3915$ & N & - \\ \hline
$1$ & $-1.68 \pm 0.23  $& $3908$ & N & - \\ \hline
$10$ & $0.81^{+0.74}_{-0.79}  $& $3915$ & N & - \\ \hline
$12.5$ & $2.44^{+0.34}_{-0.35}   $& $3910$ & N & - \\ \hline
$15$ & $-3.20^{+0.15}_{-0.19}   $& $3910$ & N & - \\ \hline
$17.5$ & $2.72^{+0.28}_{-0.30}   $& $3910$& N & -  \\ \hline
\end{tabular}
\end{center}
\end{table}

\begin{table}[ht!]
\begin{center}
\renewcommand{\arraystretch}{1.3}
\caption{Model II. Mean values and $1 \sigma$ error for $b = \sinh B$
  considering different fixed negative values of parameter
  $A$. $\chi^2$ is calculated for the whole data set. For the fiducial
  model, $\chi^2_{\rm min}=3909$. Column 4 refers to the viability of
  the model (collapse times happen during inflation), and column 5
  shows whether there is any restriction to the viability in the range
  shown by column 2.}
\label{tablam2neg}
\begin{tabular}{|c|c|c|c|c|}
\hline 
  $A$ & Mean value and 1 $\sigma$ error & $\chi^2_{\rm min}$ &   Viability & Restricted\\ \hline 
$-10^9$ & $1.55\pm 0.30 $& $3908$ & Y & N\\ \hline 
$-7 \times 10^8$ & $1.54 \pm 0.21 $& $3908$  &Y & N\\ \hline 
$-3 \times 10^8$ & $-1.56 \pm 0.30 $& $3909$ &Y & N\\ \hline 
$-10^8$ & $-1.64^{+0.21}_{-0.22} $& $3908$ & Y & N\\ \hline 
$-10^7$ & $-1.55^{+0.30}_{-0.29} $& $3909$& Y & N\\ \hline  
$-10^6$ & $-1.44^{+0.31}_{-0.30} $& $3910$ & Y & N\\ \hline 
$-5 \times 10^5$ & $1.78 \pm 0.21 $& $3911$ & Y & N\\ \hline 
$-10^5$ & $-0.17 \pm 0.46 $& $3916$& Y & N\\ \hline 
$-5 \times 10^4$ & $-1.99 \pm 0.20 $& $3908$ & Y & N\\ \hline 
$-10^4$ & $1.36 \pm 0.23 $& $3908$ & Y & N\\ \hline 
$-10^3$ & $1.49 \pm 0.23 $& $3908$ & Y & N\\ \hline 
$-500$ & $1.57 \pm 0.30 $& $3908$& Y & N\\ \hline 
$-300$ & $-2.00 \pm 0.20 $& $3911$   & Y & N\\ \hline  
$-300$ & $1.96 \pm 0.20 $& $3911$    & Y & N\\ \hline 
$-100$ & $-1.58 \pm 0.28 $& $3908$    & Y & N\\ \hline 
$-80$ & $-2.08^{+0.13}_{-0.18} $& $3913$  & Y & N\\ \hline 
$-50$ & $-1.15 \pm 0.34 $& $3913$     & Y & N\\ \hline 
$-30$ & $-1.79^{+0.21}_{-0.20} $& $3909$  & Y & N\\ \hline 
$-10$ & $1.47 \pm 0.26 $& $3908$      & Y & N\\ \hline 
$-5$ & $-1.58 \pm 0.20 $& $3908$      & Y & N\\ \hline 
$-1$ & $-2.01 \pm 0.21 $& $3909$      & Y & N\\ \hline 
$-10^{-1}$ & $-2.22^{+0.76}_{-0.65} $ & $3915$ & Y & N\\ \hline 
$-10^{-1}$ & $2.03^{+1.05}_{-0.84}  $ & $3914$ & Y & Y\\ \hline 
$-10^{-2}$ & $-2.09^{+0.97}_{-0.76} $ & $3914$& Y & N\\ \hline 
$-10^{-2}$ & $2.14^{+0.73}_{-0.93} $ & $3914$ & N & -\\ \hline 
$-10^{-3}$ & $-2.12^{+0.93}_{-0.74} $& $3914$ & Y & N\\ \hline 
$-10^{-3}$ & $2.11^{+0.75}_{-0.97} $ & $3914$ & N & -\\ \hline 
$-10^{-4}$ & $-2.13^{+0.92}_{-0.73} $& $3914$ & Y & N \\ \hline
$-10^{-4}$ & $2.14^{+0.72}_{-0.92} $& $3914$& N & -\\\hline
\end{tabular}
\end{center}
\end{table}

\begin{table}[ht!]
\begin{center}
\renewcommand{\arraystretch}{1.3}
\caption{Model II. Mean values and $1 \sigma$ error for $b = \sinh B$
  considering different fixed positive values of parameter
  $A$. $\chi^2$ is calculated for the whole data set. For the fiducial
  model, $\chi^2_{\rm min}=3909$. Column 4 refers to the viability of
  the model (collapse times happen during inflation), and column 5
  shows whether there is any restriction to the viability in the range
  shown by column 2.}
\label{tablam2pos}
\begin{tabular}{|c|c|c|c|c|}
\hline 
  $A$ & Mean value and 1 $\sigma$ error & $\chi^2_{\rm min}$ &   Viability & Restricted\\ \hline 
$10^{-8}$ & $-2.09^{+0.96}_{-0.76} $& $3914$& Y & N\\ \hline
$10^{-8}$ & $2.14^{+0.71}_{-0.90} $& $3914$& N & -\\ \hline
$10^{-7}$ & $-2.06^{+0.97}_{-0.77} $& $3914$& Y & N\\ \hline
$10^{-7}$ & $2.10^{+0.96}_{-0.75} $& $3914$& N & -\\ \hline
$10^{-6}$ & $-2.09^{+1.00}_{-0.77} $& $3914$& Y & N\\ \hline
$10^{-6}$ & $2.12^{+0.76}_{-0.97} $& $3914$& N & -\\ \hline
$10^{-5}$ & $-2.00^{+1.05}_{-0.81} $& $3914$& Y & N\\ \hline
$10^{-5}$ & $2.14^{+0.72}_{-0.90} $& $3914$& N & -\\ \hline
$10^{-4}$ & $-2.14^{+0.87}_{-0.71}  $& $3914$& N & -\\ \hline
$10^{-4}$ & $2.11^{+0.75}_{-0.98} $& $3914$& N & - \\ \hline
$10^{-3}$ & $-2.12^{+0.93}_{-0.74} $& $3914$& N & -\\ \hline
$10^{-3}$ & $2.11^{+0.95}_{-0.74} $& $3914$& N & -\\ \hline
$10^{-2}$ & $-2.12^{+0.95}_{-0.75} $& $3914$ & N & -\\ \hline
$10^{-2}$ & $2.15^{+0.73}_{-0.95} $& $3914$& N & -\\ \hline
$10^{-1}$ & $-2.43^{+0.57}_{-0.51} $& $3915$& N & -\\ \hline
$10^{-1}$ & $0.00^{+0.75}_{-0.76} $& $3915$& N & -\\ \hline
$10^{-1}$ & $2.42^{+0.47}_{-0.51}$& $3914$ & N & -\\ \hline
$10$ & $-1.47\pm 0.25 $& $3908$& N & -\\ \hline
$100$ & $1.58\pm 0.27 $& $3908$& N & -\\ \hline
$10^3$ & $1.50 \pm 0.22 $& $3908$& N & -\\ \hline
$10^4$ & $-1.33 ^{+0.32}_{-0.34} $& $3911$& N & -\\ \hline
$10^5$ & $0.19^{+0.44}_{-0.46} $& $3916$ & N & -\\ \hline
$10^6$ & $1.46^{+0.29}_{-0.30} $& $3910$& N & -\\ \hline
$10^7$ & $1.53 \pm 0.31 $& $3909$& N & -\\ \hline
$10^8$ & $1.65 \pm 0.21 $& $3908$& N & -\\ \hline
$10^9$ & $-1.53^{+0.29}_{-0.30} $& $3908$& N & -\\ \hline
\end{tabular}
\end{center}
\end{table}

Within Model I, a preliminary analysis of the modifications of the CMB
fluctuation power spectrum associated with the collapse models
indicates that for large values of the parameter $A$, there is no
dependence on the value of $B$, and also that there is a very small
difference between those models and the fiducial one. This last
feature is related to the fact that the fiducial model involves a
different value of the spectral index $n_s$ (see discussion in Section
\ref{effects}). These results are, indeed, reflected in the statistical
analyses by the fact that for $|A| < 20$, nontrivial bounds on $B$ can
be found, while for larger values of $A$, any value of $B$ gives a
good fit to the data. Furthermore, the values of the corresponding
cosmological parameters obtained from the statistical analyses for
models with $|A| > 20$ is within the $1 \sigma$ bounds established by
the standard analysis of WMAP collaboration made without considering
the collapse scheme.
In order to confirm the persistence of this behavior, we have explored
a large range of A values, from $-10^9$ to $10^9$, changing this value
in each step by one order of magnitude, and found results consistent
with the previous conclusion.
In Table~\ref{tablam1} we show the bounds on $B$ obtained for fixed
values of $|A|<20$ for Model I.

A similar analysis performed for Model II, shows that we should not
expect the same behavior   as the one obtained  for Model I. Indeed,
Fig.~\ref{cttmodelo2} shows (for two different values of $A$) that the
prediction of Model II on the $C_\ell$'s depends  rather strongly on the value of
$B$. Results from the statistical analysis for fixed values of $A$ are
shown in Table~\ref{tablam2neg} and \ref{tablam2pos}.

As mentioned previously in this section, we could not perform a
statistical analysis with $A$ and $B$ varying jointly as free
parameters, due to the fact that there are several maxima of the
probability function. Indeed, Tables~\ref{tablam2neg} and
~\ref{tablam2pos} show that, for several values of $A$, we obtain two
different maxima for which the difference between the value of
$\chi^2_{min}$ does not exceed the value of $1$ and thus one cannot
discriminate between them. There is even one case ($A=0.1$) in Model
II (see Table~\ref{tablam2pos}) in which there are three maxima.
However, we should emphasize an important difference between the
statistical analysis performed for $A=-300$ in Model II and the ones
corresponding to the other values mentioned above. In the first case,
we have performed two different statistical analyses with different
initial values of $b=\sinh(B)$ when running COSMOMC and obtained
convergence of the Markov chains to one maximum at each time. The
values of the errors have been calculated with the GETDIST program, as
is usually done in this kind of analyses. On the other hand, for
values of $|A|<1$, we have performed one statistical analysis for each
value of $A$ and obtained a marginalized likelihood with two peaks,
for the parameter $B$. We have tried to change the initial value of
$b=\sinh(B)$ in order to obtain convergence to a single value at the
time, with no success.  As the GETDIST program determines the
confidence interval assuming that there is just one maximum of the
likelihood function, we need some alternative method to calculate
those. We have thus computed the confidence interval for values of
$|A|<1$ in Model II by estimating the limits of integration for which
the integral of the marginalized likelihood function yields $68\%$ of
the value of the same integral over that peak. Therefore, it is not
surprising that the $1 \sigma$ errors calculated for $A<1$ are $4$
times larger than those calculated by the GETDIST program.

\begin{figure}[ht!]
\begin{center}
\includegraphics[scale=0.33,angle=-90]{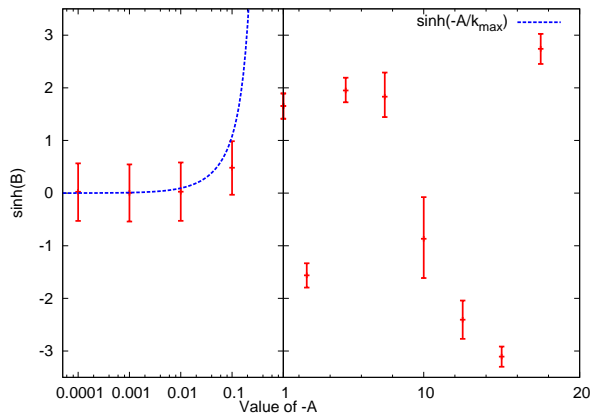}
\end{center}
\caption{Results for Model I: Bounds on $b = {\rm sinh} (B)$ obtained
  for fixed negative values of $A$. The area below the (blue) dotted line ${\rm
    sinh} (B) < {\rm sinh}(-\frac{A}{k_{\rm max}})$ indicates the
  region where the collapse of all modes happens during the
  inflationary period.}
\label{Bmodelo1}
\end{figure}

\begin{figure}[ht!]
\begin{center}
\includegraphics[scale=0.33,angle=-90]{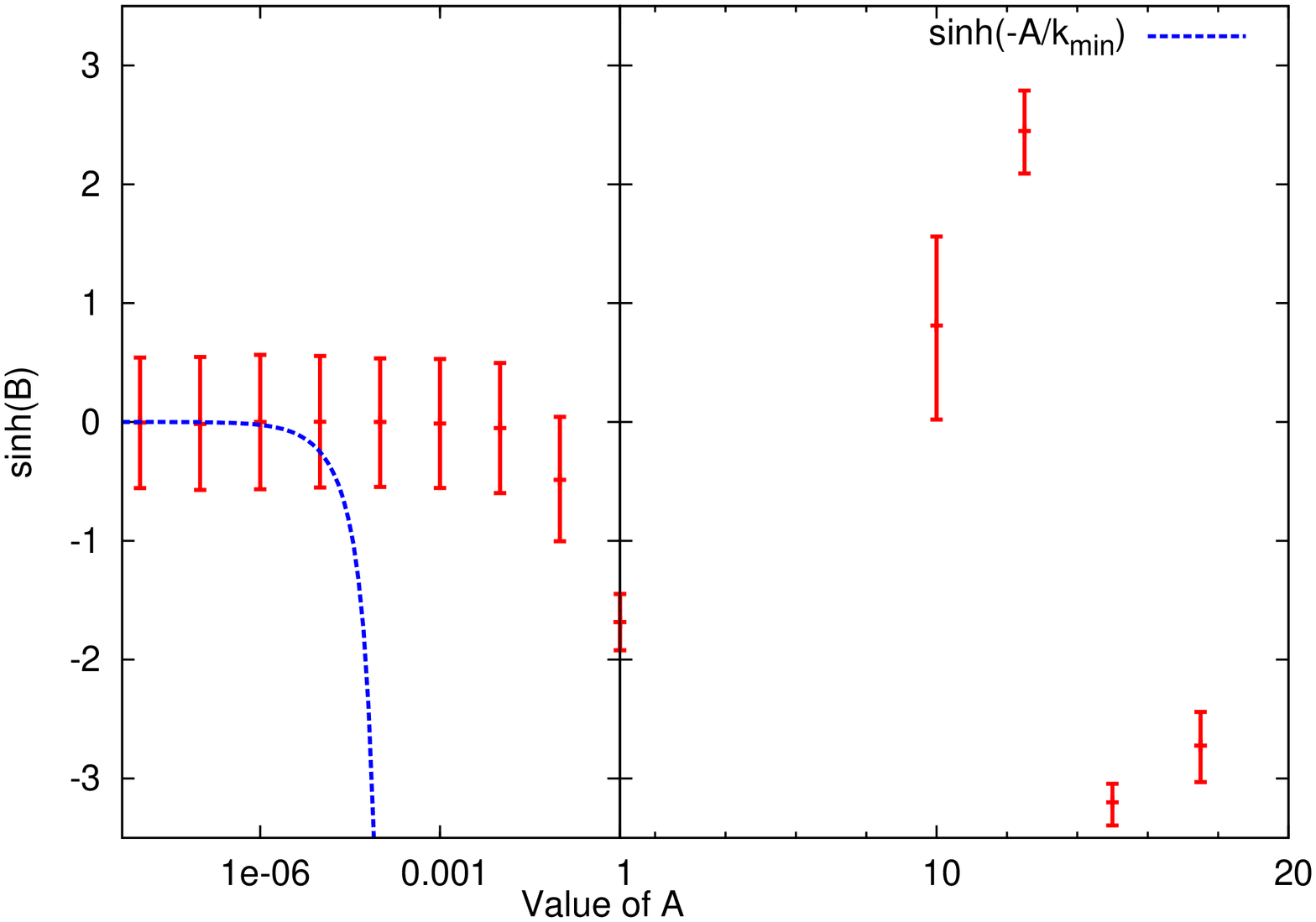}
\end{center}
\caption{Results for Model I: Bounds on $b = {\rm sinh} (B)$ obtained
  for fixed positive values of $A$. The area below the (blue) dotted line ${\rm
    sinh} (B) < {\rm sinh}(-\frac{A}{k_{\rm min}})$ indicates the
  region where the collapse of all modes happens during the
  inflationary period. }
\label{Bmodelo1_pos}
\end{figure}

\begin{figure}[ht!]
\begin{center}
\includegraphics[scale=0.33,angle=-90]{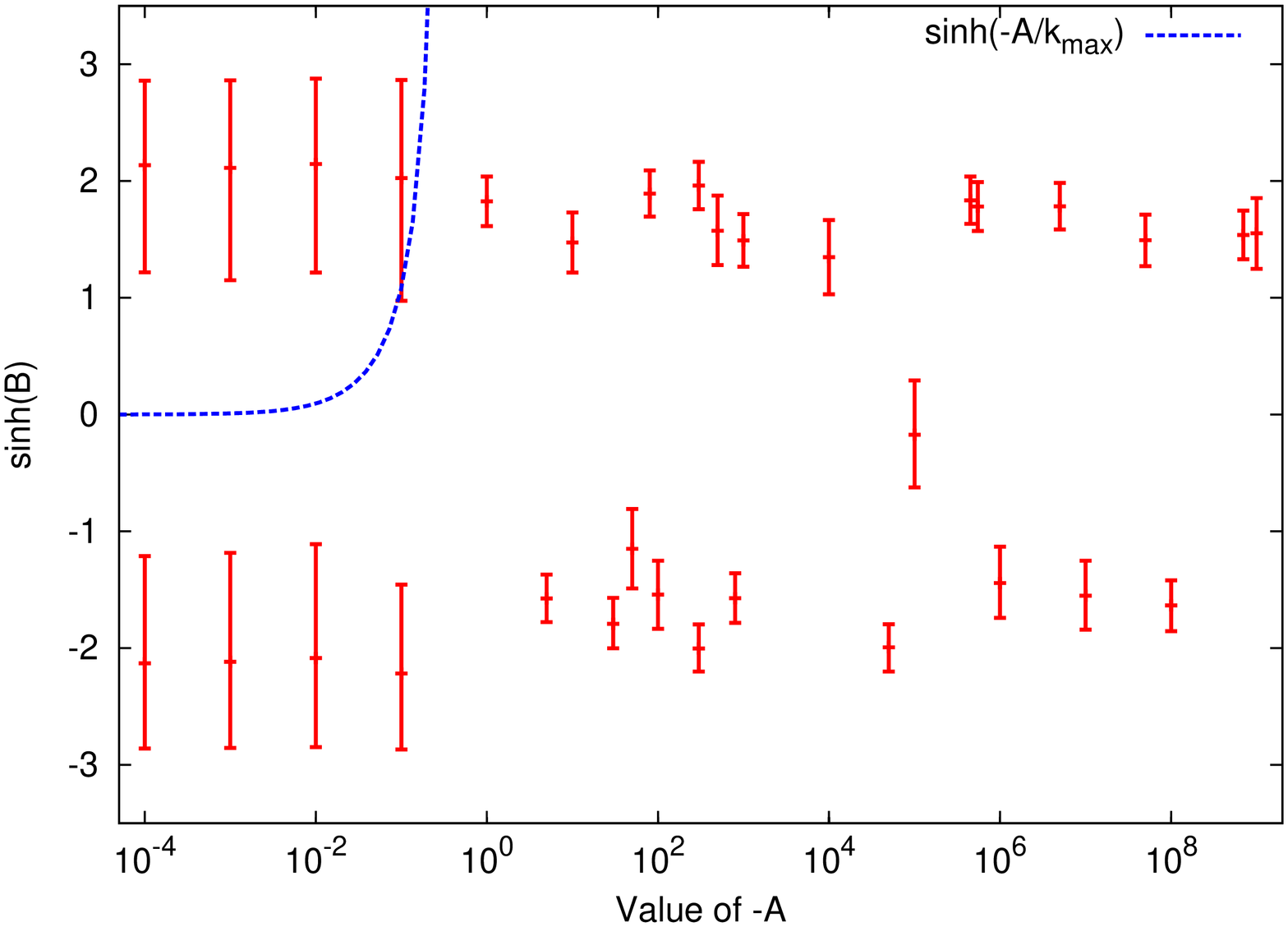}
\end{center}
\caption{Results for Model II: Bounds on $b = \sinh (B)$ obtained for
  fixed negative values of $A$. The area below the (blue) dotted line $\sinh(B)
  < \sinh(\frac{-A}{k_{\rm max}})$ indicates the region where the
  collapse of all modes happens during the inflationary period. }
\label{Bmodelo2}
\end{figure}

\begin{figure}[ht!]
\begin{center}
\includegraphics[scale=0.33,angle=-90]{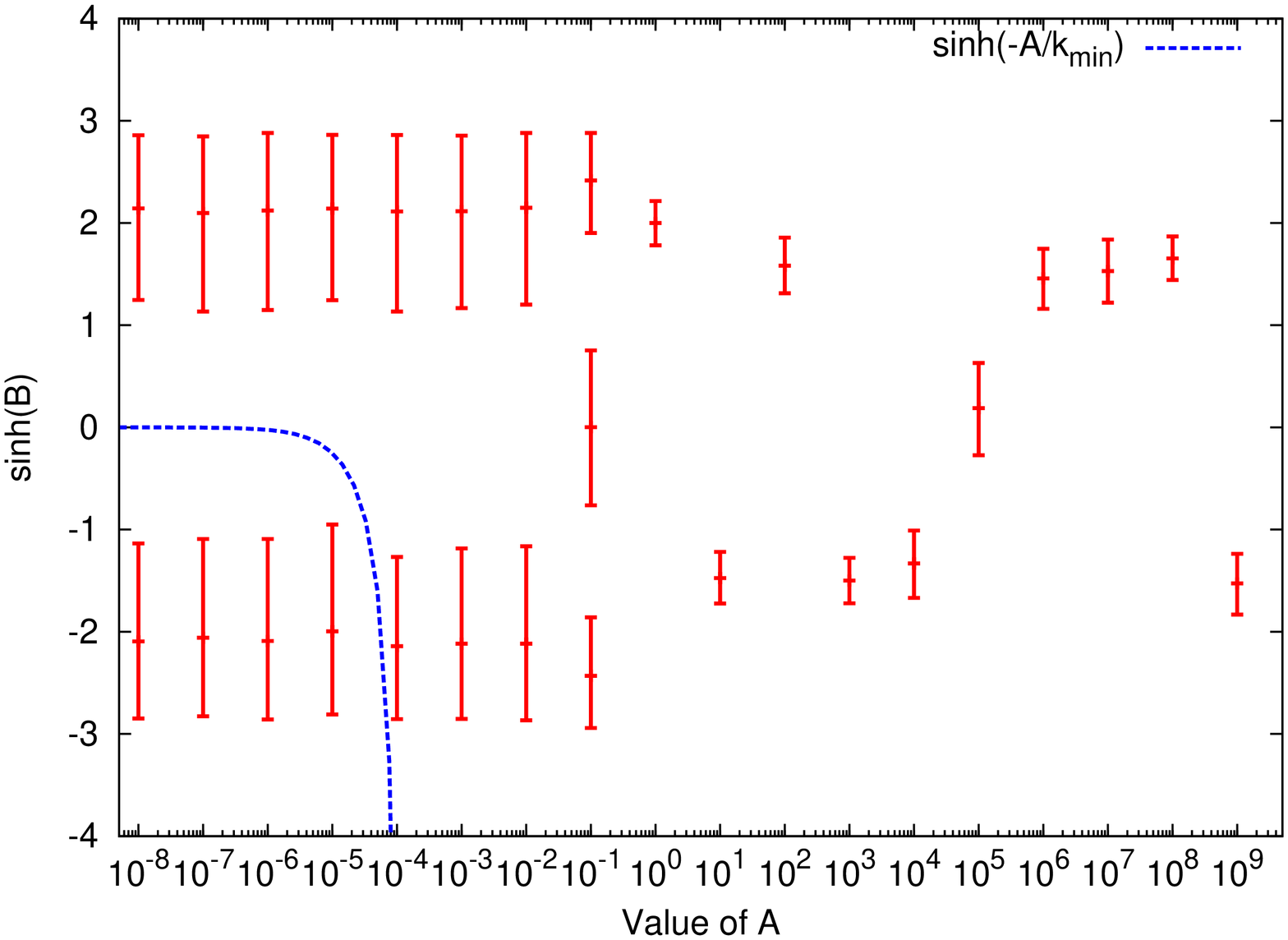}
\end{center}
\caption{Results for Model II: Bounds on $b = \sinh B$ obtained for
  fixed positive values of $A$. The area below the (blue) dotted line $\sinh(B)
  < \sinh(-\frac{A}{k_{\rm min}})$ indicates the region where the
  collapse of all modes happens during the inflationary period.}
\label{Bmodelo2_pos}
\end{figure}

Next, let us recall the discussion of Sect. \ref{priors} where for a
given value of $A$, we calculated the corresponding range of values of
$B$ that ensure that the time of collapse for the mode $k$ occurs
during the inflationary period.  Then, by considering the values of
$A$ for which the collapse schemes are tested (see
Tables~\ref{tablam1}, \ref{tablam2neg} and \ref{tablam2pos}),
Eq.~(\ref{Bphys}) implies that for $A>0$ we must have $B <
-\frac{A}{k_{{\rm min}}}$ and that for $A<0$ the condition is $B <
-\frac{A}{k_{{\rm max}}}$.Therefore, we have added the line
corresponding to $B = -\frac{A}{k_{{\rm max}}}$ in
Figs.~\ref{Bmodelo1} and ~\ref{Bmodelo2} and the line corresponding to
$B = -\frac{A}{k_{{\rm min}}}$ in Fig.~\ref{Bmodelo1_pos} and
~\ref{Bmodelo2_pos} above which the solutions found are nonphysical
and thus excluded. It should be noted that there are some cases where
only parts of the solution are physically viable. In order to
facilitate this discussion we have added column $4$ in
Tables~\ref{tablam1}, \ref{tablam2neg} and \ref{tablam2pos} which
indicates if the model is viable or not, and column $5$ in the same
tables indicates the cases where there is a further restriction on the
range of values of $B$ for the viability of the solution.  In the rest
of this section we will discuss only those results that are relevant
for our model, and exclude those nonviable values mentioned in the
above paragraph.  Recall that, unfortunately, and as it has been
already mentioned, the present analysis does not allow us to determine
any possible degeneration, as far as the fit is concerned, between the
model parameters $A$ and $B$.
However, from Fig.~\ref{Bmodelo1} we can see that the allowed values
of $B$ - within $1\sigma$ error- for $A > - 20$ turn out to lie in the
range $|\sinh(B)|<3.2$ for Model I; recall that for $A < - 20$, the
resulting CMB spectrum is the same for all values of $B$ and very
similar to the fiducial model and therefore, any value of $B$ among
those tested in this paper ($B = -10^9 \cdots 10^9)$) provides a good
fit to the data. On the other hand, for Model II, the allowed values
of $B$ - within $1\sigma$ error- lie in the range $|\sinh(B)|<3$ for
Model II for all values of $A$ studied in this paper. This allows us
to set the bounds $B < 1.88 \,\ {\rm Mpc}$ for $A < -20$ in Model I,
and $B < 1.81 \,\ {\rm Mpc}$ for Model II. On the other hand, it is
also interesting to note that for all the studied cases with $|A|<1$
in Model I, the confidence interval includes the case $B = 0$ with a
maximal allowed departure from that value of $0.6$.  In contrast, for
the cases $|A|>1$ in Model I and for almost all cases considered in
Model II, the value $B=0$ is excluded within $2 \sigma$, except for
the case $A=-10^5$ for Model II and $A=10$ for Model I.

\begin{figure}[ht!]
\begin{center}
\includegraphics[scale=0.33,angle=-90]{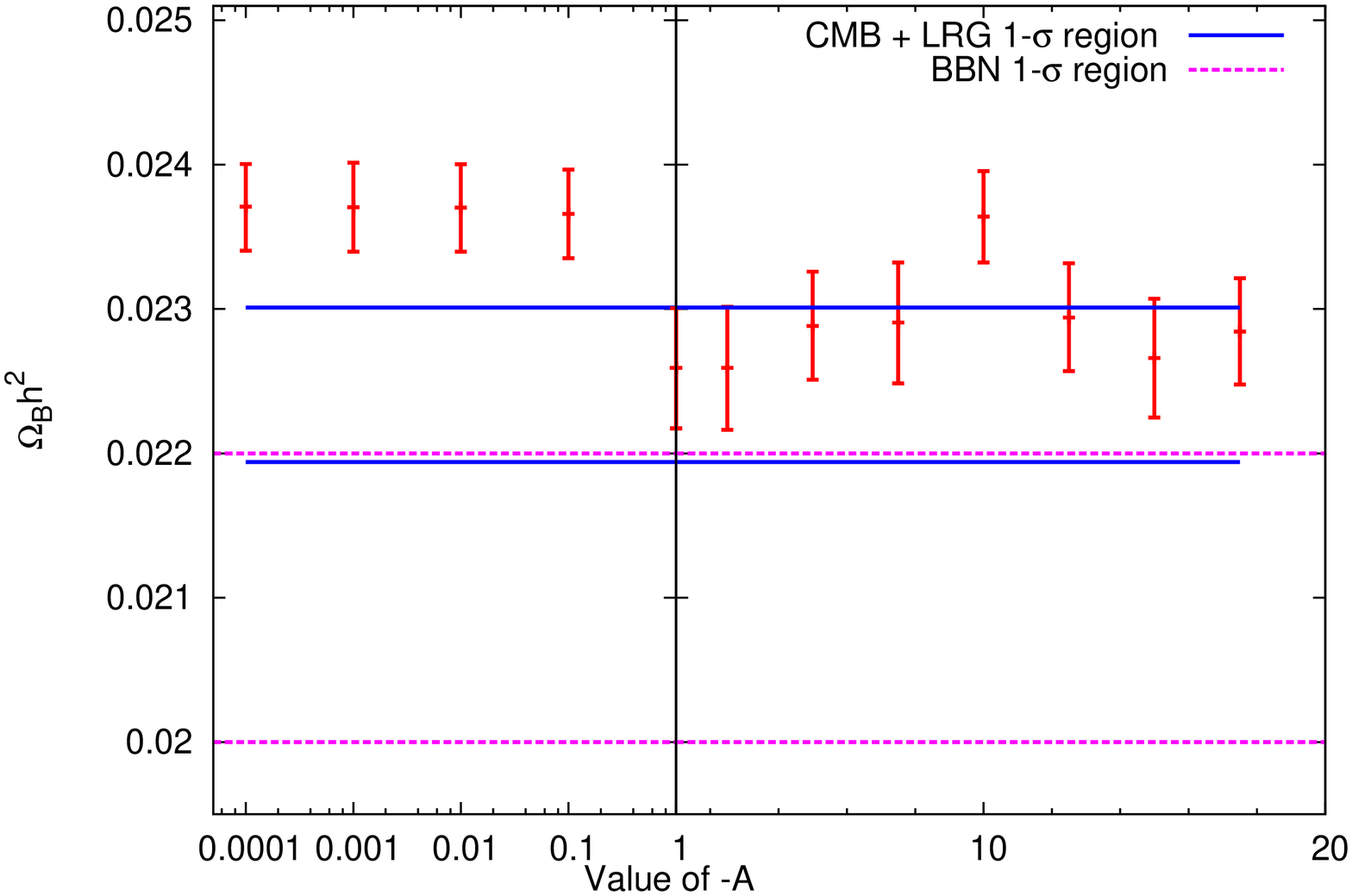}
\includegraphics[scale=0.33,angle=-90]{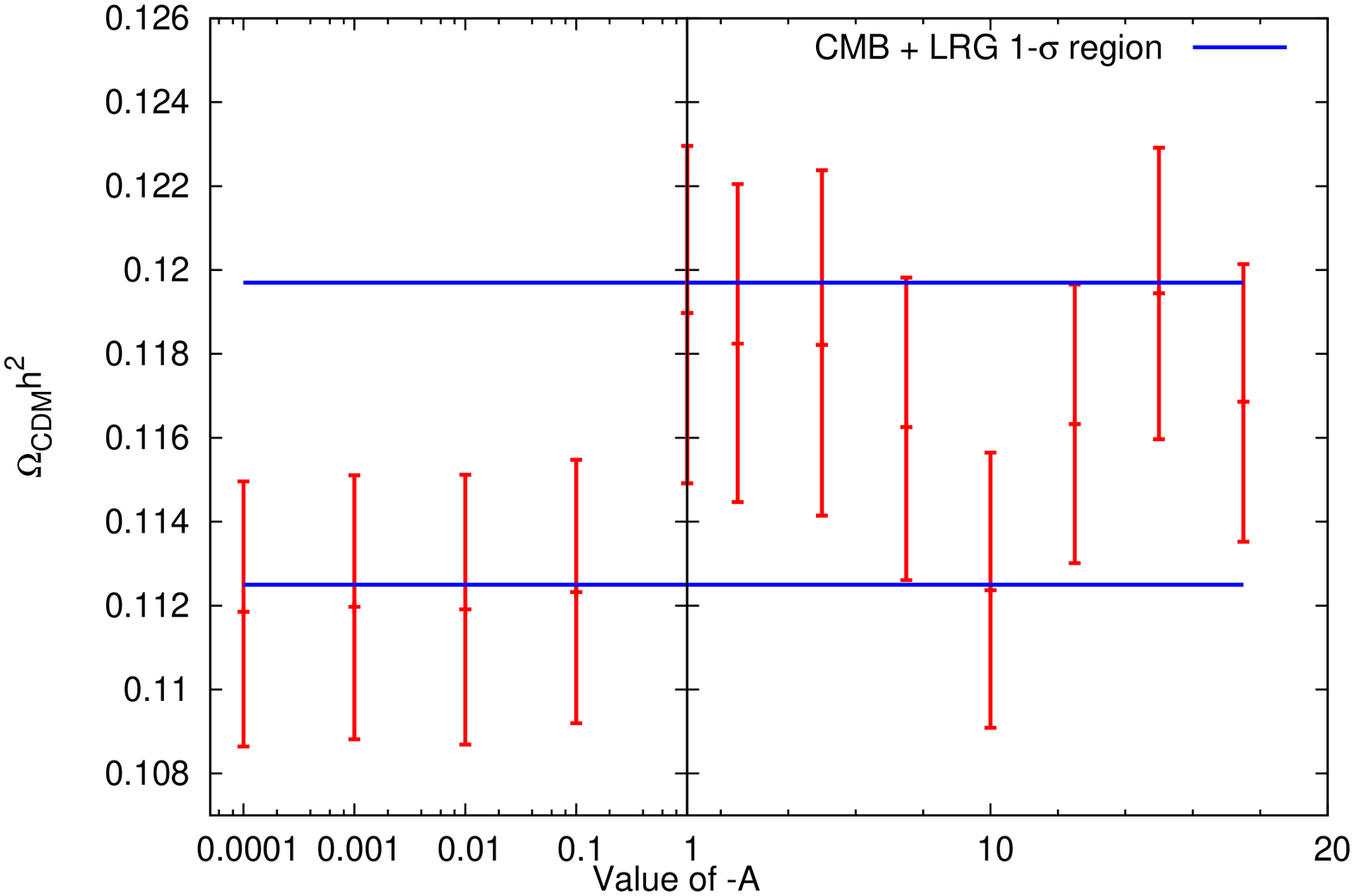}
\vspace{0.3cm}
\includegraphics[scale=0.33,angle=-90]{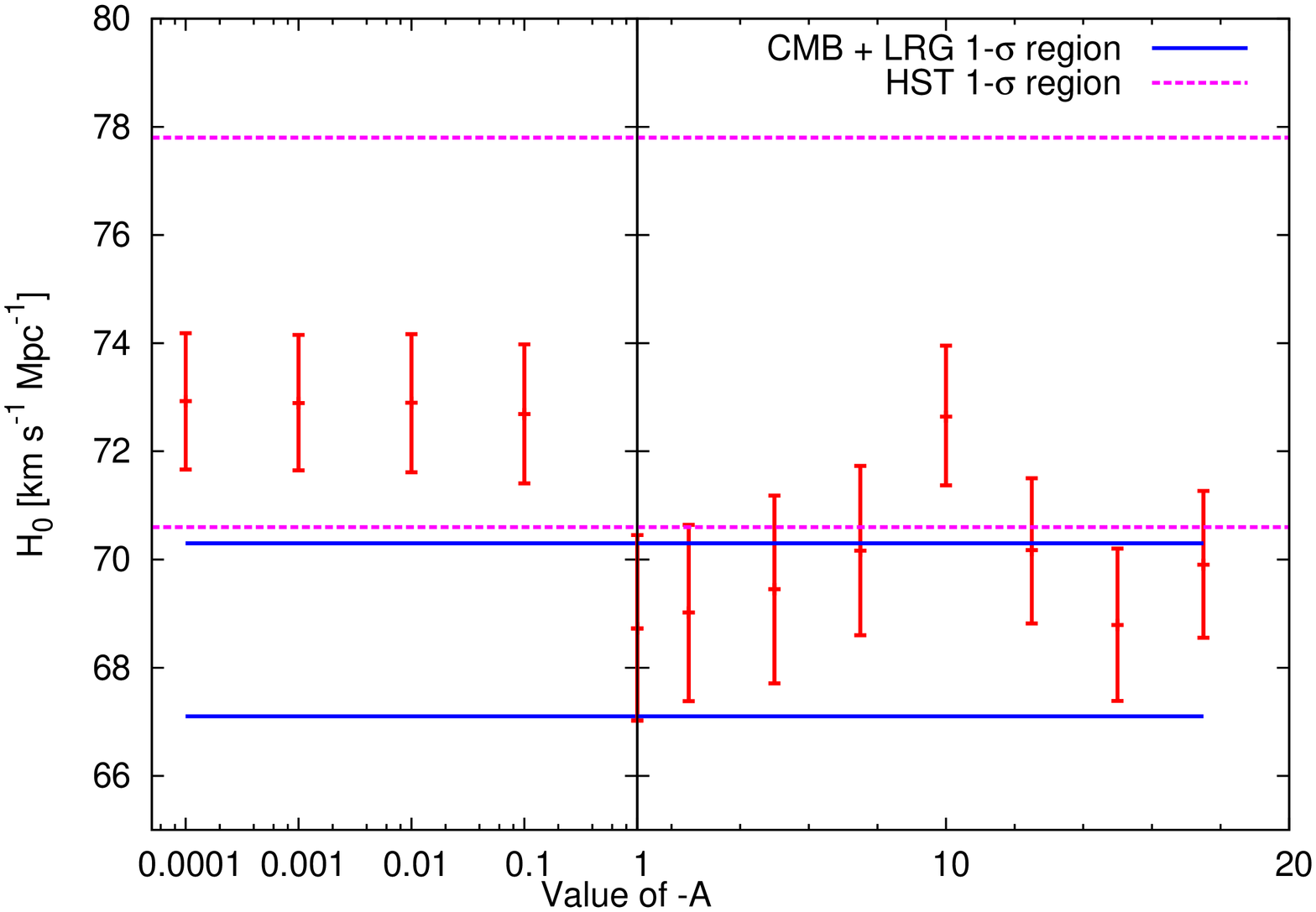}
\includegraphics[scale=0.33,angle=-90]{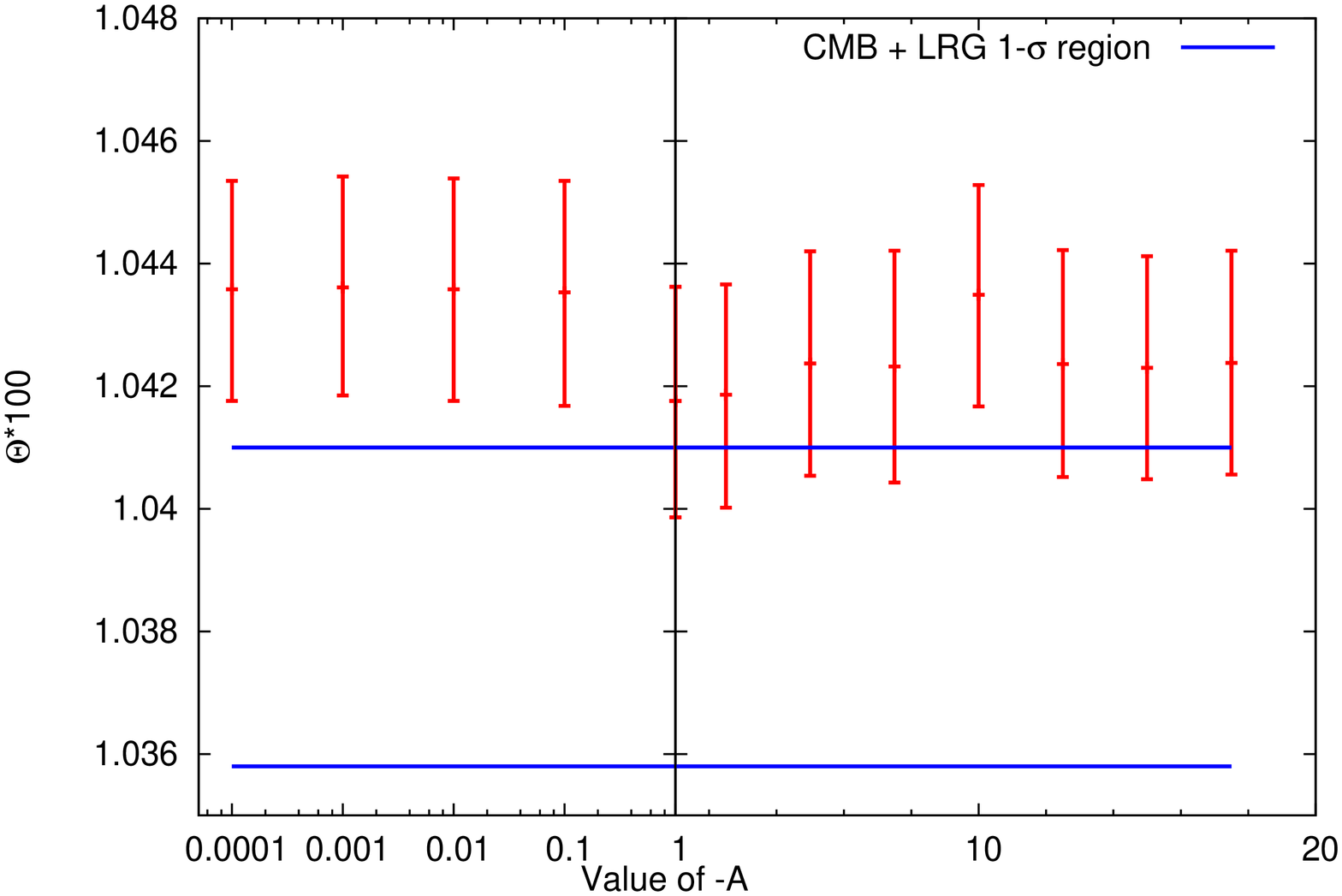}
\vspace{0.3cm}
\includegraphics[scale=0.33,angle=-90]{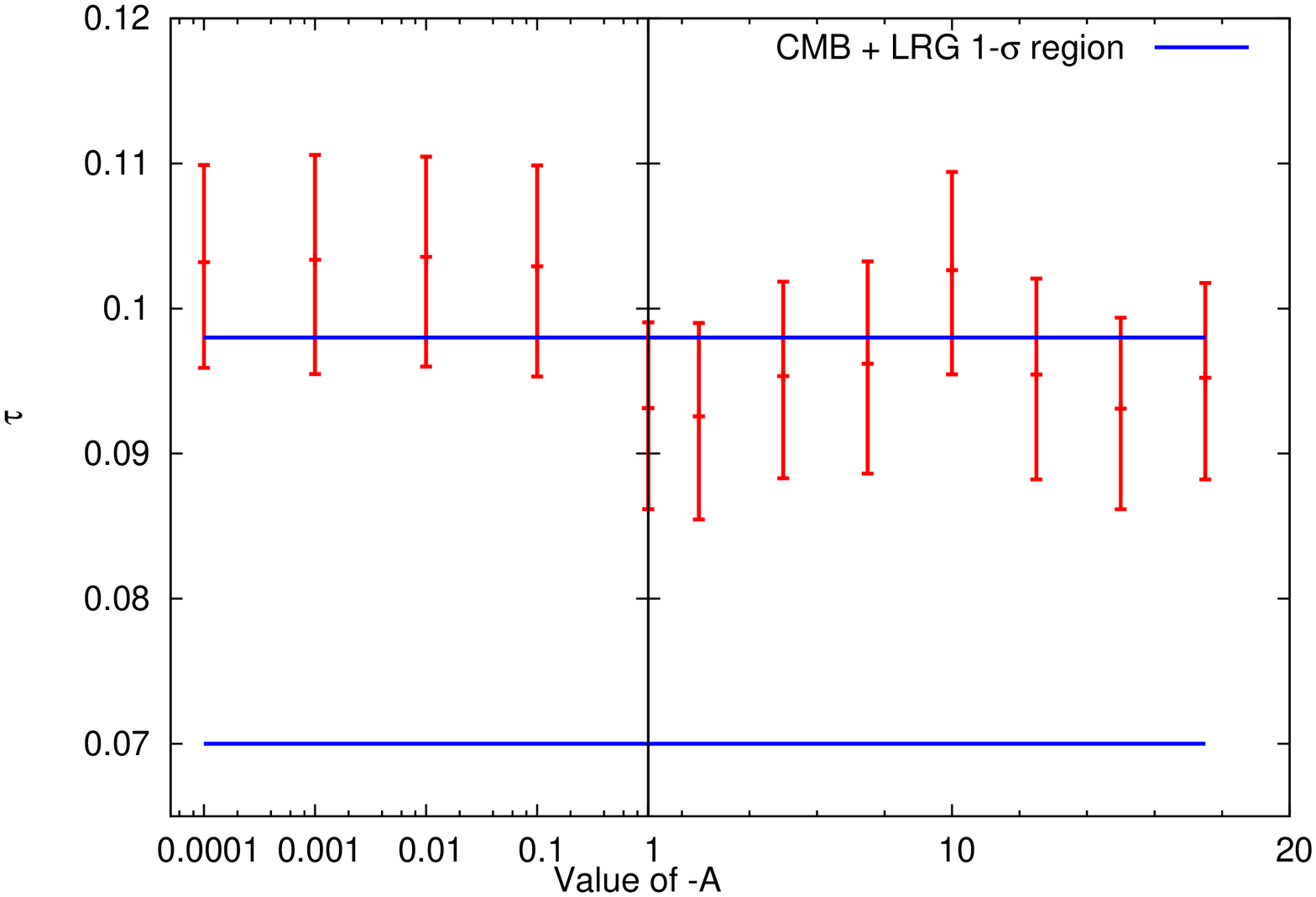}
\includegraphics[scale=0.33,angle=-90]{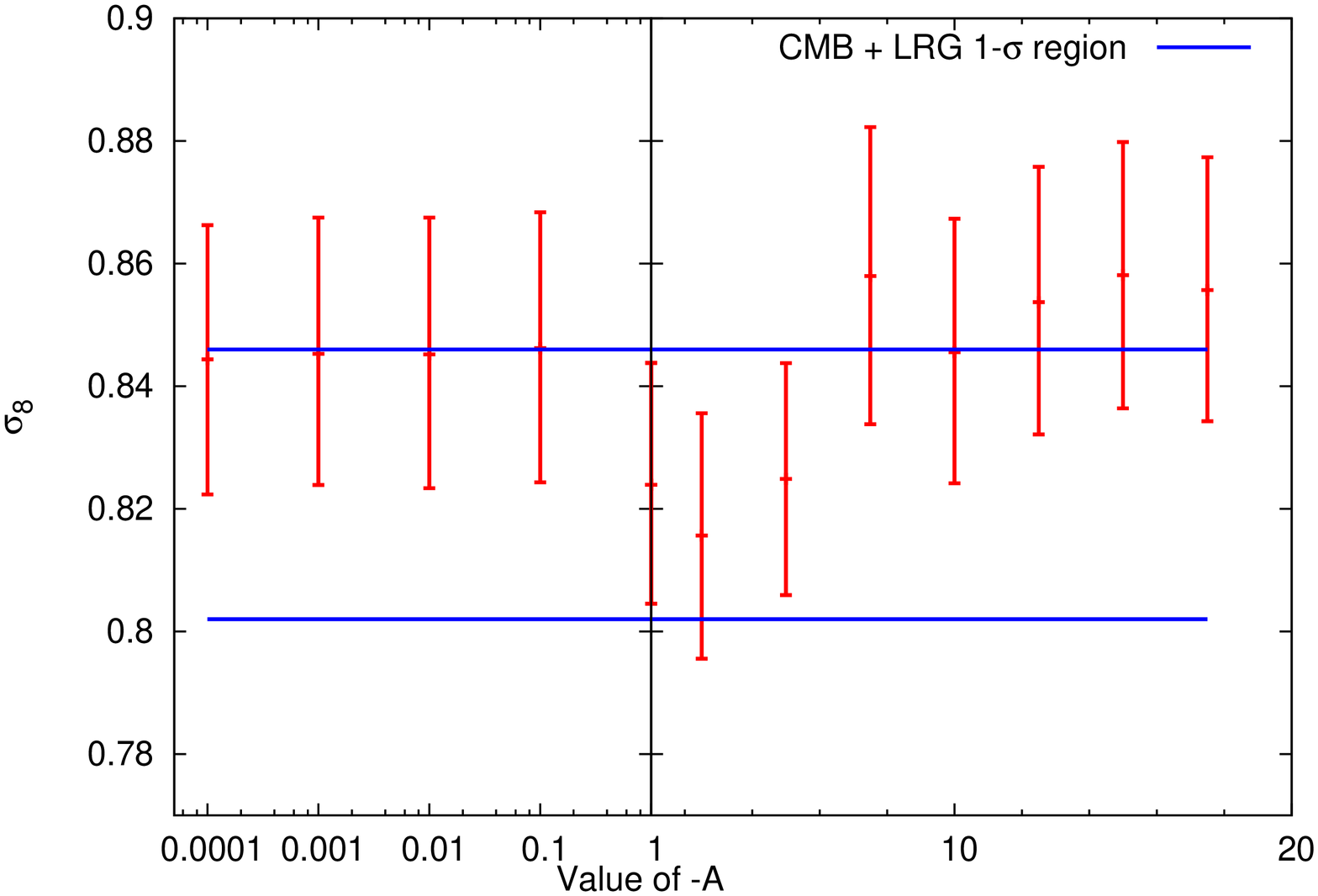}
\end{center}
\caption{Results for Model I: Best-fit parameter values and $1\sigma$
  errors for the cosmological parameters obtained for fixed negative values of
  $A$. Comparison with the values obtained by the WMAP collaboration
  (and other data sets, where relevant), is also shown. For convenience,
  the value of $-A$ is plotted in the $x$-axis. Values of $-20<A<-1$ belong to case i) described in the text, while values of $|A|< 1$ correspond to case ii).}
\label{cosmo-params_m1}
\end{figure}

\begin{figure}[ht!]
\begin{center}
\includegraphics[scale=0.33,angle=-90]{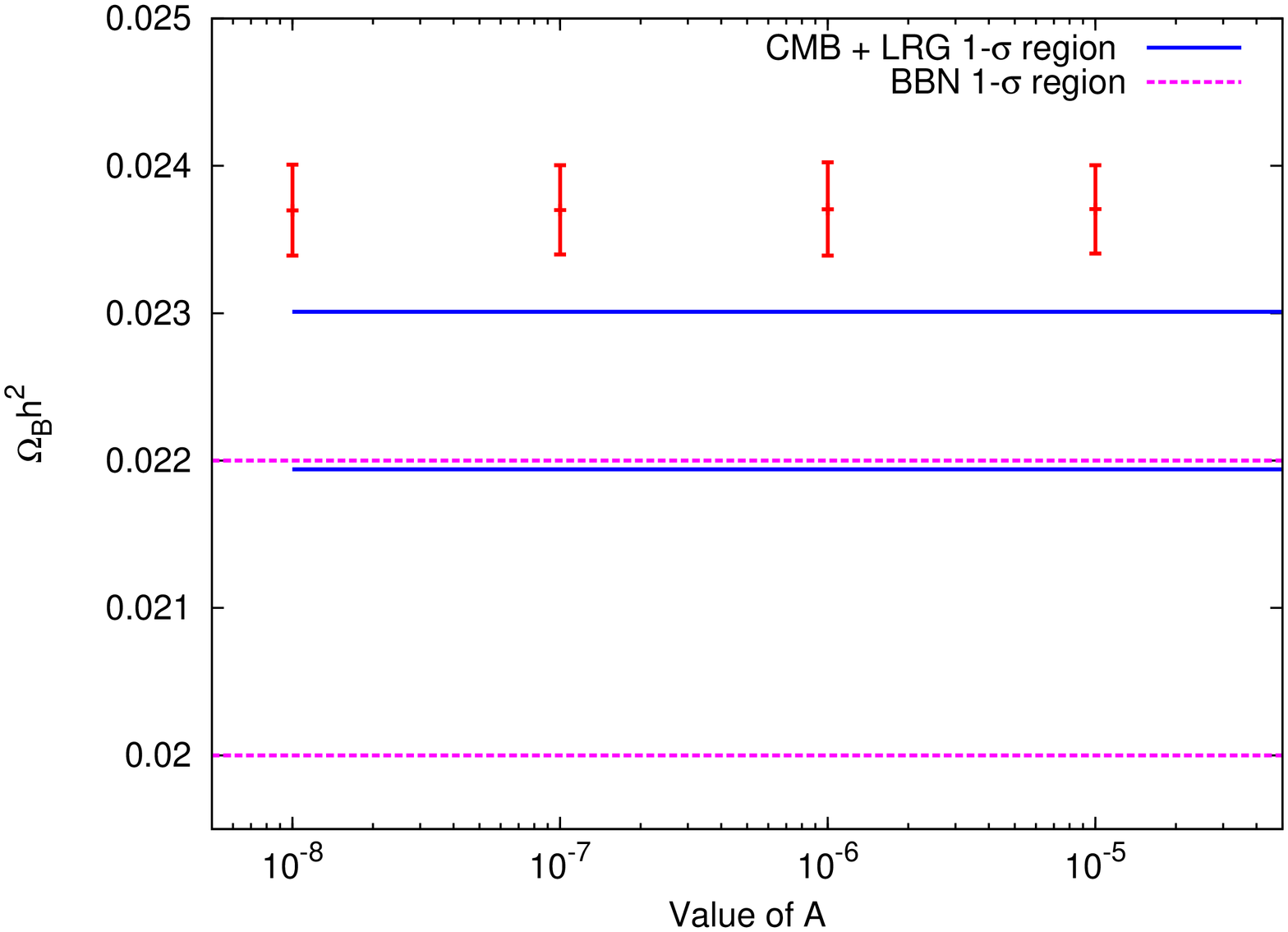}
\includegraphics[scale=0.33,angle=-90]{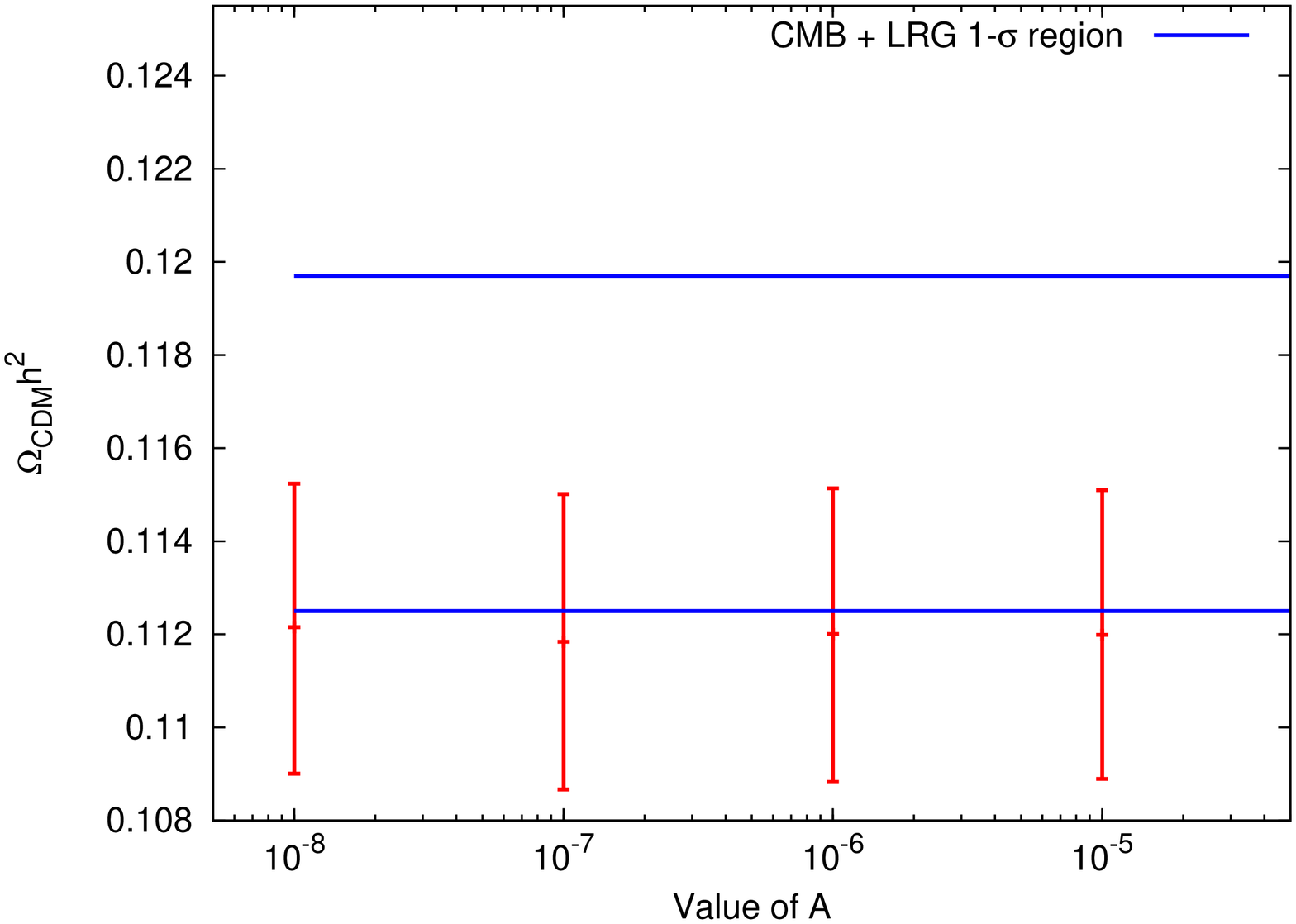}
\vspace{0.3cm}
\includegraphics[scale=0.33,angle=-90]{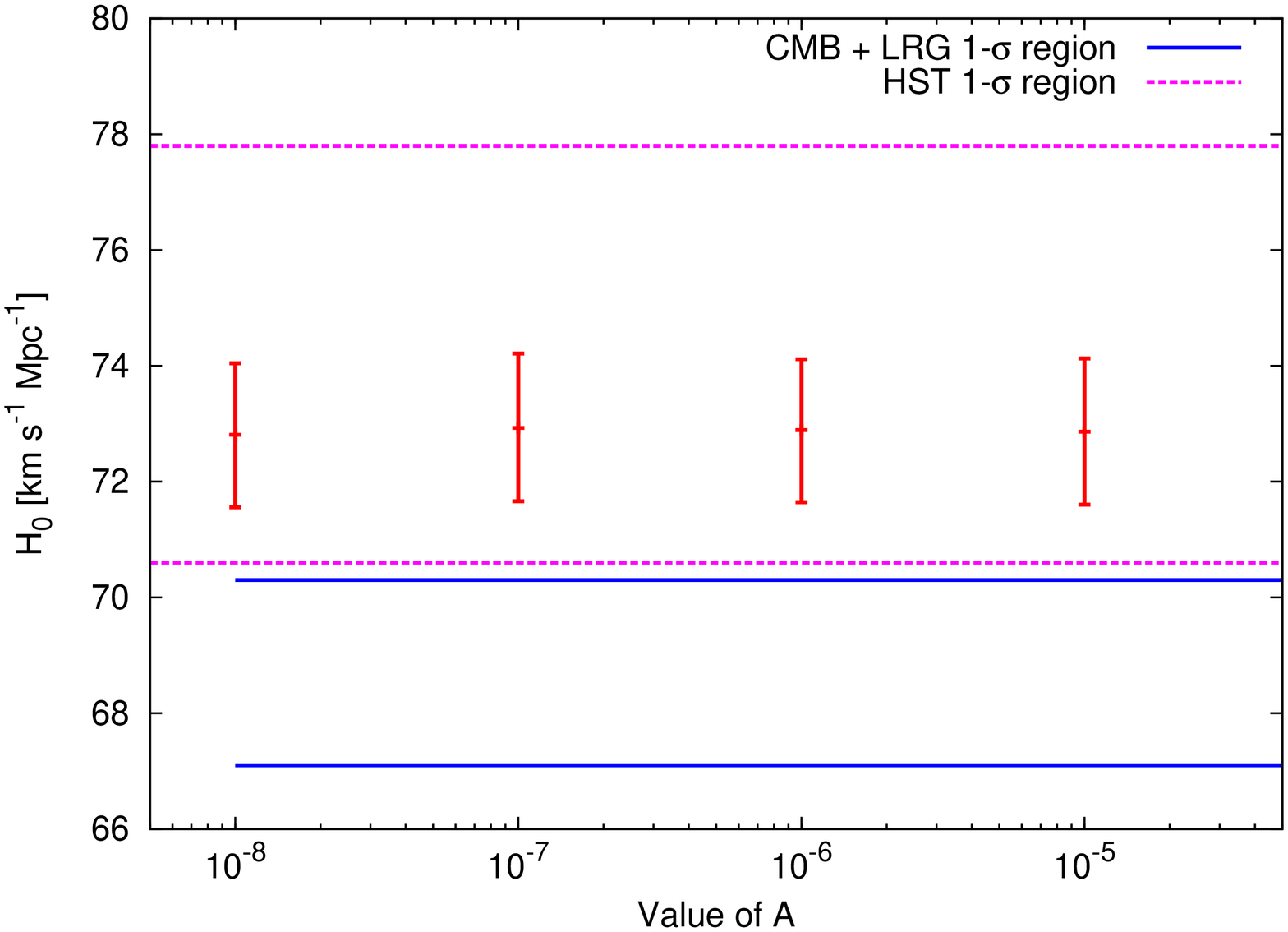}
\includegraphics[scale=0.33,angle=-90]{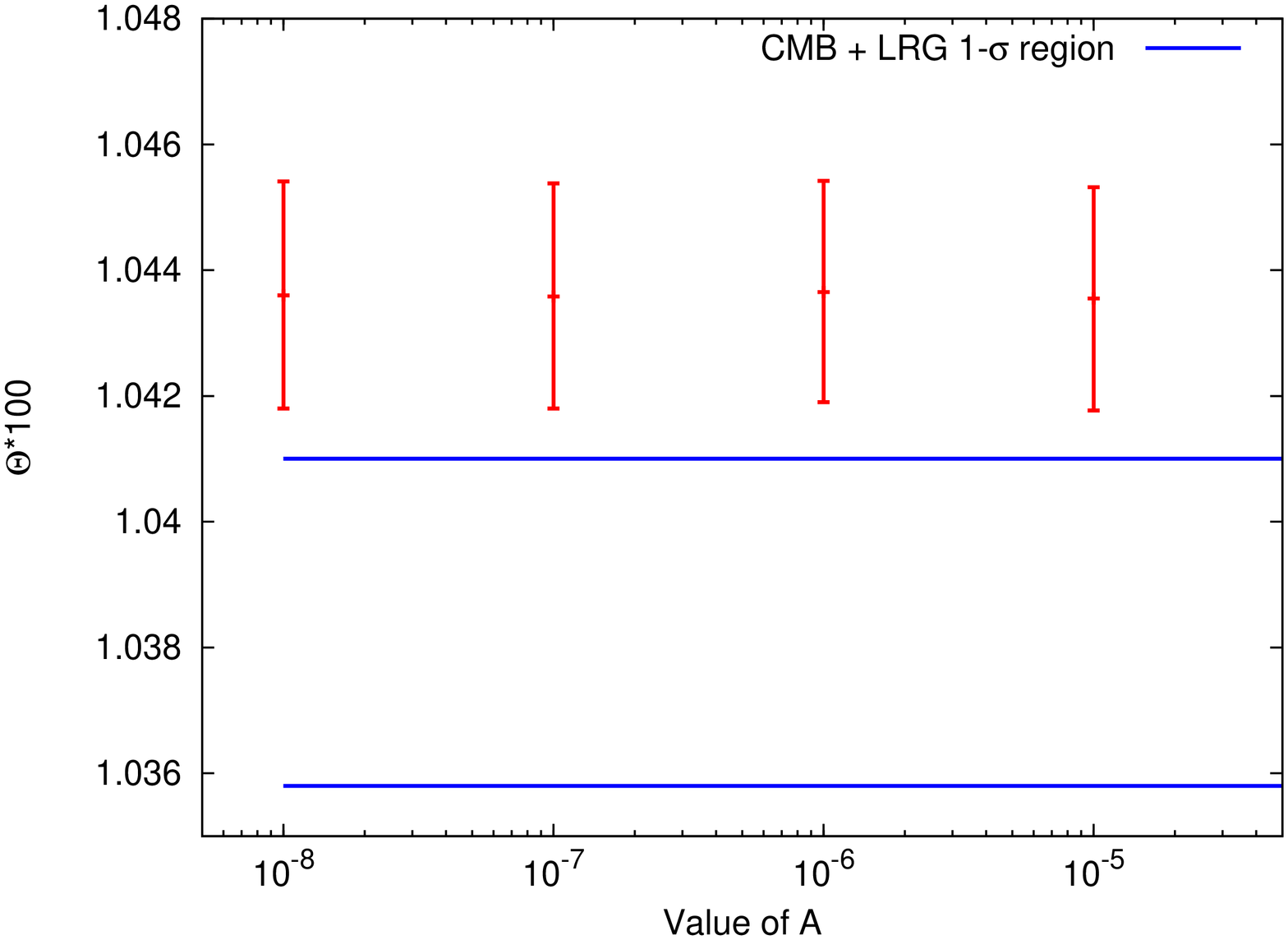}
\vspace{0.3cm}
\includegraphics[scale=0.33,angle=-90]{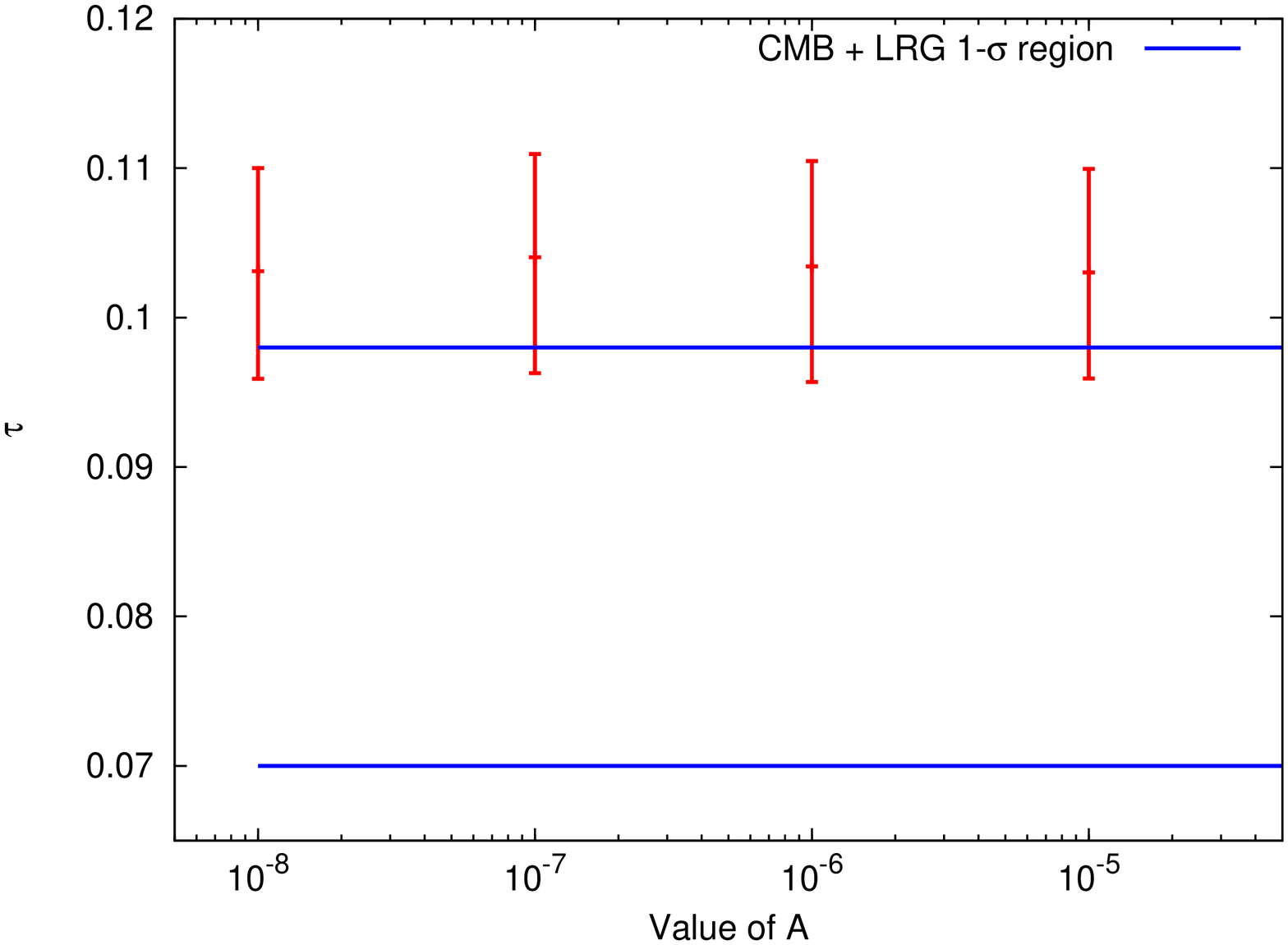}
\includegraphics[scale=0.33,angle=-90]{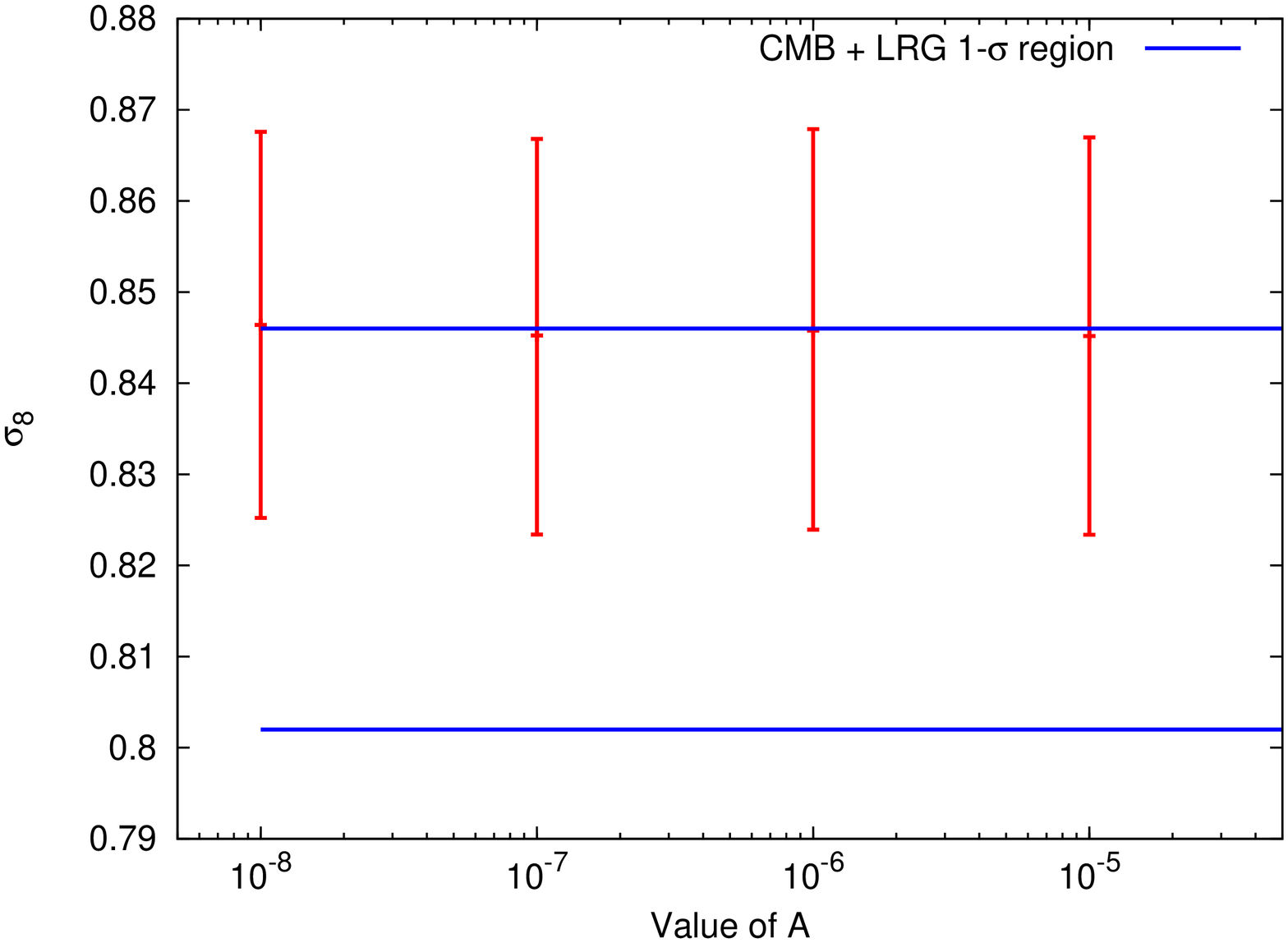}
\end{center}
\caption{Results for Model I: Best-fit parameter values and $1\sigma$
  errors for the cosmological parameters obtained for fixed positive values of
  $A$. Comparison with the values obtained by the WMAP collaboration
  (and other data sets, where relevant), is also shown. All of these
  values correspond to case ii) described in the text.}
\label{cosmo-params_m1_pos}
\end{figure}

\begin{figure}[ht!]
\begin{center}
\includegraphics[scale=0.33,angle=-90]{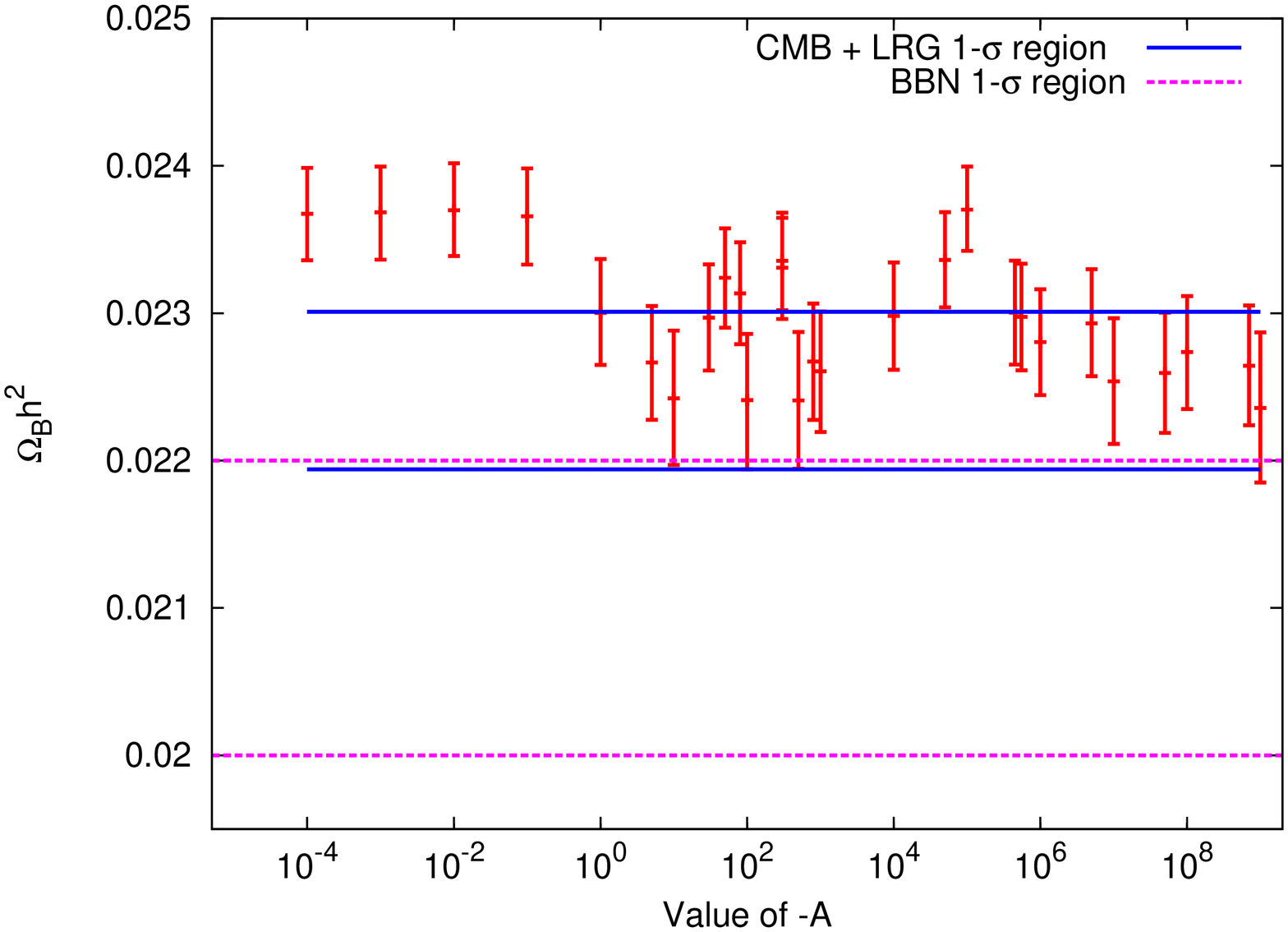}
\includegraphics[scale=0.33,angle=-90]{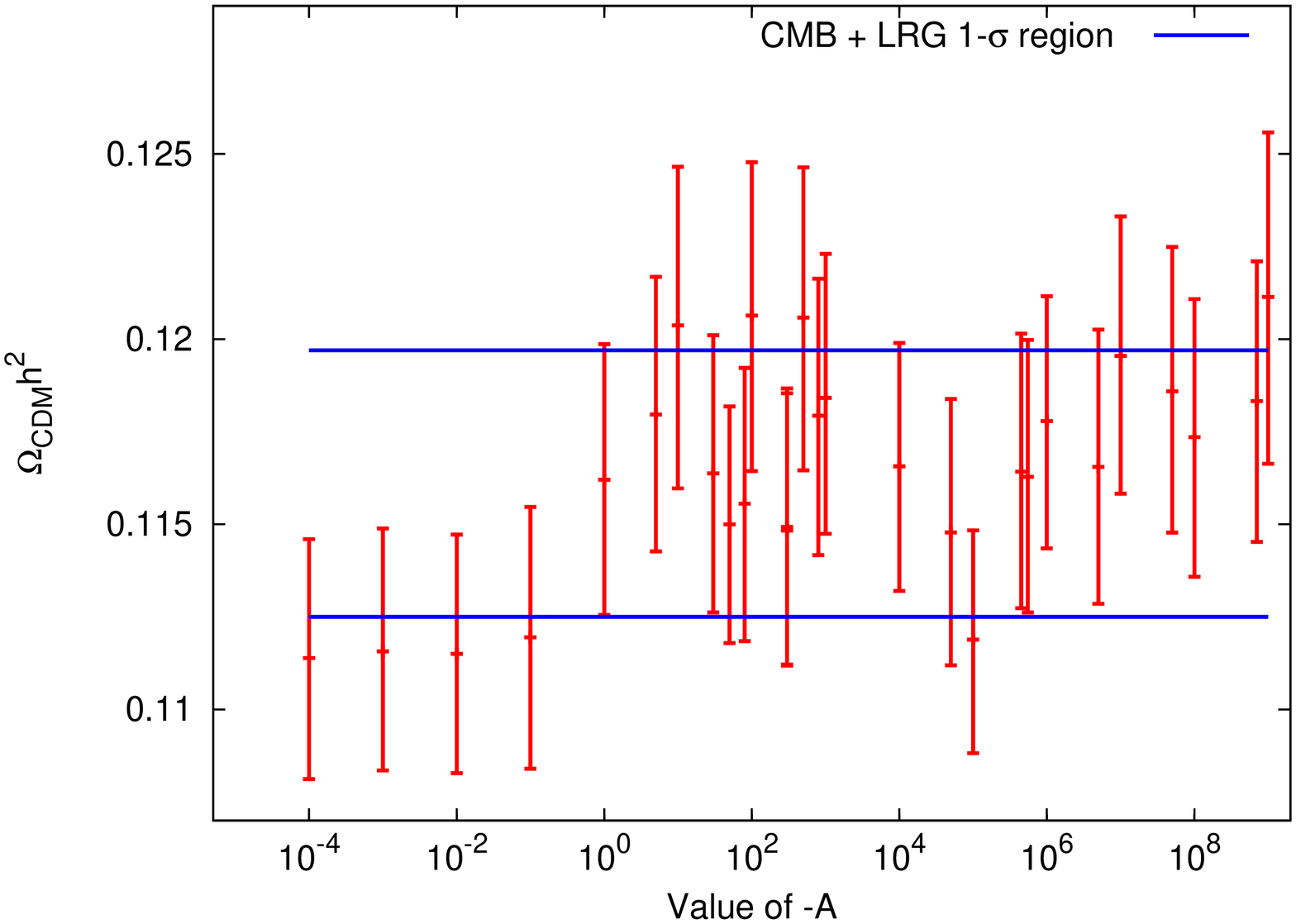}
\vspace{0.3cm}
\includegraphics[scale=0.33,angle=-90]{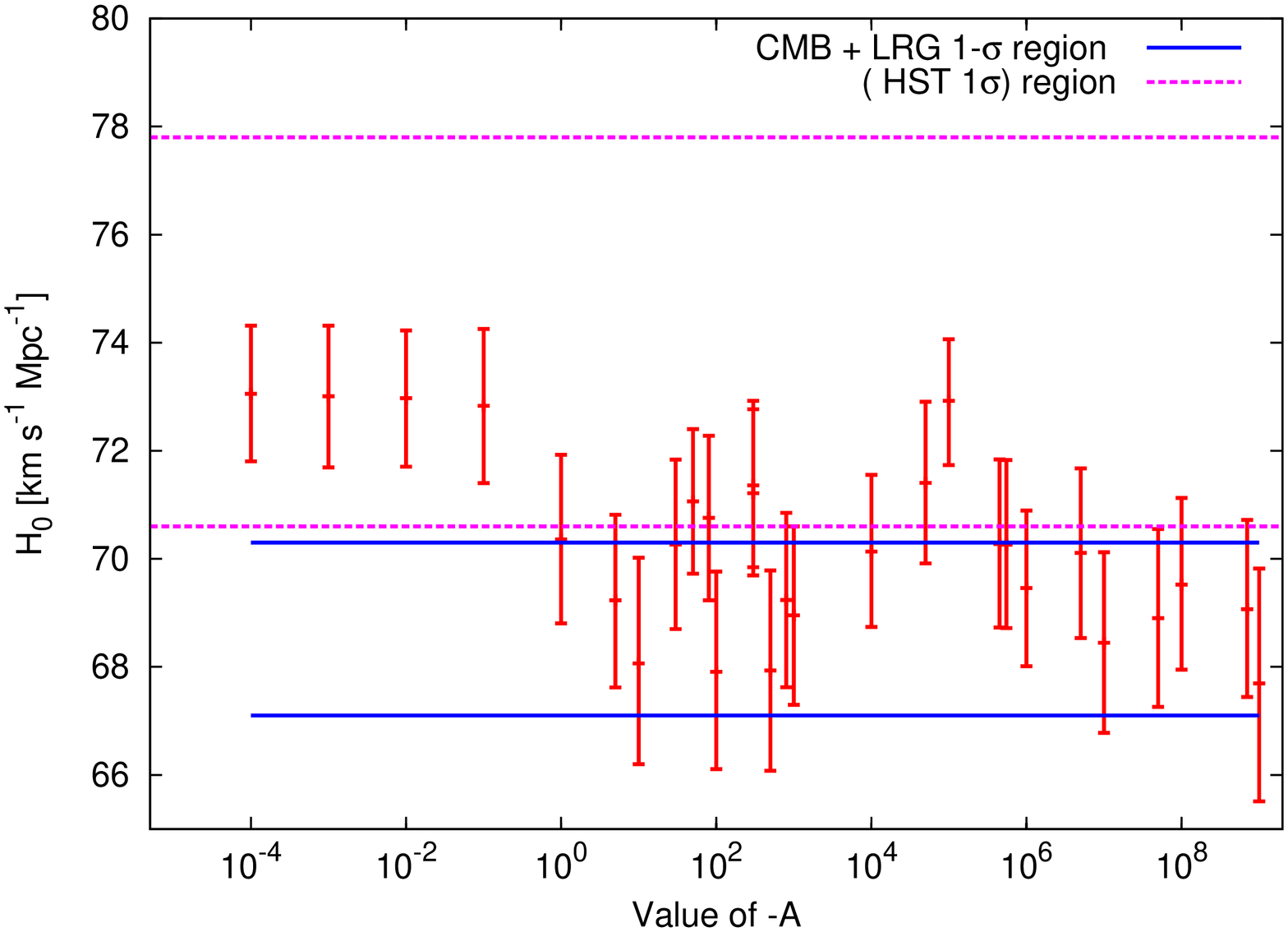}
\includegraphics[scale=0.33,angle=-90]{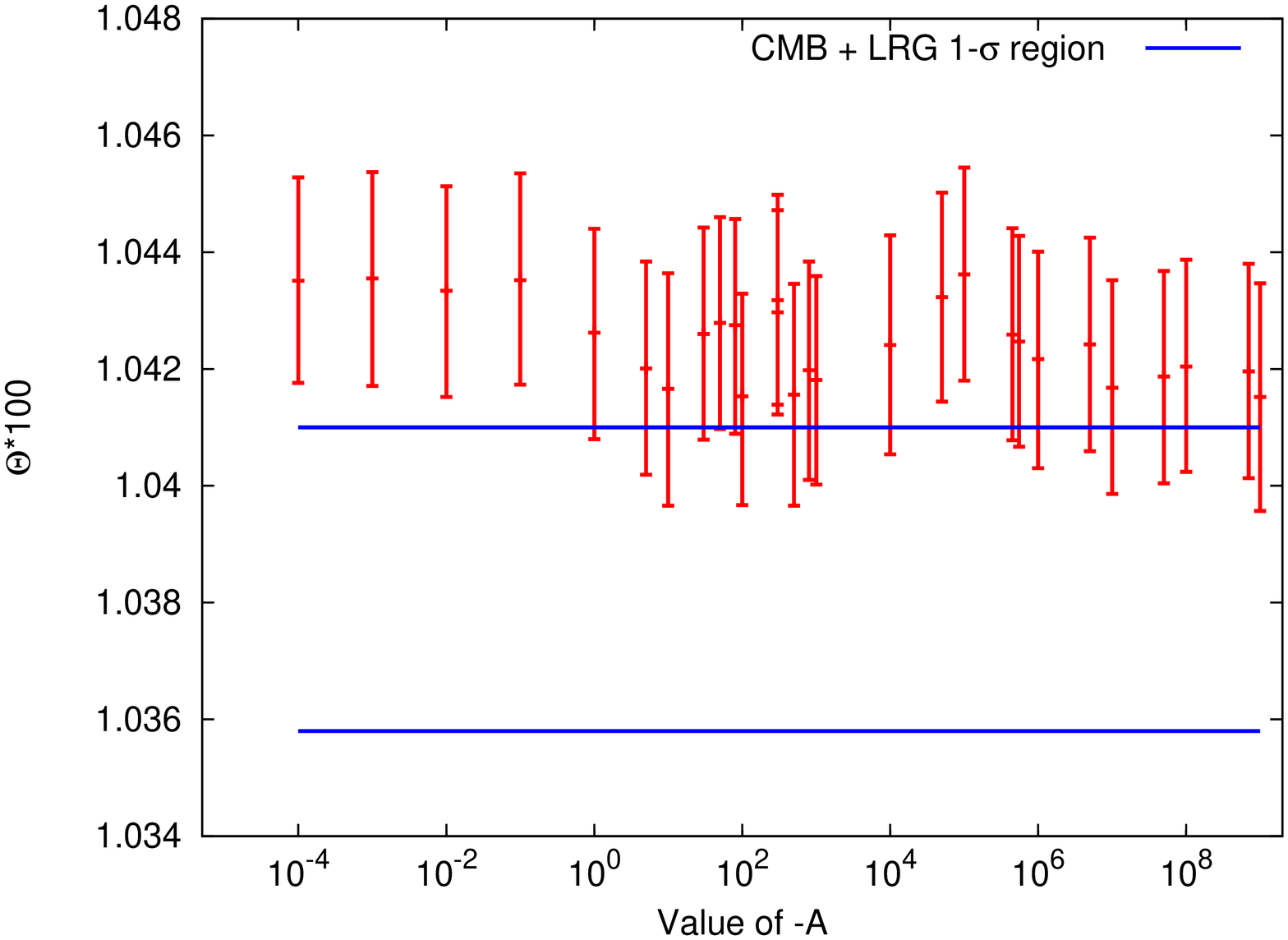}
\vspace{0.3cm}
\includegraphics[scale=0.33,angle=-90]{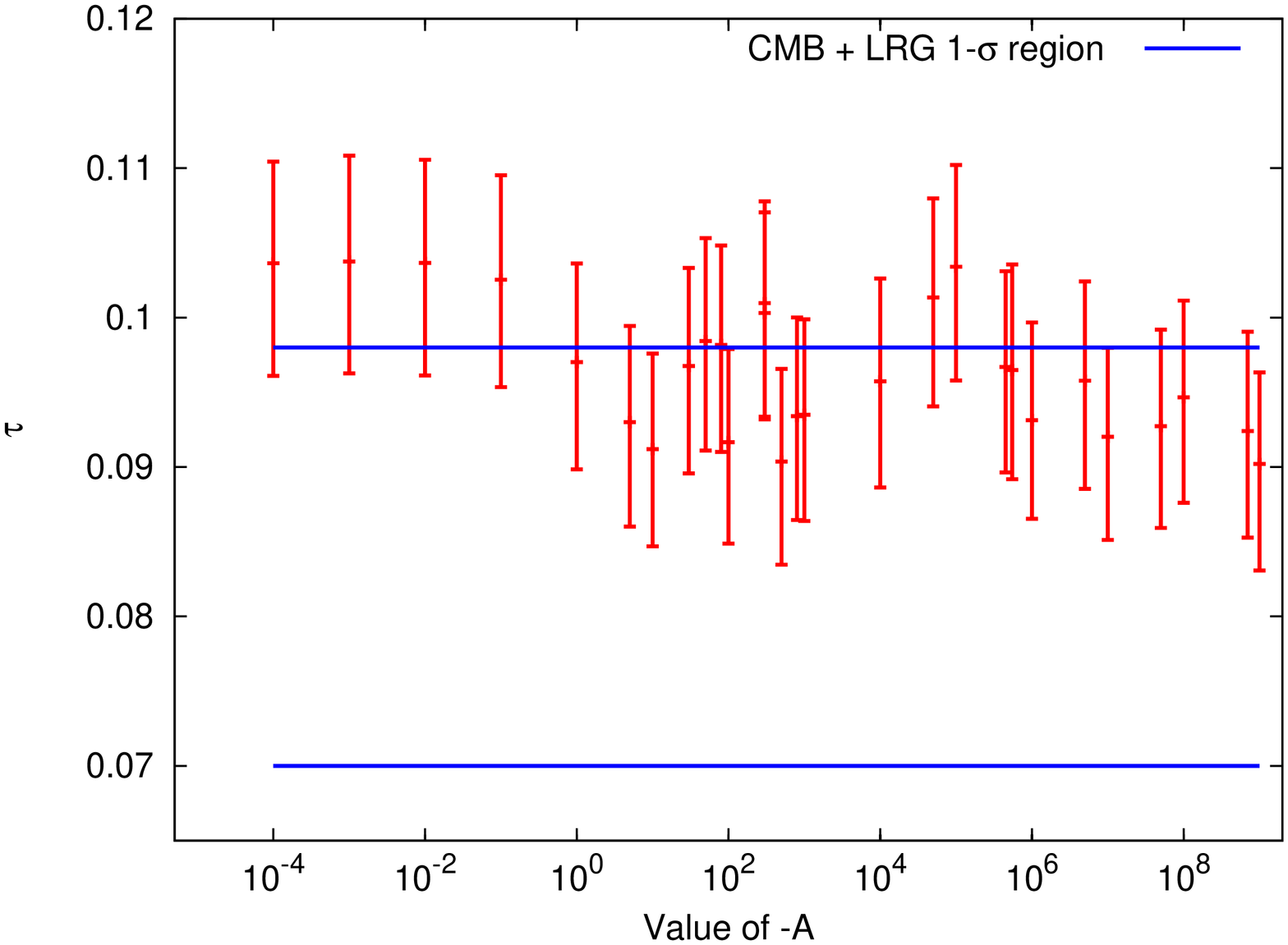}
\includegraphics[scale=0.33,angle=-90]{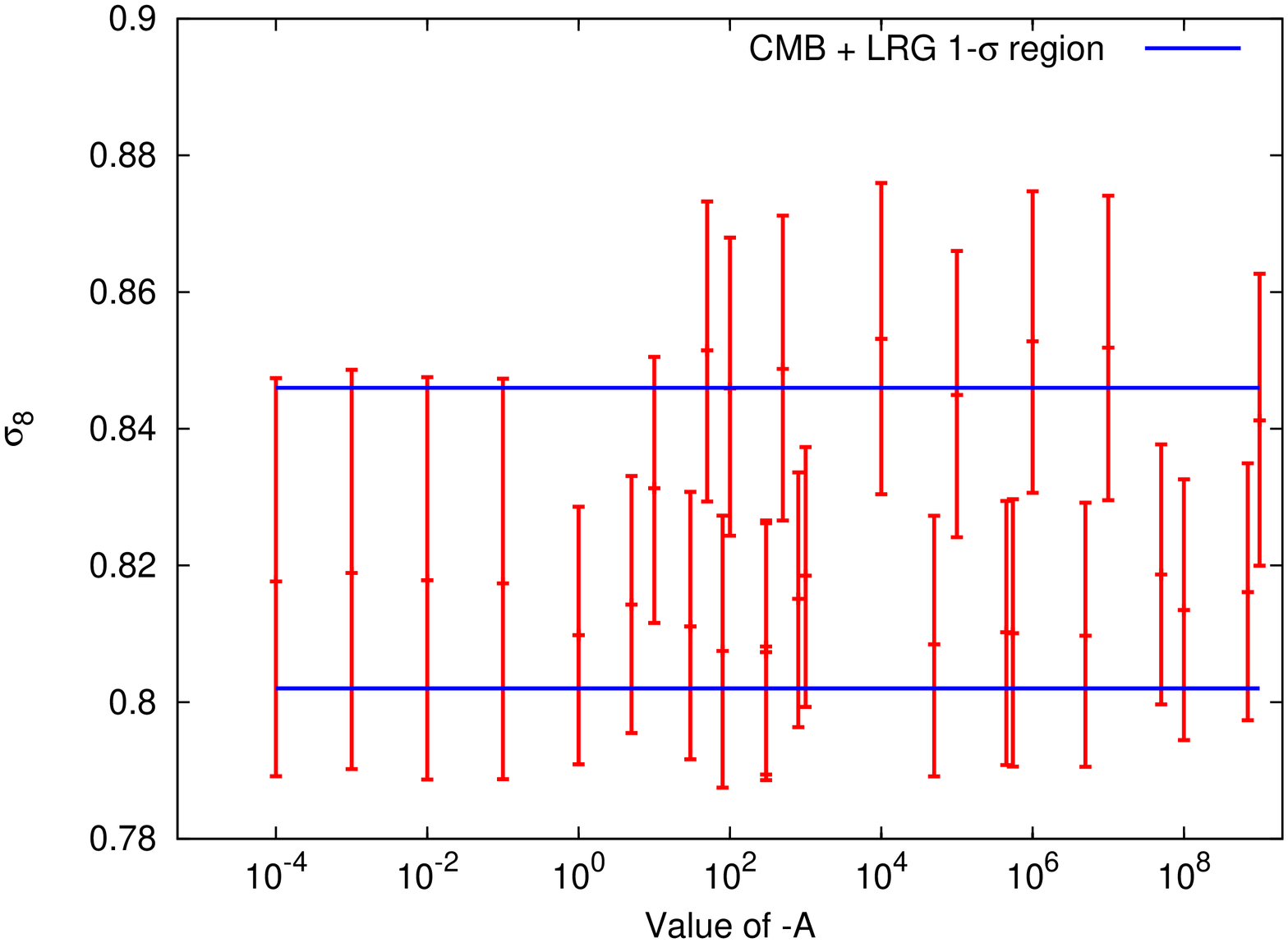}
\end{center}
\caption{Results for Model II: Best-fit parameter values and $1\sigma$
  errors for the cosmological parameters obtained for fixed negative
  values of $A$. Comparison with the values obtained by the WMAP
  collaboration (and other data sets, where relevant), is also
  shown. For convenience, the value of $-A$ is plotted in the
  $x$-axis. Values of $|A|> 1$ belong to case i) described in the text,
  while values of $|A|< 1$ correspond to case ii).}
\label{cosmo-params_m2}
\end{figure}

\begin{figure}[ht!]
\begin{center}
\includegraphics[scale=0.33,angle=-90]{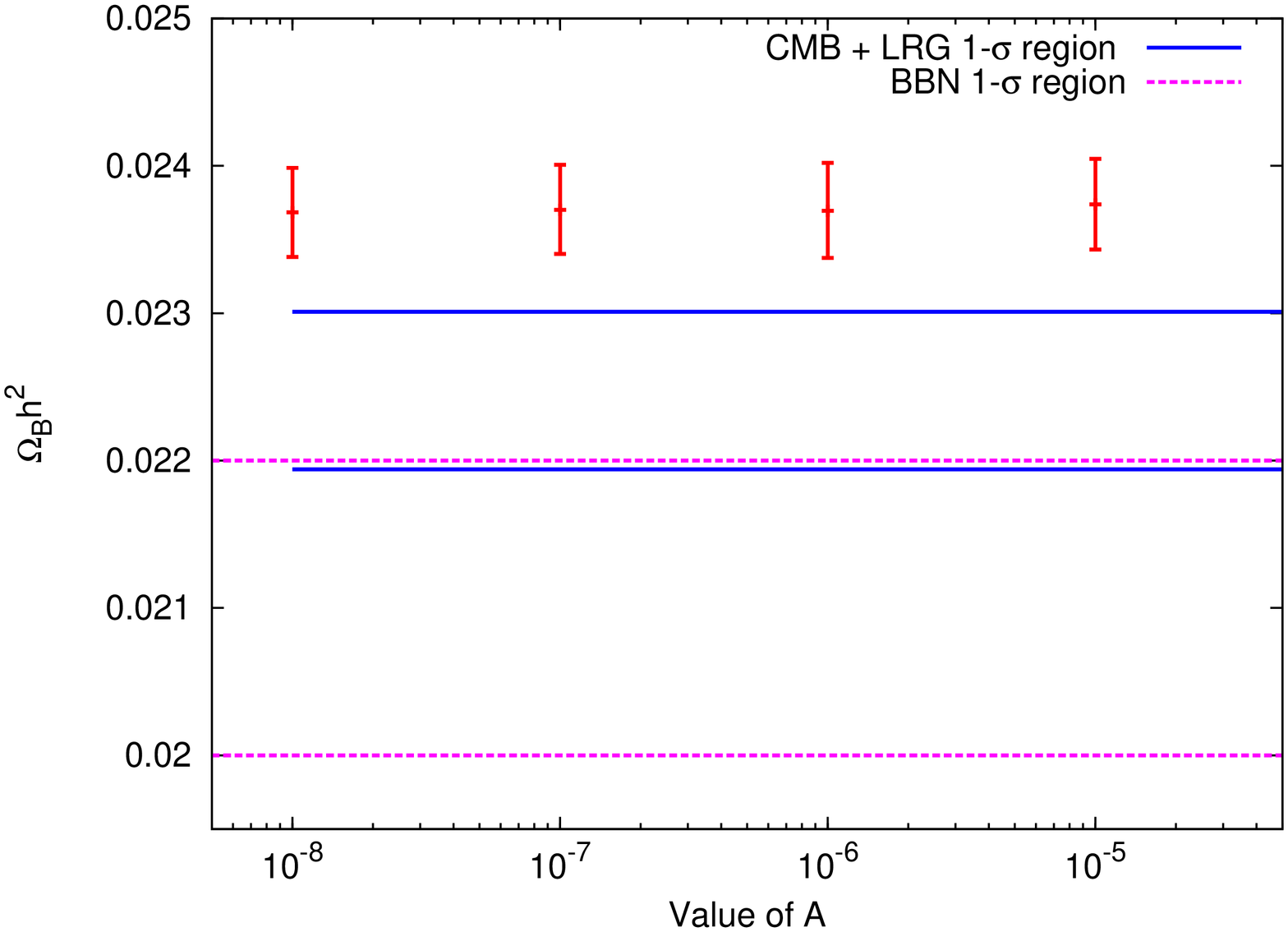}
\includegraphics[scale=0.33,angle=-90]{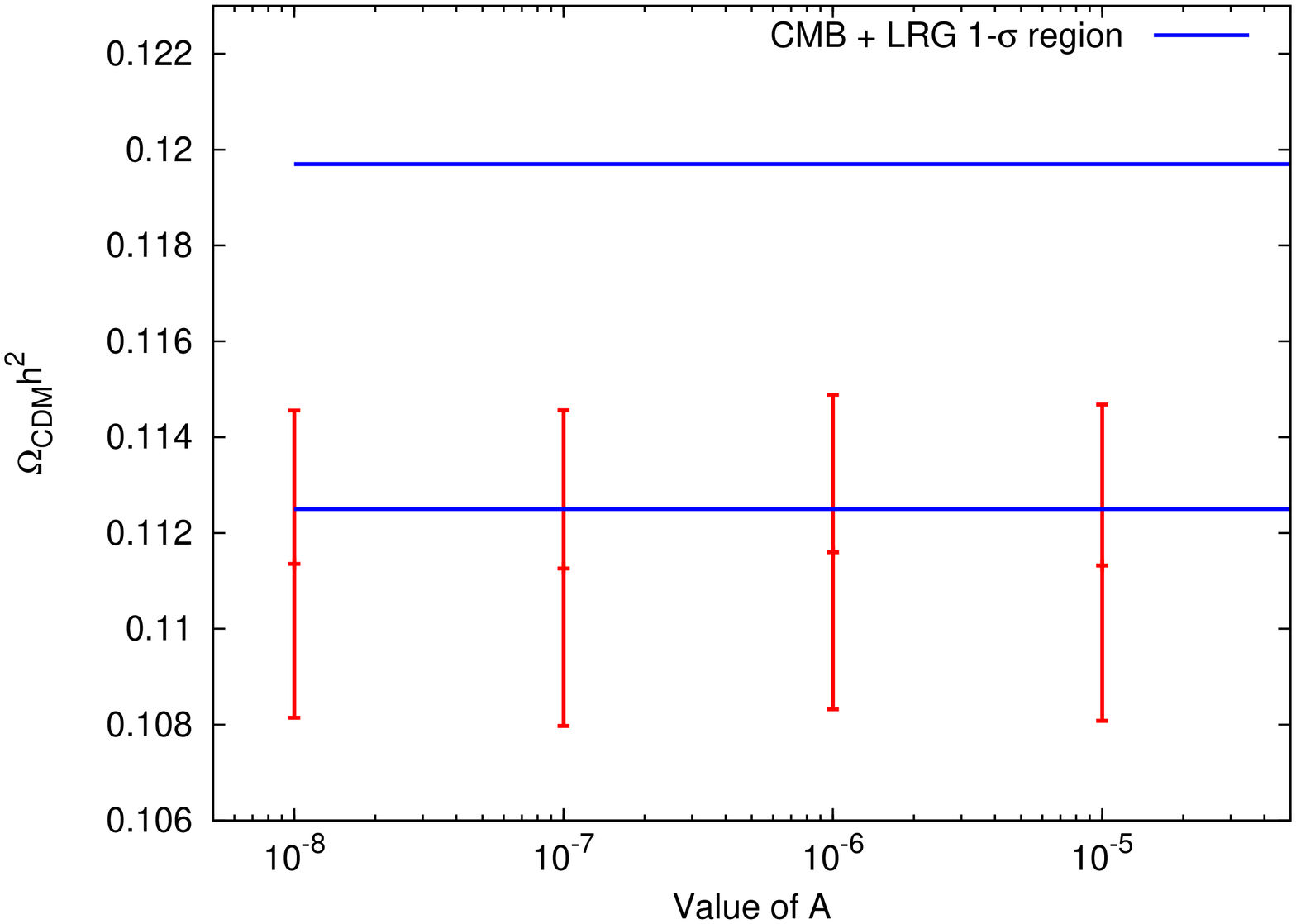}
\vspace{0.3cm}
\includegraphics[scale=0.33,angle=-90]{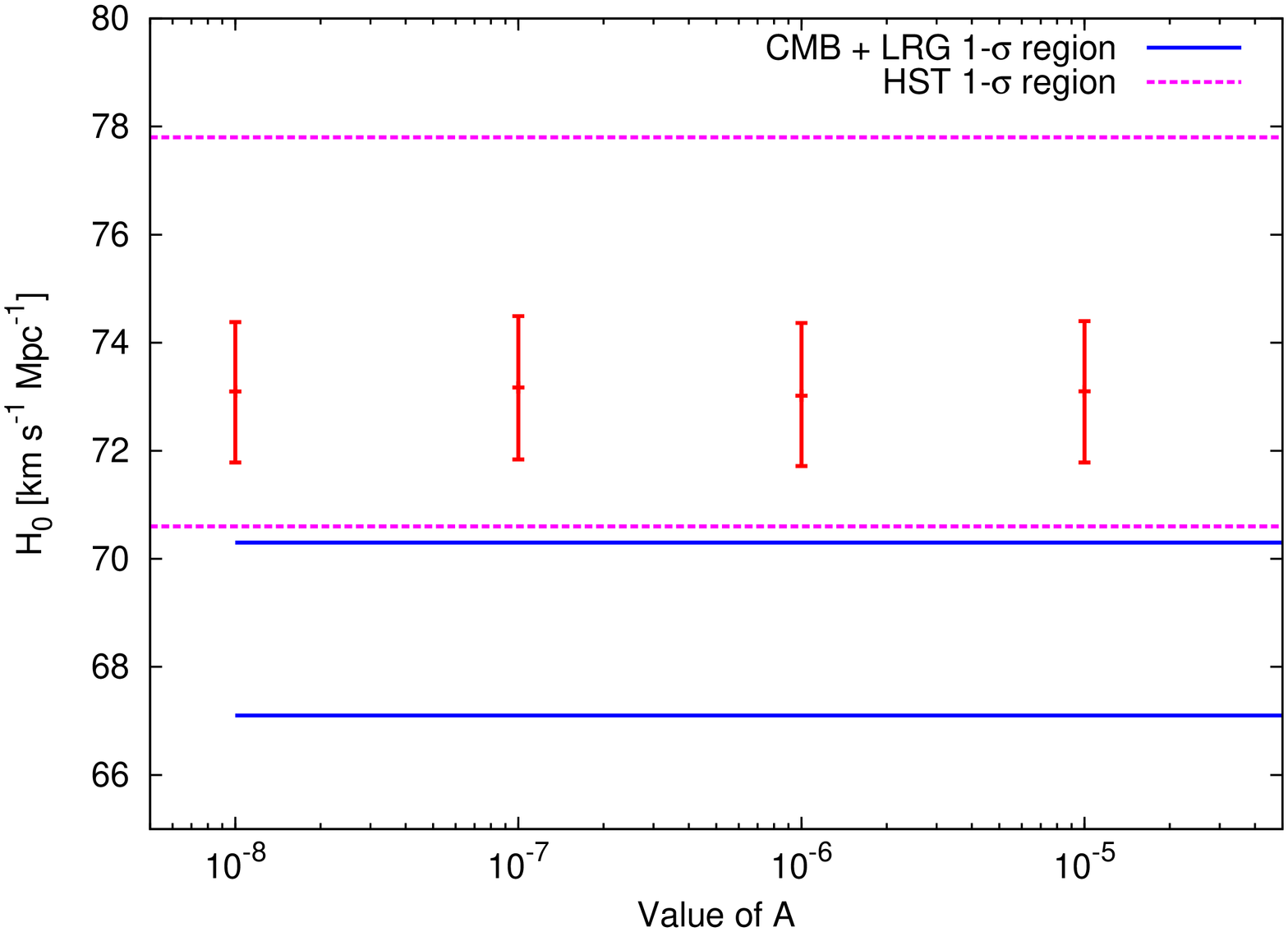}
\includegraphics[scale=0.33,angle=-90]{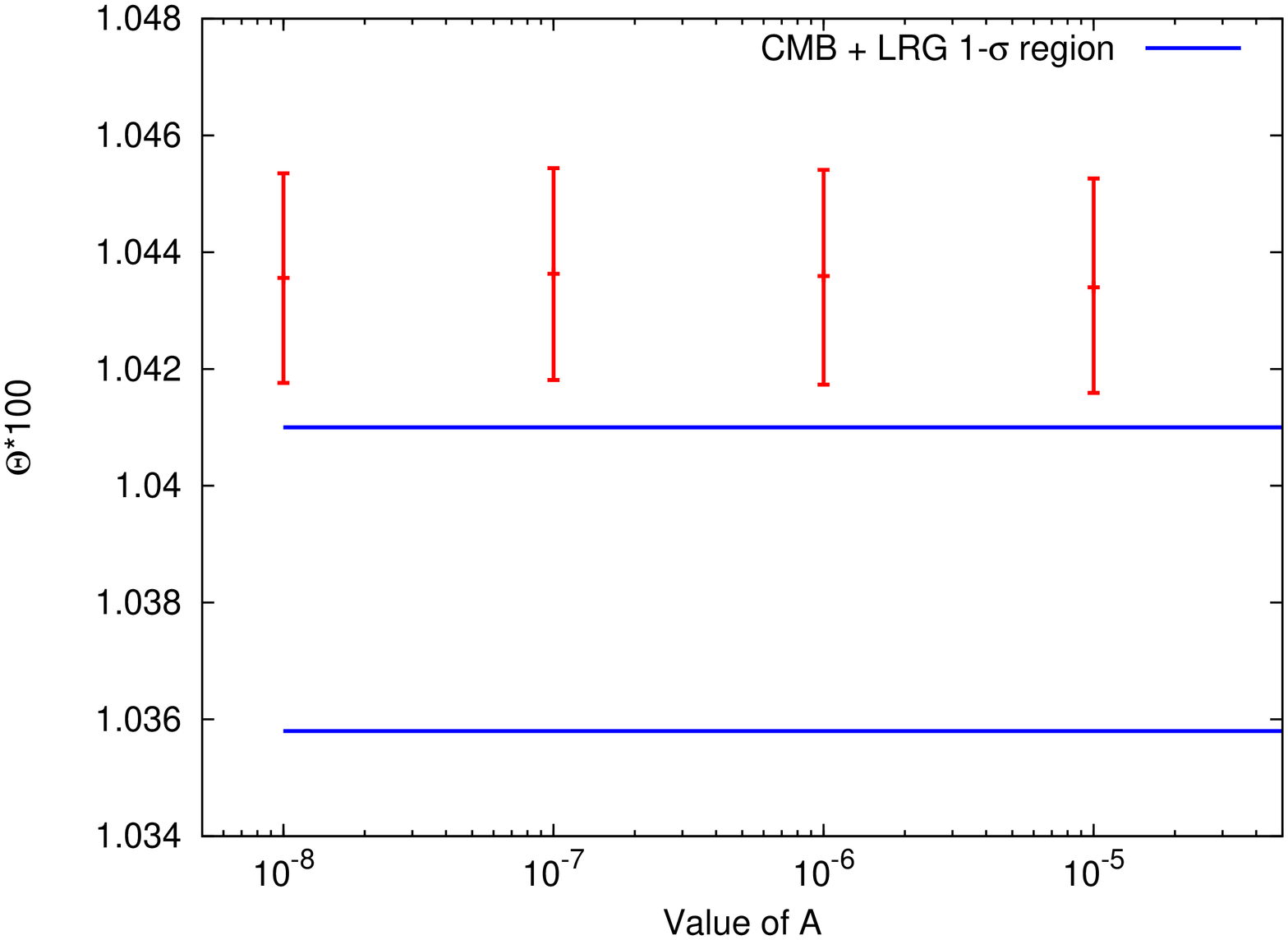}
\vspace{0.3cm}
\includegraphics[scale=0.33,angle=-90]{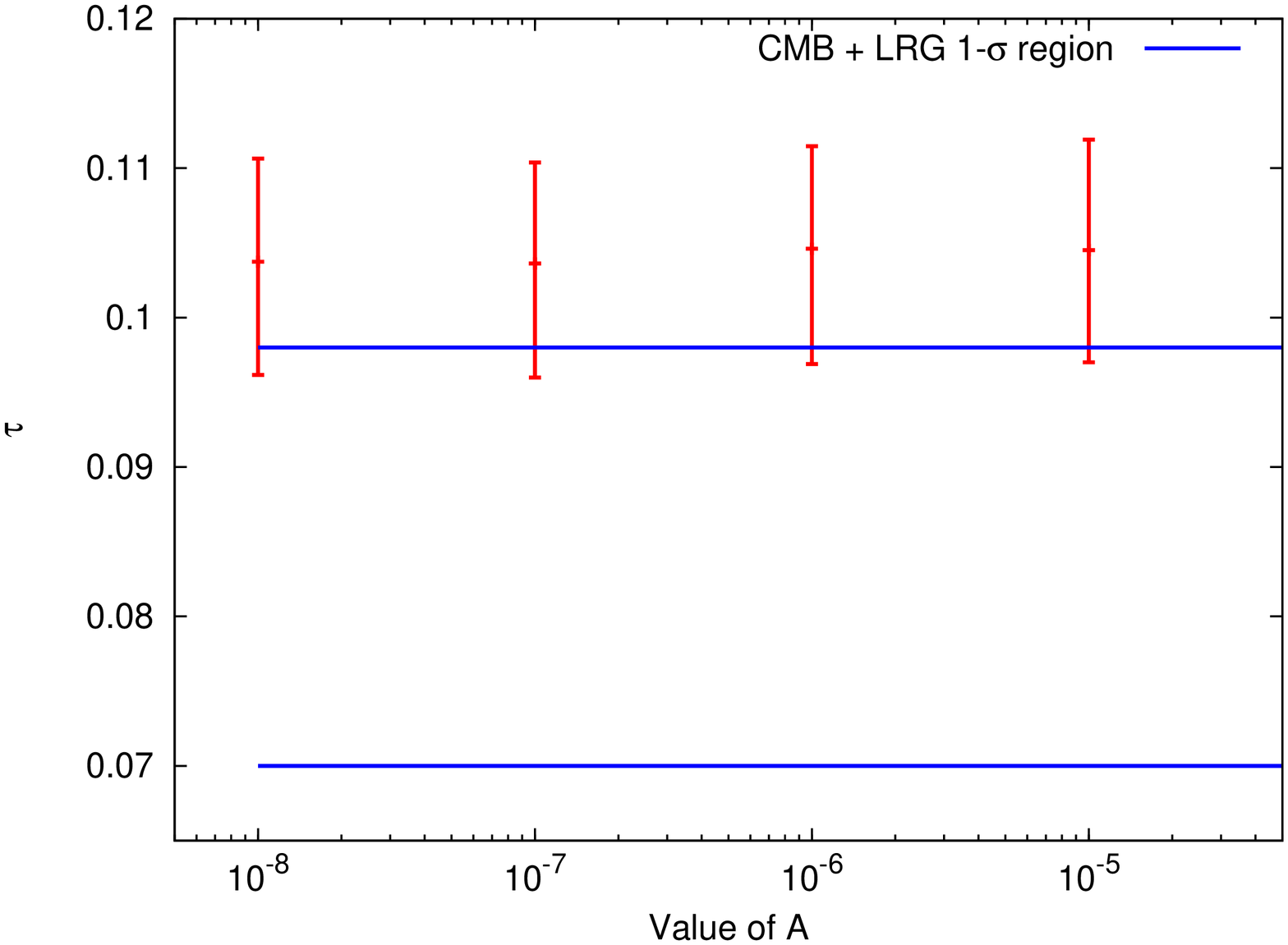}
\includegraphics[scale=0.33,angle=-90]{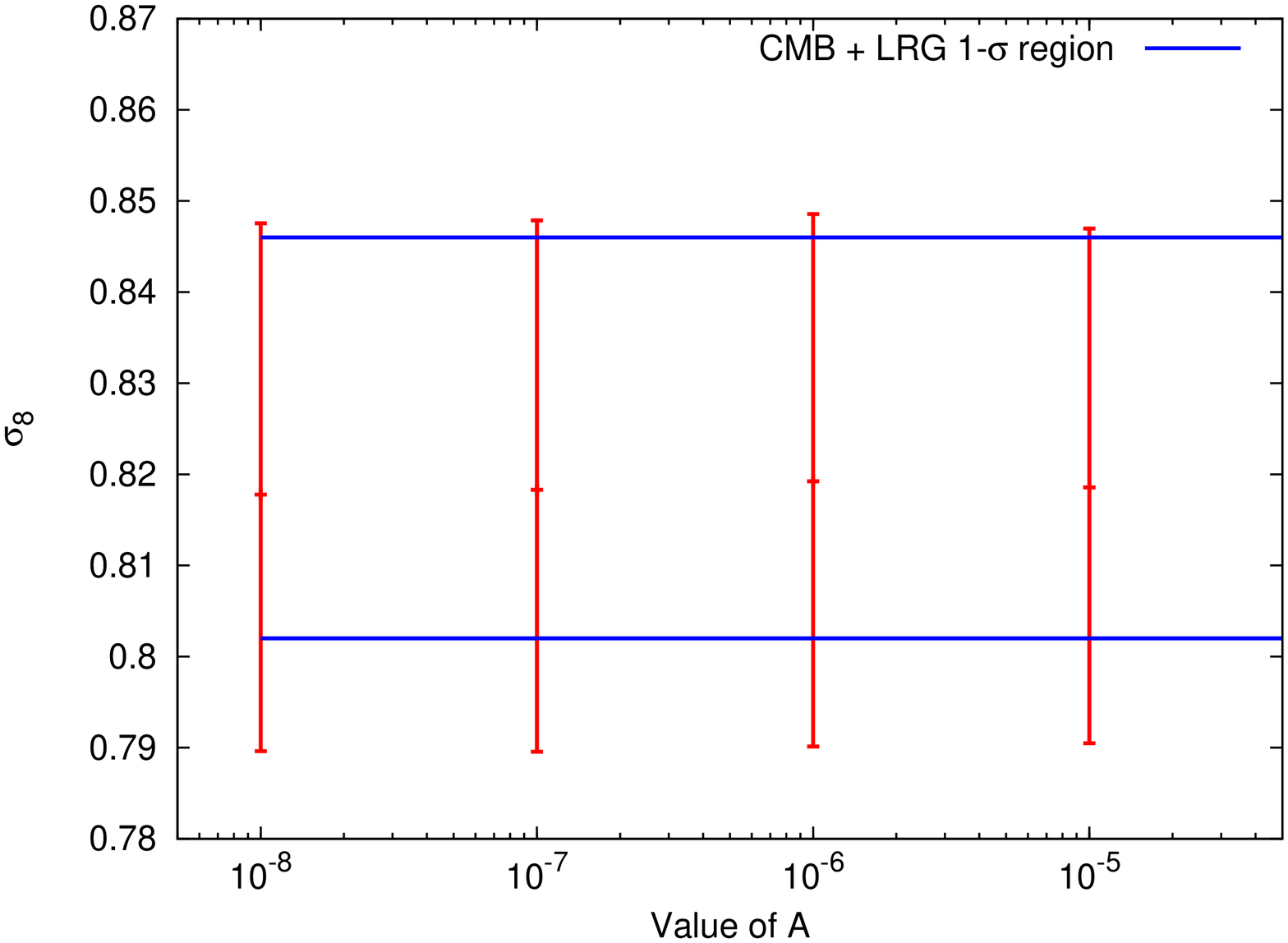}
\end{center}
\caption{Results for Model II: Best-fit parameter values and $1\sigma$
  errors for the cosmological parameters obtained for fixed positive
  values of $A$. Comparison with the values obtained by the WMAP
  collaboration (and other data sets, where relevant), is also
  shown. All of these values correspond to case ii) described in the
  text.}
\label{cosmo-params_m2_pos}
\end{figure}

The results of the statistical analysis for the cosmological
parameters are shown in Figs.~\ref{cosmo-params_m1},
\ref{cosmo-params_m1_pos}, \ref{cosmo-params_m2} and
\ref{cosmo-params_m2_pos} for Models I and II.  We can distinguish two
behaviors: i) Model I with $-20<A<-1$; Model II with $A<-1$; ii)
Models I with $|A|<1$; Model II with $|A|< 1$. Let us remind that for
$|A|>20$ in Model I we recover the scale-invariant HZ spectrum and
therefore the constraints on the cosmological parameters are those
estimated by the WMAP collaboration \cite{wmap7}.  For models included
in case i), most of the estimated values for the cosmological
parameters within the collapse models are in agreement with those
obtained by the WMAP collaboration using the 7-year data release and
considering a standard $\Lambda$CDM model. The values obtained for
$\Theta$ are marginally consistent with those obtained by the WMAP
collaboration. However, there is a better agreement in the results
obtained for $H_0$, which is derived from $\Theta$ and other
cosmological parameters.  Let us now discuss some exceptions.  For the
case $A = -10$ in Model I, the values obtained for $\Omega_b h^2$ and
$H_0$ are at variance with the values obtained by the WMAP
collaboration within $2\sigma$, while there is agreement within
$3\sigma$. For the case $A=-10^5$ in Model II, the values obtained for
$\Omega_b h^2$ and $H_0$ are at variance with the values obtained by
the WMAP collaboration within $1\sigma$, while there is an agreement
within $2\sigma$.  On the other hand, it is important to discuss
consistency of our results with independent data such as the baryon
density inferred from BBN \cite{MP09,Steigman10}\footnote{Recall that
  have put a gaussian prior on the baryon density using constraints
  inferred from BBN.} and the constraint on the Hubble constant
presented in Ref.~\cite{Riess09}. It should be noted that the $1
\sigma$ BBN region for $\Omega_b h^2$ shown in
Figs.~\ref{cosmo-params_m1}, \ref{cosmo-params_m1_pos},
\ref{cosmo-params_m2} and \ref{cosmo-params_m2_pos} refers to the
constraint obtained using the observational abundances of deuterium,
which is the most stringent one. However, one should keep in mind that
by considering the $^4{\rm He}$ abundance, the constraint on the
baryon density softens substantially, and in fact it leads to a region
which is consistent with the constraint obtained with the CMB
data. Indeed, the values we obtained for $\Omega_b h^2$ are marginally
consistent with the value inferred from deuterium abundance BBN
constraints within $1 \sigma$, while there is full consistency with
value inferred from the primordial abundance of $^4{\rm He}$. On the
other hand, the values of $H_0$ obtained in our study are marginally
consistent with the constraint extracted from Hubble Space Telescope
(HST) data.

Let us now discuss models included in case ii). The values of
$\Omega_b h^2$ obtained are consistent with the values obtained by the
WMAP collaboration in the context of the standard cosmological model
within $2 \sigma$. On the other hand, there is disagreement with the
deuterium inferred BBN constraint on $\Omega_b h^2$ within $3 \sigma$,
while there is consistency with the value inferred from the $^4{\rm
  He}$ abundance. The values of $H_0$ obtained are at variance with
the ones obtained by the WMAP collaboration within $1 \sigma$, while
there is agreement within $2 \sigma$ and with the constraints obtained
with the HST data. The values of $\Theta$ also differ at
1$\sigma$. Other values of cosmological parameters obtained do agree
with the ones obtained by the WMAP collaboration in the context of the
standard inflationary scenario.  On the other hand, for both models,
we found that there is no agreement with the values of $A_s$ obtained
in almost all of the cases of fixed values of $A$ explored. This can
be understood as due to the fact that the effective amplitude of the
power spectrum of density fluctuations depends on $A$ and $A_s$, so
the parameters act at a certain level as degenerated
parameters. However, there is agreement between the values for
$\sigma_8$ \footnote{$\sigma_8$ is the rms mass fluctuation amplitude in spheres of size 8 $h^{-1}$ Mpc.} obtained and those of the fiducial model.  Note, however,
that this is a derived parameter which depends on the effective
amplitude. It should also be noted that the models included in case i)
are preferred by the data as compared to the models included in case
ii), as can be inferred from the value of $\chi^2_{\rm min}$ of the
respective statistical analyses listed in Tables~\ref{tablam1},
\ref{tablam2neg} and \ref{tablam2pos}.


When analyzing the correlations between the cosmological parameters
$\Omega_b h^2$, $\Omega_c h^2$ and $\theta$, $\tau$ and $B$, we find
that for both Models I and II the values of the correlation
coefficient $\rho_{B \theta}$, $\rho_{B \Omega_b h^2}$ and $\rho_{B
  H_0}$ are positive for positive best-fit values of $B$, while
$\rho_{B \Omega_c h^2}$ is positive for negative best-fit values of
$B$. Furthermore, the correlation is strong for $\Omega_b h^2$,
$\Omega_c h^2$ and $H_0$ while for $\theta$ and $\tau$ the correlation
coefficient is less than $0.3$ for all the cases studied in this
paper.

\section{Summary and Conclusions}
\label{discusion}

In this paper, we have studied the phenomenological predictions of the collapse models developed  in Refs. \cite{Perez2006,Sudarsky08,Unanue2008} and incorporated  the effect of late-time plasma physics. We have
calculated the prediction for the CMB fluctuation spectrum in the case
where the collapse of the $k$ modes of the scalar field during the
inflationary period is included. Furthermore, we have performed a
statistical analysis in order to compare the prediction of the
collapse models with recent data from the CMB fluctuation spectrum and
the matter power spectrum obtained by the SDSS collaboration, setting
bounds on the model parameters $A$ and $B$ that characterize the
collapse times of the scalar field modes. Results from the statistical
analyses were discussed in Sect. \ref{resultados} and can be
summarized as follows:

\begin{itemize} 
\item{For $|A|>20$ in Model I, any value of $B$ provides a  good fit to the data within the range studied in this paper ($B= -10^9 \cdots 10^9$).}
\item{For $|A| < 20$ in Model I, there is a nontrivial range of
  values of $B$ that provide a good fit to the data. However, values
  of $A$ lying in the range: $10^{-4}<A<20$ are not viable since the
  collapse of the modes occurs after the inflationary
  period. Furthermore, the data indicate that the values in the range
  $-20<A<-1$ are preferred.}
\item{The best fit values of $B$ obtained for $|A|<20$ in Model I lie in the range   $B < 1.88 \,\ {\rm Mpc}$.}
\item{In Model II, there is a nontrivial range of values of $B$ that provide a good fit to the data for all values of $A$ tested ($A= -10^{9} \cdots 10^9$). However, values of $A$ lying in the range: $10^{-4}<A<10^9$ are not viable since the collapse of the modes occurs after the inflationary period. Furthermore, the data indicate that the values in  the range  $|A|>1$ are preferred.}
\item{The best fit values of $B$ obtained for all values of $A$ studied in this paper for Model II  lie in the range   $B < 1.88 \,\ {\rm Mpc}$.}
\item{The value obtained for the cosmological parameters is consistent within $3 \sigma$ with those obtained by the WMAP collaboration assuming a standard inflationary scenario and also with bounds established by BBN and the value of $H_0$ obtained with the HST.}
\end{itemize}

This analysis allows us to  compare the value of the scale factor at the collapse time
$a(\eta_k^c)$, with the traditional value of the scale factor at ``horizon
crossing'' which is often set to  mark the  {\it quantum to classical transition} in the
standard explanation of inflation: $a_k^H$. The ``horizon crossing''
occurs when the length corresponding to the mode $k$ has the same size
that the Hubble radius $H_I^{-1}$ (in comoving modes, $k=aH_I$)
therefore, $a_k^H \equiv a(\eta_k^H) = \frac{k}{H_I} = \frac{3k}{8\pi
  G V}$. Thus the ratio of the value of the scale factor at horizon
crossing for mode $k$ and its value at collapse time is
\begin{equation}
  \frac{a^H_k}{a^c_k} = {k\eta_k^c(k)} = A + B  k.
\end{equation}

The results discussed in Sect. \ref{resultados} show that two cases
are consistent with the data: the one for which the collapse time of
inflaton field modes is previous to the time of ``horizon crossing''
of all modes of the inflaton field, and the opposite case. However,
the data seem to favor the case in which the collapse happens before
 the modes cross the horizon.

As we found that there are good fits (for suitable ranges of $B$) for
all values of the parameter $A$, we must conclude that the existing
data do not lead to a preferred scale for the times of collapse of the
wave function for the relevant modes.  However, for each value of $A$,
interesting bounds can be placed on the value of the parameter $B$
which in this approach characterizes the modifications of the
primordial spectral shape with respect to the HZ conventional flat
scale-free spectrum.

\section*{{\bf Acknowledgments}}

The authors are grateful to Gabriel Le\'on for useful discussions about the collapse models. Numerical calculations were performed with the KANBALAM facility located at UNAM. The authors would like to thank the people of DGSCA-UNAM for computational and technical support. 
 S. J. L. is supported by PICT 2007-02184 from Agencia Nacional de Promoci\'{o}n Cient\'{i}fica y Tecnol\'{o}gica, Argentina and by PIP N 11220090100152 from Consejo Nacional de Investigaciones Cient\'{\i}ficas y T\'{e}cnicas, Argentina. 
This work has been partially funded by project AYA2010-21766-C03-02 of
 the Spanish Ministry of Science and Innovation (MICINN) and  by  Grant  No  101712  from CONACYT (M\'exico).

\bibliography{bibliografia3}
\bibliographystyle{apsrev}

\end{document}